\documentclass[aps,prb,twocolumn,superscriptaddress,10pt]{revtex4}

\usepackage{amsmath, amssymb, mathtools}
\usepackage[version=4]{mhchem}
\usepackage{graphicx}
\usepackage{epsfig}
\usepackage{dcolumn}
\usepackage{bm}
\usepackage{blindtext}
\usepackage{natbib}
\usepackage{siunitx}
\usepackage{multirow}
\usepackage{bbm}
\usepackage[usenames,dvipsnames]{xcolor}
\usepackage{amsthm}
\usepackage{booktabs}
\usepackage{tikz}
\usetikzlibrary{calc}

\usepackage{hyperref}

\def\nn{\nonumber}

\usepackage{mathtools}

\def\be{\begin{equation}} \def\ee{\end{equation}}
\def\bea{\begin{eqnarray}} \def\eea{\end{eqnarray}}

\begin{document}
\title{
Renormalization group analysis for bosonization coefficients in half-odd-integer Kitaev  spin chains 
}

\author{Jianxun Li}
\affiliation{School of Physics, Nankai University, Tianjin, 300071, China }

\author{Chao Xu}
\affiliation{Kavli Institute for Theoretical Sciences, University of Chinese Academy of Sciences, Beijing 100190, China }

\author{Wang Yang}
\email{wyang@nankai.edu.cn}
\affiliation{School of Physics, Nankai University, Tianjin, 300071, China }

\begin{abstract}

Based on a renormalization-group (RG) analysis, we study the bosonization formulas for spin-$S$ Kitaev-Gamma and Kitaev-Heisenberg-Gamma chains in the parameter regime $K<0$, $\Gamma>0$, and $J>0$, with $S$ being a half-odd integer.
We find that the effects associated with the breaking of emergent continuous symmetries in the bosonization formulas scale as $1/S$ in the large-$S$ limit, in qualitative agreement with density-matrix renormalization-group (DMRG) results for Kitaev-Gamma chains.
For Kitaev-Heisenberg-Gamma chains, the low-energy excitations are characterized by incommensurate gapless wave vectors, and symmetry analysis yields ten independent bosonization coefficients, five of which are predicted by the RG analysis to be independent of the Heisenberg coupling up to linear order.
Our results provide useful input for determining magnetic ordering tendencies in two-dimensional Kitaev spin models within a quasi-one-dimensional approach.

\end{abstract}
\maketitle

\section{Introduction}

Kitaev materials are a major topic in frustrated quantum magnetism because of their potential to host exotic quantum states \cite{Kitaev2006,Balents2010,WitczakKrempa2014,Rau2016,Winter2017,Zhou2017,Savary2017,Trebst2022,Chou2025}. 
Representative spin-$1/2$ candidates include $\alpha$-RuCl$_3$,
Na$_2$IrO$_3$, and several polymorphs of Li$_2$IrO$_3$
\cite{Kocsis2022,Ran2022,Zhao2022,Wolf2022,Imamura2024,Bastien2022,Breznay2017,Shen2022Li2IrO3,Halloran2022,Perreault2015,Sanders2022,Kruger2023,Shen2025Li2RhO3,Lee2022K2IrCl6,Ishikawa2022,Onuorah2024,Kao2026VBr3,Voleti2024}.
Beyond the spin-$1/2$ setting, higher-spin Kitaev materials have also attracted growing interest. 
Examples include Co-based honeycomb cobaltates with high-spin $d^7$ ions, as well as Cr-based van der Waals compounds and related monolayers, where Kitaev-type interactions have been proposed for spin-$3/2$ moments \cite{Xu2018,Xu2020,Lee2020,Stavropoulos2021,Jin2022}.
Since real materials generally deviate from the ideal Kitaev limit, generalized Kitaev models are widely used as effective descriptions \cite{Jackeli2009,Rau2014,Chaloupka2010,Chaloupka2013,Chaloupka2015,Cen2022,Liu2022Exchange,Ran2017,Wang2017,Zhao2025}. 
Besides the bond-dependent Kitaev exchange, these models typically include off-diagonal Gamma interactions, Heisenberg exchange, and further-neighbor couplings.
Among them, the Kitaev-Gamma and Kitaev-Heisenberg-Gamma models are
canonical minimal extensions of the Kitaev model
\cite{Rau2014,Chaloupka2015,Stavropoulos2018,Wang2019PKSL,
Knolle2018,Buessen2021,Zhang2021,Rayyan2021,Nanda2020,Nanda2021,
Stavropoulos2024};
other extensions and related perturbed Kitaev models have also been
studied extensively
\cite{WangLiu2023,Joy2022,Fukui2023,Singh2023,Freitas2024,Chen2023,
Jin2024,Keskiner2023,Luo2025,Janssen2025,Schwenke2025,Li2026}.

In recent years, one-dimensional and quasi-one-dimensional generalized Kitaev models have attracted considerable attention \cite{Sela2014,Agrapidis2019,Yang2020PRL,Yang2020PRR,Yang2022PRB_RG,Yang2021PRBChain,Yang2022BondAlt,Sorensen2021,Churchill2024,Sorensen2024,Gohlke2024,Yang2024Nonsymmorphic,Yang2025DM,Yang2026_Efield,Zhuang2026}. 
Besides offering a tractable route to understanding aspects of two-dimensional Kitaev physics, they also exhibit a variety of exotic phenomena in their own right.  
Previous studies have uncovered rich physics in these systems, including emergent conformal symmetry, spontaneous breaking of nonsymmorphic symmetries, nonlocal string order, spin nematicity, soliton excitations, topological order, and topological phase transitions
\cite{Sela2014,Agrapidis2019,Yang2020PRL,Yang2020PRR,
Yang2021PRBChain,Yang2022BondAlt,Sorensen2021,Sorensen2024,
Yang2024Nonsymmorphic,Yang2025DM}.
These results establish one-dimensional generalized Kitaev systems as an important platform for exploring the interplay among frustration, symmetry, topology, and quantum fluctuations.

In the spin-$1/2$ Kitaev-Gamma model, a gapless phase has been identified in the region with $K<0$, whose low-energy physics is described by an emergent SU(2)$_1$ Wess-Zumino-Witten (WZW) model \cite{Yang2020PRL}. 
Accordingly, the effective low-energy theory exhibits an emergent continuous SU(2) symmetry, and operators that break this emergent symmetry are irrelevant in the sense of renormalization-group (RG).
It is important to emphasize that the exact symmetries of the system are discrete and the continuous SU(2) symmetry emerges only in the low energy limit.
Nevertheless, the exact discrete nonsymmorphic symmetries of the underlying lattice model still leave an imprint on the bosonized expressions of the spin operators:
The bosonization formulas only respect the exact microscopic symmetries of the model rather than the emergent SU(2) symmetry \cite{Yang2020PRL,Yang2025DM}. 
Such SU(2)-breaking nonsymmorphic bosonization formulas play an important role in determining the magnetic ordering tendencies in two dimensions within the coupled-chain approach.

The general structure of the bosonization formulas subject to nonsymmorphic symmetry can be fixed by symmetry considerations. 
However, symmetry alone only imposes constraints among the bosonization coefficients, and cannot determine their signs or magnitudes. 
In the spin-$1/2$ Kitaev-Gamma model, these coefficients were further estimated using an RG analysis, yielding predictions that are in qualitative agreement with numerical results \cite{Yang2020PRL}. 
The central idea is that, at the microscopic lattice scale, SU(2)-breaking interactions in the Hamiltonian induce wavefunction renormalization of the spin operators. 
Such effects originate from short-wavelength physics and can therefore be captured by the RG flow in the ultraviolet regime \cite{Yang2020PRL}. 

\begin{figure}
\includegraphics[width=0.35\textwidth]{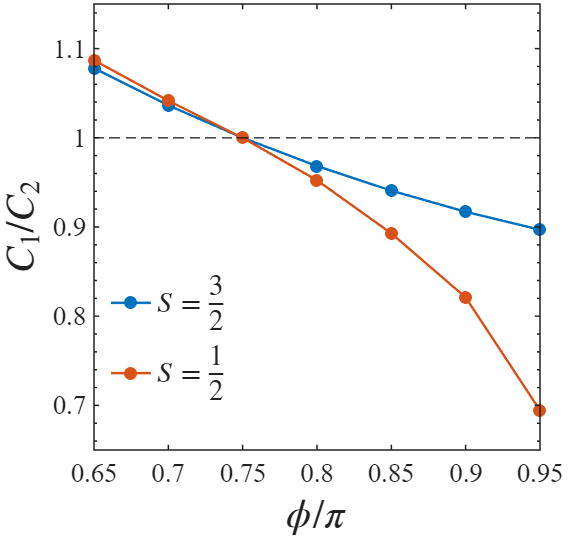}
\caption{$C_1/C_2$ as a function of $\phi$ for $S=3/2$  (blue curve) and $S=1/2$ (orange curve) in Kitaev-Gamma chains,
where $K=\cos(\phi)$, $\Gamma=\sin(\phi)$.
DMRG numerics are performed on systems of $L=144$ sites under periodic boundary conditions.
Bond dimension $m$ and truncation error $\epsilon$ in DMRG numerics are taken as $m=3000$, $\epsilon=10^{-7}$.
}
\label{fig:C12_phi}
\end{figure}

In this work, we extend the above RG analysis of the nonsymmorphic bosonization coefficients in the $K<0$ region of the spin-$1/2$ Kitaev-Gamma model in two directions.
Explicit expressions of the bosonization coefficients are obtained up to linear order in $|K|-|\Gamma|$ and $J$,
which apply in the vicinity of the hidden SU(2) symmetric point $(K=\pm \Gamma,J=0)$ \cite{Yang2020PRL}.
The first direction  concerns the generalization to general spin values for half-odd-integer spins. 
RG analysis shows that the degree of the breaking of the emergent continuous symmetry exhibits a characteristic $1/S$ scaling with the spin quantum number $S$.
The breaking of emergent SU(2) symmetry in Kitaev-Gamma chains can be characterized by the difference between the ratio $C_1/C_2$ of the two bosonization coefficients and unity.  
As can be seen from Fig. \ref{fig:C12_phi}, our large-scale  density matrix renormalization group (DMRG) numerical results 
show that the deviation of $C_1/C_2$ away from unity (black dashed line) in the spin-$3/2$ case (blue line) is less significant than the spin-$1/2$ case (orange line),
which is in qualitative agreement with the RG predictions. 

The second extension concerns the Kitaev-Heisenberg-Gamma model. 
Previous studies have shown that, in the parameter regime $(K<0,\Gamma>0,J>0)$, this model possesses an emergent U(1) symmetry and its low-energy physics is described by the Luttinger liquid theory \cite{Yang2020PRR,Yang2024Nonsymmorphic,Yang2026_Efield}. 
Using the exact microscopic symmetries, we first determine the general form of the corresponding nonsymmorphic bosonization formulas, which involve incommensurate gapless wave vectors and ten independent bosonization coefficients.
We then employ an RG approach to predict the signs and magnitudes of these coefficients by computing the wavefunction renormalization effects induced by short-wavelength fluctuations. 
In particular, we find that half of these coefficients are independent of the Heisenberg coupling to linear order. 
However, we are not able to reliably  extract the ten bosonization coefficients in DMRG numerics, possibly because of large finite size effects. 
Compared with the Kitaev-Gamma model, the Kitaev-Heisenberg-Gamma model is more realistic, and the coupling regime $(K<0,\Gamma>0,J>0)$ is relevant to iridate Kitaev materials \cite{Chaloupka2015,Stavropoulos2018,Buessen2021,Rayyan2021}. 
It is therefore expected that our results can provide useful input for understanding magnetic ordering tendencies in iridates within the coupled-chain approach.

The rest of the paper is organized as follows. 
Sec. \ref{Model Hamiltonians} defines the Hamiltonians of the systems. 
Sec. \ref{sec:review_spin_S} gives a brief review on how the SU(2)$_1$ low energy theory is derived for antiferromagnetic half-odd-integer Heisenberg spin chain.
Sec. \ref{sec:Low_KG} shows that the low energy physics of  spin-$S$ Kitaev-Gamma chains can be described by the SU(2)$_1$ WZW model for $K<0$ and half-odd-integer spin values.
In Sec. \ref{sec:RG_KG}, the SU(2) breaking bosonization coefficients in half-odd-integer-spin Kitaev-Gamma chains are analyzed. 
Sec. \ref{sec:Low_KHG} shows that the low energy physics of  spin-$S$ Kitaev-Heisenberg-Gamma chains can be described by the Luttinger liquid theory
and derives  the nonsymmorphic incommensurate bosonization formulas for the Kitaev-Heisenberg-Gamma chains. 
In Sec. \ref{sec:RG_KHG}, the U(1) breaking bosonization coefficients in half-odd-integer-spin Kitaev-Heisenberg-Gamma chains are analyzed. 
Sec. \ref{sec:summary} summarizes the main results of the paper. 

\section{Model Hamiltonians }
\label{Model Hamiltonians}

In this section, we define  the model Hamiltonians for spin-$S$ Kitaev-Gamma and  Kitaev-Heisenberg-Gamma chains,
and briefly summarize the main RG results in this work.

\begin{figure}
\includegraphics[width=0.45\textwidth]{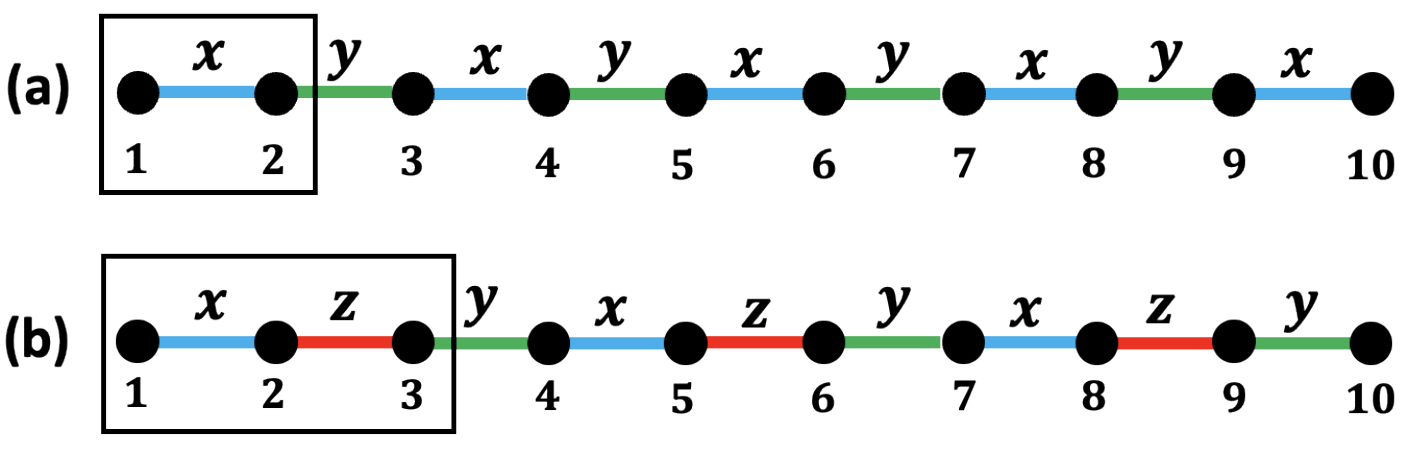}
\caption{Bond structures (a) before and (b) after the six-sublattice rotation.
The rectangular boxes denote unit cells.
}
\label{fig:bonds}
\end{figure}

\subsection{Hamiltonians and sublattice rotation}

\subsubsection{Spin-$S$ Kitaev-Gamma chain}

The spin-$S$ Kitaev-Gamma chain is defined as
\bea
H_{K\Gamma} = \sum_{\langle ij\rangle\in\gamma\text{-bond}}
\bigl[
  K S_i^\gamma S_j^\gamma
  + \Gamma (S_i^\alpha S_j^\beta + S_i^\beta S_j^\alpha)
\bigr] ,
\label{eq:KG_Ham}
\eea
in which $\gamma\in\{x,y\}$ denotes the spin direction associated with the bond connecting nearest-neighbor sites $i$ and $j$ shown in Fig. \ref{fig:bonds} (a);
$\alpha,\beta$ are the two spin directions other than $\gamma$;
and $\vec{S}_i$ are spin-$S$ operators at site $i$.
It is convenient to parametrize $K,\Gamma$ as 
\bea
K&=&\cos(\phi),\nn\\
\Gamma&=&\sin(\phi).
\eea
A global spin rotation by $\pi$ around the $z$-axis flips the sign of $\Gamma$ without changing the physics. 
As a result, there is the equivalence relation
\bea
(K,\Gamma)\simeq (K,-\Gamma),
\label{eq:equiv_KG}
\eea
where ``$\simeq$" denotes ``unitarily equivalent".
Because of Eq. (\ref{eq:equiv_KG}), it is enough to restrict to the parameter region $\Gamma>0$.
We will be interested in the $(K<0,\Gamma>0)$ region, corresponding to $\phi\in(\pi/2,\pi)$.

A useful unitary transformation, dubbed six-sublattice rotation $U_6$, 
is defined as
\begin{align}
\text{sublattice }1:\quad
& (S_i^{\prime x},S_i^{\prime y},S_i^{\prime z})
  = (-S_i^{x},-S_i^{y},S_i^{z}),\notag \\
\text{sublattice }2:\quad
& (S_i^{\prime x},S_i^{\prime y},S_i^{\prime z})
  = (S_i^{x},-S_i^{z},S_i^{y}),\notag \\
\text{sublattice }3:\quad
& (S_i^{\prime x},S_i^{\prime y},S_i^{\prime z})
  = (-S_i^{y},S_i^{z},-S_i^{x}),\notag \\
\text{sublattice }4:\quad
& (S_i^{\prime x},S_i^{\prime y},S_i^{\prime z})
  = (S_i^{y},S_i^{x},-S_i^{z}),\notag\\
\text{sublattice }5:\quad
& (S_i^{\prime x},S_i^{\prime y},S_i^{\prime z})
  = (S_i^{z},-S_i^{x},-S_i^{y}),\notag \\
\text{sublattice }6:\quad
& (S_i^{\prime x},S_i^{\prime y},S_i^{\prime z})
  = (-S_i^{z},S_i^{y},S_i^{x}),
 \label{eq:U6}
\end{align}
in which $S_i^\alpha$ and $S_i^{\prime\alpha}$ denote the spin operators before and after the $U_6$ transformation, respectively;
and $i\equiv j\pmod 6$, where $i$ is the site index and $j\in\{1,\dots,6\}$ is the sublattice index. 
The $U_6$ transformation maps the Hamiltonian $H_{K\Gamma}$ in Eq. (\ref{eq:KG_Ham}) to the following transformed form
\bea
H^\prime_{K\Gamma} = \sum_{\langle ij\rangle\in\gamma\text{-bond}}
[
  -K S_i^{\prime\gamma} S_j^{\prime\gamma}
 +\Gamma (S_i^{\prime\alpha} S_j^{\prime\alpha} + S_i^{\prime\beta} S_j^{\prime\beta})
],
\label{eq:KG_Ham_prime}
\eea
in which the pattern of bond $\gamma$ has a three-site periodicity shown in Fig. \ref{fig:bonds} (b).

\subsubsection{Spin-$S$ Kitaev-Heisenberg-Gamma chain}

The spin-$S$ Kitaev-Heisenberg-Gamma chain is defined as
\bea
H_{KJ\Gamma} &=& H_{K\Gamma}+\sum_{<ij>\in \gamma\,\text{bond}} J\vec{S}_i\cdot \vec{S}_j,
\label{eq:KHG_Ham}
\eea
in which the pattern for bond $\gamma$ is shown in Fig. \ref{fig:bonds} (a). 
The six-sublattice rotation transforms the Hamiltonian to the following form 
\begin{flalign}
&H^\prime_{KJ\Gamma}=H^\prime_{K\Gamma} -\nn\\
&\sum_{<ij>\in \gamma\,\text{bond}}J(S_i^{\prime\gamma} S_j^{\prime\gamma}+S_i^{\prime\alpha} S_j^{\prime\beta}+S_i^{\prime\beta} S_j^{\prime\alpha})\big],
\label{eq:6rotated_KJG}
\end{flalign}
in which the pattern for bond $\gamma$ is shown in Fig. \ref{fig:bonds} (b). 
It is convenient to parametrize $K,J,\Gamma$ as 
\bea
J&=&\cos(\theta),\nn\\
K&=&\sin(\theta)\cos(\phi),\nn\\
\Gamma&=&\sin(\theta)\sin(\phi).
\eea
Again, a global spin rotation by $\pi$ around $z$-direction establishes the following equivalence relation, 
\bea
(K,J,\Gamma)\simeq (K,J,-\Gamma).
\label{eq:equiv_KJG}
\eea

\subsection{Nonsymmorphic bosonization formulas}

In this section, we briefly summarize the main RG results on nonsymmorphic bosonization formulas, leaving detailed derivations  for later sections. 

\subsubsection{Kitaev-Gamma chains}

The bosonization formulas for spin-$S$ Kitaev-Gamma chains in the six-sublattice rotated frame take the following form ($1\leq j \leq 3$)
\bea
\frac{1}{a} S_{j+3n}^{\prime\alpha}
&=& D^{\alpha}_{j}\,\big(J_L^{\prime\alpha}(x)+J_R^{\prime\alpha}(x)\big)  \nonumber \\
&&\quad +(-1)^{j+n}\,C^{\alpha}_{j}\,\frac{1}{\sqrt a}\,
  i\operatorname{tr}\big[g(x)\sigma^\alpha\big],  
  \label{S_c}
\eea
in which $J_L^{\prime\alpha}(x)$ and $J_R^{\prime\alpha}(x)$ ($\alpha\in\{x,y,z\}$) are left and right WZW currents,
$g(x)$ is the WZW primary field,
$a$ is the lattice constant, $x=(j+3n)a$ is the spatial coordinate in the continuum limit,
and the $C^\alpha_j,D^\alpha_j$ coefficients are given by
\begin{flalign}
&D_1^z=D_2^y=D_3^x(=D_1)\nn\\
&D_1^x=D_1^y=D_2^x=D_2^z=D_3^y=D_3^z(=D_2)\nn\\
&C_1^z=C_2^y=C_3^x(=C_1)\nn\\
&C_1^x=C_1^y=C_2^x=C_2^z=C_3^y=C_3^z(=C_2).
\label{eq:CD_relations}
\end{flalign}
Close to the high-symmetry point $K=-\Gamma$ within the
$(K<0,\Gamma>0)$ region, the RG analysis predicts to linear order in an expansion over $|K|-\Gamma$ that
\bea
\frac{C_1}{C_2}&=&1-\ln b_s\frac{0.1\pi}{S+1}  \frac{\Delta_\Gamma}{\Gamma} \nn\\
\frac{D_1}{D_2}&=&1+\ln b_s\frac{0.21\pi}{S+1}  \frac{\Delta_\Gamma}{\Gamma} ,
\label{eq:1overS_CD_predict}
\eea
in which 
\bea
\Delta_\Gamma=-K-\Gamma=|K|-\Gamma, 
\eea
$b_s\sim 3$ is an RG stopping scale, and the  $O(\Delta_\Gamma^2)$ terms are neglected. 

\subsubsection{Kitaev-Heisenberg-Gamma chains}

The bosonization formulas for spin-$S$ Kitaev-Heisenberg-Gamma chains in the six-sublattice rotated frame are given by
\begin{flalign}
&(S^{\prime x}_{i+3n}~S^{\prime y}_{i+3n}~S^{\prime z}_{i+3n})=\nn\\
&(\mathcal{J}^{ x}~\mathcal{J}^{ y}~\mathcal{J}^{ z})Q(x)F_iO^{-1}
+(-)^{i+n}(\mathcal{N}^{ x}~\mathcal{N}^{ y}~\mathcal{N}^{ z})Q(x)E_iO^{-1},
\label{eq:abelian_LL1_matrix}
\end{flalign}
in which $\mathcal{J}^{\alpha},\mathcal{N}^{\alpha}$ ($\alpha=x,y,z$) are defined as
\bea
\mathcal{J}^{ x}&=&\frac{1}{a}\cos(\sqrt{4\pi} \varphi) \cos(\sqrt{\pi} \theta),\nn\\
\mathcal{J}^{ y}&=&\frac{1}{a}\cos(\sqrt{4\pi} \varphi) \sin(\sqrt{\pi} \theta),\nn\\
\mathcal{J}^{ z}&=&-\frac{1}{\sqrt{\pi}}\nabla \varphi,
\label{eq:bosonize_J}
\eea
and 
\bea
\mathcal{N}^{ x}&=&\frac{1}{a}\cos(\sqrt{\pi}\theta),\nn\\
\mathcal{N}^{ y}&=&\frac{1}{a}\sin(\sqrt{\pi}\theta),\nn\\
\mathcal{N}^{ z}&=&\frac{1}{a}\sin(\sqrt{4\pi}\varphi),
\label{eq:bosonize_N}
\eea
where the fields $\theta,\varphi$ satisfy $[\varphi(x),\theta(x^\prime)]=\frac{i}{2}\text{sgn}(x^\prime-x)$. 

In Eq. (\ref{eq:abelian_LL1_matrix}), the $3\times 3$ position-dependent matrix $Q(x)$ is defined as
\bea
Q(x)=\begin{pmatrix}
\cos(q_{\text{ic}}x) & -\sin(q_{\text{ic}}x) & 0\\
\sin(q_{\text{ic}}x) & \cos(q_{\text{ic}}x) & 0\\
0 & 0 & 1
\end{pmatrix},
\label{eq:define_Q_mat_a}
\eea
in which $x=(i+3n)a$ where $i+3n$ is the numbering of the lattice site and $a$ is the lattice constant, and $q_{\text{ic}}$ is an incommensurate wave vector determined by the microscopic parameters of the system as
\bea
q_{\text{ic}}=4\pi\sqrt{\frac{3}{2}}\frac{J}{v}[(C_2)^2-(C_1)^2],
\eea
where $v$ is spin velocity, and $C_1$, $C_2$ are the bosonization coefficients in the Kitaev-Gamma model. 
The constant matrices $E_i,F_i$ ($i=1,2,3$) are given by
\bea
E_2&=&\left(\begin{array}{ccc}
\lambda_C & 0 & \sigma_C\\
0 & \lambda_C+\delta_C & 0\\
\rho_C&0&\nu_C
\end{array}
\right),\nn\\
F_2&=&\left(\begin{array}{ccc}
\lambda_D & 0 & \sigma_D\\
0 & \lambda_D+\delta_D & 0\\
\rho_D&0&\nu_D
\end{array}
\right),
\label{eq:E2_F2}
\eea
and
\bea
E_1&=&M_z^{-1}E_2M_z,\nn\\
E_3&=&M_zE_2M_z^{-1},\nn\\
F_1&=&M_z^{-1}F_2M_z,\nn\\
F_3&=&M_zF_2M_z^{-1},
\label{eq:EF_mat_13}
\eea
in which $M_z$ is the $3\times 3$ rotation matrix around $z$-direction by $-2\pi/3$, i.e., 
\bea
M_z=\left(\begin{array}{ccc}
-\frac{1}{2} & \frac{\sqrt{3}}{2} & 0\\
-\frac{\sqrt{3}}{2} & -\frac{1}{2} & 0\\
0&0&1
\end{array}
\right).
\label{eq:Mz_mat}
\eea
The matrix $O$ in Eq. (\ref{eq:abelian_LL1_matrix}) is given by
\begin{equation}
O =
\begin{pmatrix}
-\frac{1}{\sqrt6} & -\frac{1}{\sqrt2} & \frac{1}{\sqrt3} \\
\sqrt{\dfrac{2}{3}} & 0 & \frac{1}{\sqrt3} \\
-\frac{1}{\sqrt6} & \frac{1}{\sqrt2} & \frac{1}{\sqrt3}
\end{pmatrix}.
\label{eq:def_O_mat}
\end{equation}

Close to the high-symmetry point $\Delta_\Gamma=0$, $J=0$, and for
sufficiently small positive $J/\Gamma$, the RG analysis predicts to linear
order in $\Delta_\Gamma/\Gamma$ and $J/\Gamma$ that
\bea
\lambda_C&=&b\left[1- \ln b_s\frac{\pi}{S+1} \left(\frac{0.26}{3}\frac{\Delta_\Gamma}{\Gamma}-0.10\frac{J}{\Gamma}\right) \right]\nn\\
\sigma_C&=&-b\ln b_s\frac{0.10\pi}{S+1}\frac{\sqrt{2}}{3}  \frac{\Delta_\Gamma}{\Gamma} \nn\\
\delta_C&=&b\ln b_s\frac{0.10\pi}{S+1} \frac{2}{3}\frac{\Delta_\Gamma-3J}{\Gamma}\nn\\
\nu_C&=& b\left[1-\ln b_s\frac{0.08\pi}{S+1} \left(\frac{2}{3}  \frac{\Delta_\Gamma}{\Gamma}-2\frac{J}{\Gamma}\right) \right]\nn\\
\rho_C&=&\sigma_C,
\label{eq:U1_boson_C_expr_b}
\eea
and
\bea
\lambda_D&=&b\left[1+\ln b_s\frac{\pi}{S+1} \left(0.07\frac{\Delta_\Gamma}{\Gamma}-0.21\frac{J}{\Gamma}\right) \right]\nn\\
\sigma_D&=&b\ln b_s\frac{0.21\pi}{S+1}\frac{\sqrt{2}}{3}  \frac{\Delta_\Gamma}{\Gamma}\nn\\
\delta_D&=&-b\ln b_s\frac{0.21\pi}{S+1} \frac{2}{3}\frac{\Delta_\Gamma-3J}{\Gamma} \nn\\
\nu_D&=& b\nn\\
\rho_D&=&\sigma_D,
\label{eq:U1_boson_D_expr_b}
\eea
in which $\Delta_\Gamma=|K|-\Gamma$, $b_s\sim 3$ is an RG stopping scale, $b$ is an overall infrared scale in RG flow which can be fixed by proper normalizations of the current and N\'eel fluctuation operators in Eqs. (\ref{eq:bosonize_J},\ref{eq:bosonize_N}),  and the $O(\Delta_\Gamma^2,J^2,\Delta_\Gamma J)$ terms are neglected. 
Notice in particular that up to linear order in $J$, five of the couplings $\sigma_C$, $\rho_C$, $\sigma_D$, $\rho_D$, $\nu_D$ do not depend on $J$. 

\section{Review of spin-$S$ Heisenberg chain}
\label{sec:review_spin_S}

In this section, we briefly review how  the SU(2)$_1$ low energy field theory  of spin-$S$ antiferromagnetic  (AFM) Heisenberg chain is derived for half-odd-integer spin values,
which provides a perturbative starting point for an RG analysis of bosonization coefficients in generalized Kitaev spin chains to be discussed in later sections. 
The spin value $S$ throughout this work takes half-odd-integer value unless otherwise stated. 

\subsection{Large-$U$ limit of multi-orbital Hubbard model}
\label{subsec:large_U}

We consider an electronic lattice model whose low-energy sector realizes a spin-$S$ degree of freedom on each site. To this end, we introduce
$n_c=2S$ electrons per site and define the Hamiltonian as \cite{AffleckHaldane1987}
\begin{equation}
H_f = H_t + H_U ,
\label{eq:fermion}
\end{equation}
in which the nearest-neighbor hopping $H_t$ and on-site Hund's coupling $H_U$ are given by
\begin{eqnarray}
H_t &=& -t \sum_{\langle ij\rangle}\sum_{a=1}^{n_c}\sum_{\sigma}\left( c_{ia\sigma}^\dagger c_{ja\sigma} + \mathrm{h.c.} \right)
-\mu \sum_{ia\sigma} c_{ia\sigma}^\dagger c_{ia\sigma}\nn\\
H_U &=& -U \sum_i \vec{S}_i^2 .
\end{eqnarray}
Here $c_{ia\sigma}$ annihilates an electron on site $i$ with color index $a=1,\dots,n_c$ and spin $\sigma$, and
\begin{equation}
\vec{S}_i
=
\frac12
\sum_{a=1}^{n_c}\sum_{\alpha,\beta}
c_{ia\alpha}^\dagger \vec{\sigma}_{\alpha\beta} c_{ia\beta}
\end{equation}
is the total spin operator on site $i$. In the large-$U$ limit (i.e., $U\gg t$), the on-site Hund's term dominates at zeroth order in $t/U$. For fixed occupation $n_c=2S$, the Hund energy is minimized when the total spin is maximal, so the low-energy manifold on each site is the spin-$S$ multiplet.

Treating $H_t$ perturbatively, the leading nontrivial correction arises at second order and generates an antiferromagnetic Heisenberg exchange between neighboring sites,
\begin{equation}
H_{\mathrm{eff}}
=
J\sum_{\langle ij\rangle}\vec{S}_i\cdot\vec{S}_j + \mathrm{const.},
\label{eq:H_eff_Heisenberg}
\end{equation}
in which $J$ is given by
\begin{equation}
J=\frac{4t^2}{US(4S+1)}.
\label{eq:J_expresssion}
\end{equation}
Detailed derivation of Eq. (\ref{eq:H_eff_Heisenberg}) is included in Appendix \ref{app:2nd_perturbation}.


\subsection{Weak-$U$ limit of multi-orbital Hubbard model}
\label{subsec:weak_U_Hubbard}

In this section, following Ref. \onlinecite{AffleckHaldane1987}, we briefly review how the low energy theory of the spin-$S$ AFM Heisenberg model can be inferred by analyzing the fermion model in Eq. (\ref{eq:fermion}).

Assuming no phase transition occurs from weak-$U$ ($U\ll t$) to strong-$U$ ($U\gg t$),
the low energy physics of large-$U$ limit is therefore the same as that of the weak-$U$ limit. 
As a result, in virtue of  the discussions in Sec. \ref{subsec:large_U}, the low energy field theory of the spin-$S$ AFM Heisenberg model can be obtained from the weak-$U$ case. 
Here the crucial point is that the weak-$U$ limit is amenable for a perturbative analysis.
More specifically, the low-energy theory of the free-fermion Hamiltonian $H_t$ is a one-dimensional Lorentz-invariant fixed-point theory formulated in terms of left- and right-moving chiral fermions near the Fermi points, and the interaction term $H_U$ can be treated as a perturbation around this free fixed point in a controlled manner.

At half filling and in the continuum limit, the free theory is a set of massless Dirac fermions carrying U(1) charge, SU(2) spin, and SU($n_c$) color indices. 
The energy-momentum tensor $T_L$ in the left moving sector  for the free theory is given by (similar for the right-moving sector)
\begin{flalign}
T_L=\frac{\pi v}{2n_c} J_LJ_L +\frac{2\pi v}{n_c+2} \vec{J}_L\cdot \vec{J}_L+\frac{2\pi v}{n_c+2} \sum_{A=1}^{n_c^2-1} J_L^AJ_L^A,
\label{eq:T_L}
\end{flalign}
in which $J_L$, $\vec{J}_L$ and $J_L^A$ ($A=1,...,n_c^2-1$) are WZW currents in  the charge, spin and color sectors, respectively,
and $v=ta$ ($a$ is lattice constant). 
In particular, the spin currents obey an $\mathrm{SU}(2)$ Kac--Moody algebra at level
$k=n_c=2S$.
Therefore, if one temporarily neglects the interaction terms that mix the spin sector with the charge and color sectors, the spin part of the continuum theory is  the $\mathrm{SU}(2)_{n_c}$ WZW theory. 
On the other hand, it can be shown that the couplings in the charge and color sectors flow to the strong coupling limits under RG flow, opening gaps in these two sectors. 

The crucial point is that this decoupled $\mathrm{SU}(2)_{2S}$ description is not generically stable. 
Because of the couplings between the spin sector and the charge and color sectors,
spin-rotationally invariant terms are generated that can spoil the $\mathrm{SU}(2)_{n_c}$  WZW description. 
To understand the fate of the theory, Ref. \onlinecite{AffleckHaldane1987} considered the large-$n_c$ limit and parametrized the WZW field as
$g(x)=e^{\frac{i}{2}\,\vec{\sigma}\cdot\vec{\varphi}(x)}$, 
where $g(x)$ is the WZW primary field. 
For large $n_c$, a semiclassical analysis can be performed by minimizing the potential term for $\vec{\varphi}$ in the action. 
Since the minimum occurs near $|\vec{\varphi}|=\pi$, the radial fluctuation becomes massive, while the remaining low-energy degrees of freedom are the angular variables of the unit vector
$\hat{\boldsymbol{n}}=\vec{\varphi}/|\vec{\varphi}|$.
Freezing $|\vec{\varphi}|$, the $\mathrm{SU}(2)_{n_c}$ WZW action yields  the $O(3)$ nonlinear sigma model with a $\theta$-term,
\begin{equation}
\mathcal{L}_{\sigma}
=
\frac{1}{2g}\,(\partial_\mu \hat{\boldsymbol{n}})^2
+i\theta\,Q[\hat{\boldsymbol{n}}],
\label{eq:nonlinear_sigma}
\end{equation}
in which $g=\pi/S$, and the quantized term $Q[\hat{\boldsymbol{n}}]$ is given by
\begin{equation}
Q[\hat{\boldsymbol{n}}]
=
\frac{1}{8\pi}
\int d^2x\,
\epsilon^{\mu\nu}\,
\hat{\boldsymbol{n}}\cdot
\left(\partial_\mu \hat{\boldsymbol{n}}\times \partial_\nu \hat{\boldsymbol{n}}\right).
\end{equation}
The value of $\theta$ in Eq. (\ref{eq:nonlinear_sigma}) is $\theta=\pi k = 2\pi S$,
which is $\pi$ ($\mathrm{mod}\ 2\pi$) for half-odd-integer spins,
and $0$ ($\mathrm{mod}\ 2\pi$) for integer spins. 

The above analysis shows that all half-odd-integer-spin AFM Heisenberg chains flow to the same theory, namely, the $O(3)$ nonlinear sigma model with a $\theta=\pi$ term.  
Since the spin-$1/2$ case is solvable by Bethe ansatz and is known to be gapless with $\mathrm{SU}(2)_1$ critical behavior, 
we know that the low energy theory for all half-odd-integer AFM Heisenberg spin chains is described by the $\mathrm{SU}(2)_1$ WZW model. 

\subsection{Effective Heisenberg coupling in the weak-$U$ limit}
\label{subsec:Heisenberg_weak_U}

According to the discussions in Sec. \ref{subsec:weak_U_Hubbard}, the weak-$U$ fermion model can be used to mimic the spin-$S$ Heisenberg model, in the sense of low energy theory. 
However, the effective coupling $J_w$ of the Heisenberg model in the weak-$U$ limit can be different from $J$ in Eq. (\ref{eq:J_expresssion}) in the large-$U$ limit. 
In this section, we derive the scaling of $J_w$ as a function of  the spin value $S$ in the large-$S$ limit. 

Starting from the following spin-$S$ Heisenberg model  (where $S$ is a half-odd-integer)
\bea
H_{H}=J_w\sum \vec{S}_i\cdot \vec{S}_{i+1},
\label{eq:H_H}
\eea
it is standard to derive the O(3) nonlinear sigma model with a $\theta=\pi$ term in a semiclassical approach. 
The spin wave velocity $v_H$ in the O(3) nonlinear sigma model is given by 
\bea
v_H=2J_wSa.
\label{eq:v_H}
\eea
The form of $v_H$ can be understood as follows.
The Néel stiffness scales as
$\rho_s \sim J_w S^2 a$,
where $a$ is the lattice spacing, while the transverse uniform susceptibility scales as
$\chi_\perp \sim 1/(J_w a)$.
Hence, the spin wave velocity is  determined by
$v \sim \sqrt{\rho_s/\chi_\perp}
   \sim \sqrt{(J_w S^2 a)(J_w a)}
   \sim J_w S a$.
   
On  the other hand, the spinon velocity $v_f$ in the low energy SU(2)$_1$ WZW model in the weak-$U$ fermion model can be read from Eq. (\ref{eq:T_L}) as 
\bea
v_f=\frac{2\pi ta}{n_c+2}.
\eea
Therefore, if the weak-$U$ fermion model is able to produce the same low energy theory as the Heisenberg model in Eq. (\ref{eq:H_H}),
the two velocities $v_H$ and $v_f$ have to be equal,
which gives
\bea
J_w=\frac{\pi t}{2S(S+1)},
\label{eq:Jw_expression}
\eea
where $n_c=2S$ is used. 
This shows that  $J_w$ scales as $1/S^2$ in the large-$S$ limit.

\section{Low energy theory of  spin-$S$ Kitaev-Gamma  chain}
\label{sec:Low_KG}

In this section, we derive the low energy theory of spin-$S$ Kitaev-Gamma chain  in a perturbative way,
taking the SU(2)$_1$ WZW model for spin-$S$ AFM Heisenberg chains as the unperturbed theory,
where $S$ is a half-odd-integer number.

\subsection{Symmetries}

We consider the spin-$S$ Kitaev-Gamma chain in the $U_6$ frame, with Hamiltonian $H^\prime_{K\Gamma}$ given in Eq. (\ref{eq:KG_Ham_prime}).
For whatever spin value $S$, the symmetry group $G^\prime_{K\Gamma}$ of $H^\prime_{K\Gamma}$ is discrete and the same,
which has been analyzed for the spin-$1/2$ case to be a nonsymmorphic $O_h$ group in Ref. \onlinecite{Yang2020PRL}. 

The system is invariant under $T$, $R_aT_a$, $R_II$, $R(\hat{\alpha},\pi)$ ($\alpha=x,y,z$),
where $T$ is time reversal;
$T_{a}$ is the spatial translation operation for one lattice site;
$R_a$ is global spin rotation around $(1,1,1)$-direction by $2\pi/3$;
$I$ is spatial inversion with inversion center located at site $2$;
$R_I$ is global spin rotation around $(1,0,-1)$-direction by $\pi$;
and $R(\hat{\alpha},\pi)$ ($\alpha=x,y,z$) are global spin rotations around the $x$-, $y$-, and $z$-axes by angle $\pi$.
Explicit forms of these symmetry operations are given by
\begin{eqnarray}
1.&T &:  (S_i^x,S_i^y,S_i^z)\rightarrow (-S_{i}^x,-S_{i}^y,-S_{i}^z)\nn\\
2.&R_I I&: (S_i^x,S_i^y,S_i^z)\rightarrow (-S_{10-i}^z,-S_{10-i}^y,-S_{10-i}^x)\nn\\
3.& R_aT_a&:  (S_i^x,S_i^y,S_i^z)\rightarrow (S_{i+1}^z,S_{i+1}^x,S_{i+1}^y)\nn\\
4. & R(\hat{x},\pi) &: (S_i^x,S_i^y,S_i^z)\rightarrow (S_i^x,-S_i^y,-S_i^z)\nn\\
5. & R(\hat{y},\pi) &: (S_i^x,S_i^y,S_i^z)\rightarrow (-S_i^x,S_i^y,-S_i^z)\nn\\
6. & R(\hat{z},\pi) &: (S_i^x,S_i^y,S_i^z)\rightarrow (-S_i^x,-S_i^y,S_i^z).
\label{eq:sym_Jneq0}
\end{eqnarray}

We note that as discussed in Ref. \onlinecite{Yang2020PRL}, $G^\prime_{K\Gamma}$ is a nonsymmorphic group in the sense of the following short exact sequence
\bea
1\rightarrow \langle T_{3a}\rangle \rightarrow G^\prime_{K\Gamma} \rightarrow O_h \rightarrow 1,
\eea
in which $\langle T_{3a}\rangle$ is the translational group generated by spatial translation $\langle T_{3a}\rangle$ of three lattice sites,
and $O_h$ is the full octahedra group which is the largest 3D point group. 

\subsection{SU(2)$_1$ theory for half-odd-integer-spin Kitaev-Gamma chains}

Combining $H^\prime_{K\Gamma}$ in Eq. (\ref{eq:KG_Ham_prime}) with the equivalence relation in Eq. (\ref{eq:equiv_KG}), 
it can be seen that $H^\prime_{K\Gamma}$ is exactly the spin-$S$ AFM Heisenberg model when $K=- \Gamma$ and $K<0$. 
Therefore, at the special points $K=-\Gamma$, the low energy theory of the spin-$S$ Kitaev-Gamma model is known to be SU(2)$_1$ WZW model for half-odd-integer spins.

When $K\neq -\Gamma$, $H^\prime_{K\Gamma}$ can be written as
\bea
H^\prime_{K\Gamma}=\Gamma \sum_i \vec{S}^\prime_i\cdot \vec{S}^\prime_{i+1}+\Delta_\Gamma \sum_{\langle ij\rangle\in\gamma\text{-bond}}
S_i^{\prime\gamma} S_j^{\prime\gamma},
\label{eq:H_KG_separate}
\eea
where $\Delta_\Gamma=|K|-\Gamma$. 
Notice that the first term in Eq. (\ref{eq:H_KG_separate}) is the AFM spin-$S$ Heisenberg model,
whose low energy theory for half-odd-integer spin values is known to be the SU(2)$_1$ WZW model
according to the discussions in Sec. \ref{sec:review_spin_S}.

When  $\Delta_\Gamma$ is small, the effect of the $\Delta_\Gamma$ term can be analyzed by perturbing the low energy theory of the first term in Eq. (\ref{eq:H_KG_separate}), namely, perturbing the  SU(2)$_1$ WZW model. 
Using the symmetry group $G^\prime_{K\Gamma}$, it can be shown that in the sense of RG,
all symmetry allowed relevant and marginal terms do not break the SU(2) symmetry,
and the SU(2)-breaking terms only arise at irrelevant levels. This was
derived explicitly for the spin-$1/2$ case in Ref.~\onlinecite{Yang2020PRL};
for general half-odd-integer spins, the same conclusion follows from the
identical microscopic symmetry constraints together with the SU(2)$_1$
low-energy fixed point discussed above.
Therefore, at least for small enough $\Delta_\Gamma$, the low energy theory of the half-odd-integer-spin Kitaev-Gamma model remains to be the SU(2)$_1$ WZW model. 
It is worth noting that the precise range of $\Delta_\Gamma$ for the emergence of low energy SU(2)$_1$ theory cannot be obtained from analytic analysis, but can only be determined by numerics. 
Such range is expected to vary with the spin value $S$.

\subsection{Nonsymmorphic nonabelian  bosonization coefficients}

Although the system has an emergent SU(2) symmetry at low energies, the bosonization formulas only respect the exact discrete symmetry group $G^\prime_{K\Gamma}$.
In Ref. \onlinecite{Yang2020PRL}, Eq. (\ref{S_c}) is derived for spin-$1/2$ Kitaev-Gamma chain based on the nonsymmorphic $O_h$ symmetry. 
Since the symmetry group remains the same for other spin values, Eq. (\ref{S_c}) holds for general half-odd-integer-spin Kitaev-Gamma chains as well. 
However, notice that although symmetry analysis is able to determine the relations among the bosonization coefficients in Eq. (\ref{eq:CD_relations}),
the signs and magnitudes of these coefficients cannot be obtained from a pure symmetry analysis. 
Ref. \onlinecite{Yang2020PRL} also provides an RG analysis for the values of $C_1,C_2,D_1,D_2$ coefficients for the spin-$1/2$ case,
whereas an RG calculation for general spin values is still lacking. 

\section{RG analysis of spin-$S$ Kitaev-Gamma chain}
\label{sec:RG_KG}

In this section, we perform an RG calculation to obtain the nonsymmorphic bosonization coefficients $C_1,C_2,D_1,D_2$ in Eq. (\ref{S_c}) for spin-$S$ Kitaev-Gamma model where $S$ is a half-odd integer. 
As previously discussed, the system possesses a hidden exact SU(2) symmetry at the special point $K=-\Gamma$. 
Once $K\neq -\Gamma$, however, the SU(2) symmetry is explicitly broken at the lattice scale. 
Although the low-energy fixed-point theory still exhibits an emergent SU(2) symmetry, the bosonization formulas generally contain SU(2)-breaking coefficients inherited from the underlying microscopic model. 

Since the bosonization coefficients originate from short-distance, lattice-scale physics, they cannot be determined within the low-energy SU(2)$_1$ theory itself, whose scope is limited to universal infrared properties. 
On the other hand, although the goal is to perturb the AFM spin-$S$ Heisenberg model, 
this model is a lattice model, not  a fixed-point theory in the continuum limit,  and therefore does not provide a convenient starting point for a controlled RG analysis. 
To overcome these difficulties, by virtue of the discussions in Sec. \ref{subsec:weak_U_Hubbard}, we instead employ the weak-$U$ fermionic model as a microscopic representation of the same low energy SU(2)$_1$ theory. 
We then treat  $\Delta_\Gamma=|K|-\Gamma$ as a perturbation and analyze its effect by performing an RG calculation around the free-fermion fixed point.
Throughout this section, the $U_6$ frame is used. 

\subsection{The fermionic model}

Since the Kitaev-Gamma chain in the $U_6$ frame has a three-site periodicity,
we introduce three sets of scaling fields $h_l^\alpha(\tau,n)$ ($l\in\{1,2,3\}$, $\alpha\in\{x,y,z\}$, $n\in \mathbb{Z}$),
one set for each of the three sites in the unit cells,
where $\tau$ is the time. 
The Hamiltonian of the fermionic model can be written as 
\bea
H_F&=&H_f+\delta_\Gamma \sum_{\langle ij\rangle\in\gamma\text{-bond}}S_i^{\prime\gamma} S_j^{\prime\gamma}\nn\\
&&-\sum_{l=1}^3\sum_n \sum_\alpha h_{l}^{(0)\alpha}(\tau,n) S_{l+3n}^{\prime\alpha},
\label{eq:ham_F}
\eea
in which $H_f$ is defined in Eq. (\ref{eq:fermion}), the chemical potential is properly tuned such that the system is at half filling, 
and $\delta_\Gamma$ is given by
\bea
\delta_\Gamma=\frac{\Delta_\Gamma}{2S(S+1)}.
\eea
A superscript ``$(0)$" is added to the scaling field to indicate that it is the bare field that directly couples to spin operators.
The spin-$S$ operator in the fermion representation in Eq. (\ref{eq:ham_F}) is 
\bea
S_i^{\prime\gamma}=\sum_{d=1}^{n_c}\sum_{\sigma,\sigma^\prime=\uparrow,\downarrow} c_{id\sigma}^\dagger \frac{1}{2}\sigma^\gamma_{\sigma\sigma^\prime} c_{id\sigma^\prime},
\eea
where $\gamma\in\{x,y,z\}$, and $d$ is the color index. 
Notice that $\Delta_\Gamma$ cannot be directly used in Eq. (\ref{eq:ham_F}),
which has to be scaled down by a factor of $S(S+1)$.
This is because the Heisenberg model mimicked by $H_f$ yields a Heisenberg coupling $J_w=\frac{\pi t}{2S(S+1)}$ as discussed in Sec. \ref{subsec:Heisenberg_weak_U}.
To keep the ratio between coupling constants fixed in Eq. (\ref{eq:ham_F}) in the large-$S$ limit, one has to replace $\Delta_\Gamma$ with $\delta_\Gamma$.

At low energies, what are important in the spin operators are those components which have wave vectors in the vicinity of $0$ or $\pi$.
Hence, we separate the spin operators $S_{l+3n}^{\prime\alpha}$ into a uniform part $S_{\text{u}}^{\prime\alpha}(l+3n)$ and a staggered part $S_{\text{s}}^{\prime\alpha}(l+3n)$, restricted to wave vectors around $0$ and $\pi$, respectively.
Namely,
\bea
S_{l+3n}^{\prime\alpha}=S_{\text{u}}^{\prime\alpha}(l+3n)+(-)^{l+3n}S_{\text{s}}^{\prime\alpha}(l+3n).
\label{eq:decompose_S_us}
\eea
As a result, the scaling field term in Eq. (\ref{eq:ham_F}) becomes
\begin{flalign}
&-\sum_{l=1}^3\sum_n \sum_\alpha h_{l}^{(0)\alpha}(\tau,n) S_{\text{u}}^{\prime\alpha}(l+3n)\nn\\
&-\sum_{l=1}^3\sum_n \sum_\alpha (-)^{l+3n} h_{l}^{(0)\alpha}(\tau,n) S_{\text{s}}^{\prime\alpha}(l+3n).
\end{flalign}

\begin{figure}[h]
\includegraphics[width=0.4\textwidth]{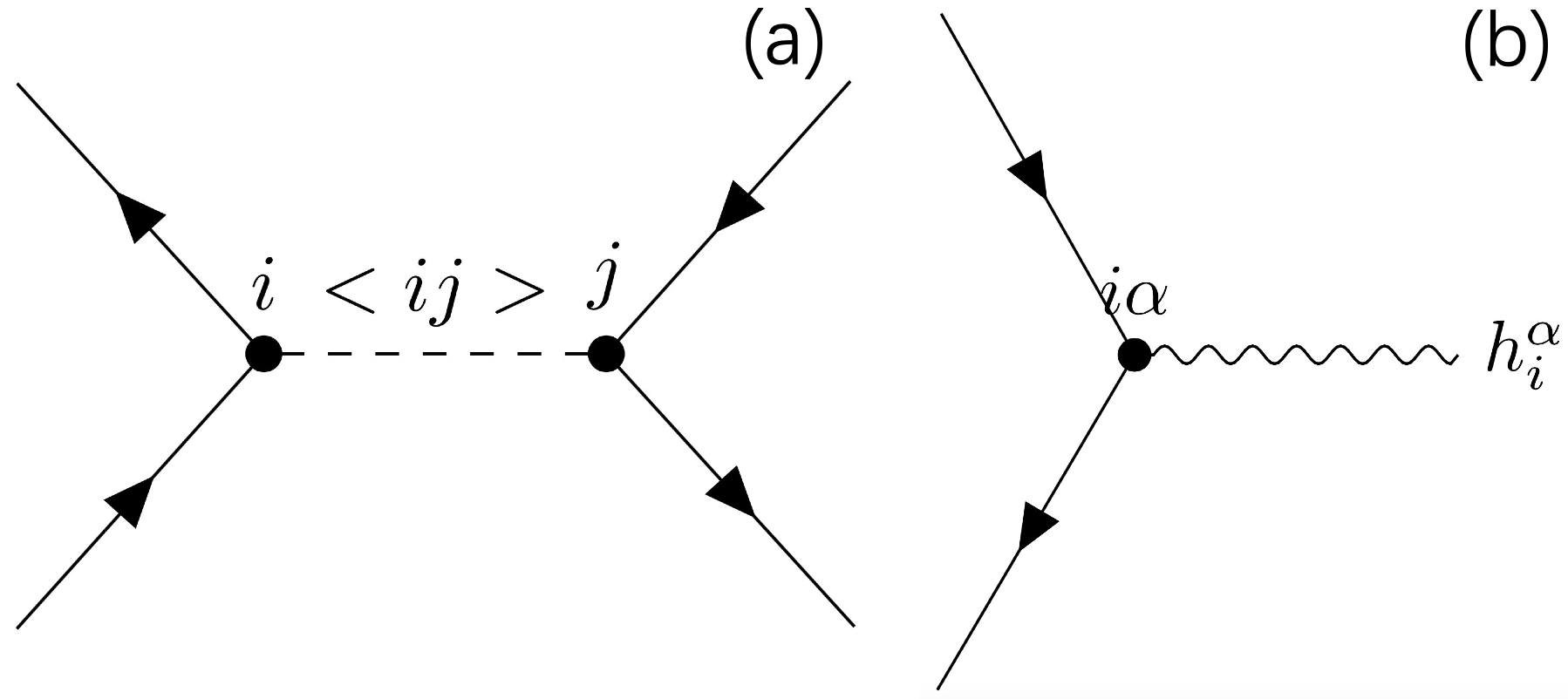}
\caption{Vertex of (a) interaction and (b) scaling field.}
\label{fig:vertex}
\end{figure}

In the path integral formulation, the partition function in imaginary time can be written as 
\bea
\mathcal{Z}=\int D[c,c^\dagger] e^{-\mathcal{S}},
\eea
where 
\bea
\mathcal{S}=\int d\tau\left(\sum_{ia\sigma} c^\dagger_{ia\sigma} \partial_\tau c_{ia\sigma} +H_F\right).
\label{eq:action}
\eea
The diagrammatic representations for the $\delta_\Gamma$ term and the scaling fields are depicted in Fig. \ref{fig:vertex} (a,b), respectively, in which the ingoing and outgoing arrowed lines represent fermion annihilation and creation operators. 
 
For Feynman diagram calculations, it is useful to introduce Fourier transformations as ($1\leq d\leq n_c$)
\bea
 c^\dagger_{d\sigma}(k)&=&\frac{1}{\sqrt{\beta L}}\int d\tau \sum_{j=1}^L  c^\dagger_{jd\sigma}(\tau) e^{i(\omega\tau-\vec{k} \cdot ja\hat{x})},
 \eea
 and
 \bea
 h^\alpha_{\text{u},l}(q)&=&\frac{1}{\sqrt{\beta L/3}}\int d\tau \sum_{n=1}^{L/3} h^\alpha_{\text{u}}(\tau,n) e^{i(\omega\tau-\vec{q}\cdot 3na\hat{x})},\nn\\
  h^\alpha_{\text{s},l}(q)&=&\frac{1}{\sqrt{\beta L/3}}\int d\tau \sum_{n=1}^{L/3} h^\alpha_{\text{s}}(\tau,n) e^{i(\omega\tau-\vec{q}\cdot 3na\hat{x})},
 \label{eq:fourier_c}
\eea
in which $L$ is system size; 
$a$ is lattice constant; 
$\beta$ is the inverse temperature;
$l\in\{1,2,3\}$ are labels for the three sites in a unit cell;
$\hat{x}$ is the spatial unit vector along the chain direction;
$k=(i\omega_n,\vec{k})$ is the two-component vector containing both the fermionic Matsubara frequency $\omega_n=(2n+1)\pi/\beta$ and the spatial wave vector $\vec{k}$;
$q=(i\omega_m,\vec{q})$ contains the bosonic Matsubara frequency $\omega_m=2m \pi/\beta$ and the spatial wave vector $\vec{q}$,
where $|\vec{q}|\ll 1$ is a small wave vector.
Notice that the sum for lattice sites in the Fourier transform of the scaling fields in Eq. (\ref{eq:fourier_c}) is performed over one third of the lattice, since we treat the three sites within each unit cell separately in the scaling field analysis,
which is crucial for obtaining the wave function renormalization effects for spin operators on different sites.

\subsection{RG flows for scaling fields}
\label{subsec:RG_KG}

The RG flow equations can be derived by gradually integrating over the fast modes in high momentum shells. 
Upon integrating over fast modes, the scaling fields are renormalized by interaction terms, including both the Hund's coupling term $U$ in $H_f$ in Eq. (\ref{eq:fermion}), as well as the $\delta_\Gamma$ term.
Since we are only interested in the SU(2) breaking effects,
we neglect the renormalizations by Hund's coupling, since it only contributes SU(2) invariant renormalizations. 
There are two contributions from the $\delta_\Gamma$ term to the renormalization of scaling fields as shown in Fig. \ref{fig:diagrams} (a,b). 
It can be demonstrated that the contribution from Fig. \ref{fig:diagrams} (b) vanishes (for details, see Appendix \ref{app:proof_vanish}), hence it is enough to focus on the diagram in Fig. \ref{fig:diagrams} (a).

As will be clear in later RG calculations, the scaling fields that couple to $S_{\text{u}}^{\prime\alpha}$ and $S_{\text{s}}^{\prime\alpha}$ in Eq. (\ref{eq:decompose_S_us}) are renormalized differently along RG flows. 
To account for such difference, we introduce $h_{\text{u},l}^\alpha(\tau,n)$ and $h_{\text{s},l}^\alpha(\tau,n)$,
so that along RG flow, the coupling to scaling fields acquires the following general form
\begin{flalign}
&-\sum_{l=1}^3\sum_{n=1}^{L/3} \sum_\alpha h_{\text{u},l}^{\alpha}(\tau,n) S_{\text{u}}^{\prime\alpha}(l+3n)\nn\\
&-\sum_{l=1}^3\sum_{n=1}^{L/3} \sum_\alpha (-)^{l+3n} h_{\text{s},l}^{\alpha}(\tau,n) S_{\text{s}}^{\prime\alpha}(l+3n).
\end{flalign}
In particular, at the beginning of the RG flow, there are the relations $h_{\text{u},l}^\alpha=h_{\text{s},l}^\alpha=h_{l}^{(0)\alpha}$.

Before presenting the results of diagrammatic calculations, we comment on the RG stopping scale. 
Denote $\Lambda_0\sim1/a$ to be the UV cutoff of the microscopic lattice. 
The RG analysis for the SU(2) breaking bosonization coefficients $C_1,C_2,D_1,D_2$ has to stop at the scale $\Lambda_s\sim 1/(3a)$,
since at $\Lambda_s$, the three sites within a unit cell become smeared and can no longer be clearly distinguished.
At energy scales lower than $\Lambda_s$, 
the three types of scaling fields can be combined into a single one.
Although there are still renormalization effects on scaling fields below $\Lambda_s$, 
these renormalizations are SU(2) invariant and do not contribute nontrivially to the SU(2) breaking effects. 

\subsubsection{RG flows for uniform components}

\begin{figure}[!htbp]
\includegraphics[width=0.3\textwidth]{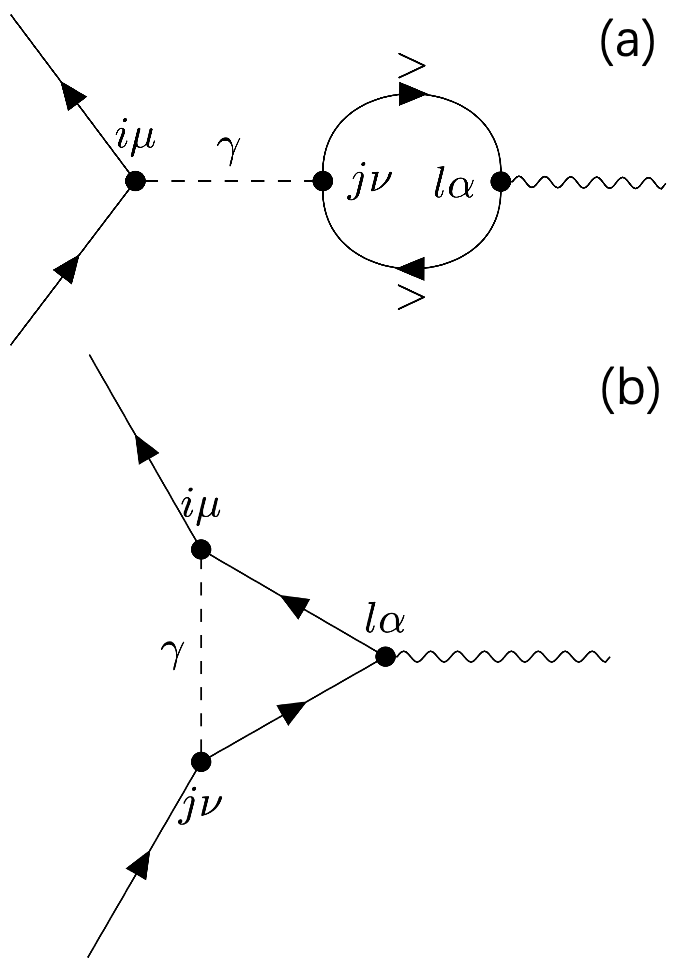}
\caption{Diagrams for the renormalizations of scaling fields.}
\label{fig:diagrams}
\end{figure}

We start with analyzing how the uniform components of the scaling fields get renormalized by the diagram in Fig. \ref{fig:diagrams} (a) when the cutoff is lowered from $\Lambda_0/b$ to $\Lambda_0/(b+\Delta b)$,
where $b<\Lambda_0/\Lambda_s$ and $\Delta b\ll 1$.
Calculations show that Fig. \ref{fig:diagrams} (a) contributes the following term
\bea
\delta_\Gamma\frac{1}{t}\lambda^{(S)}_{\text{u},jl}\delta_{\alpha\gamma} \ln b\cdot \int d\tau \sum_n h_{\text{u},l}^\alpha (\tau,n) S_{\text{u},i}^{\prime\gamma}(\tau,n),
\label{eq:diagram_1_expr}
\eea
in which there is no summation over $\gamma$ since $\gamma$ is tied with $\langle ij\rangle$, and $\lambda^{(S)}_{\text{u},jl}$ is given by
\bea
\lambda^{(S)}_{\text{u},jl}=n_c \Lambda_{jl}^{\text{u}},
\label{eq:Lambda_S}
\eea
where the factor $n_c$ originates from a summation over color index in the fermion contractions in the  loop diagram. 
The expression of $\Lambda^\text{u}_{jl}$ in Eq. (\ref{eq:Lambda_S}) is given by
\begin{flalign}
&\frac{1}{t}\Lambda^\text{u}_{jl}\ln b=\nn\\
&-\frac{a}{6}  \sum_{m=1}^3 e^{-i\frac{2\pi}{3}m(j-l)}\int_{\Lambda/(b+\Delta b)}^{\Lambda/b} d^2k G(k)G(k+\frac{2\pi }{3a} m \hat{x}),
\label{eq:Lam_jl}
\end{flalign}
in which $G(k)$ is the free fermion Green's function in imaginary time,
\bea
G(k)=\frac{1}{i\omega_n-\epsilon(\vec{k})},
\eea
where $\epsilon(\vec{k})$ is the linearized free fermion dispersion around the Fermi points.
The appearance of $\frac{2\pi }{3a} m \hat{x}$ in Eq. (\ref{eq:Lam_jl}) originates from the fact that the system is three-site periodic so that spatial momentum is conserved modulo $2\pi/(3a)$.
Detailed derivation of Eq. (\ref{eq:Lam_jl}) is included in Appendix \ref{app:Lambda_derive}.

We note that there are only two unequal parameters among $\Lambda^\text{u}_{jl}$ in Eq. (\ref{eq:flow_KG_compact}). 
As can be seen from Eq. (\ref{eq:Lambda_S}), the value of $\Lambda^\text{u}_{jl}$ is the same as $\lambda^{(S)}_{\text{u},jl}$ when $S=1/2$, 
and the result is 
\bea
\frac{1}{t}\Lambda^\text{u}_{jl}=\frac{0.14}{t} \cos(\frac{2\pi}{3} (j-l)),
\label{eq:Lambda_jl}
\eea
where $j,l\in\{1,2,3\}$.
According to Eq. (\ref{eq:Lambda_jl}), there are only two independent values of  $\Lambda^\text{u}_{jl}$, namely,
\begin{flalign}
&\Lambda^\text{u}_{11}=\Lambda^\text{u}_{22}=\Lambda^\text{u}_{33}(=\lambda^\text{u}_{0})\nn\\
&\Lambda^\text{u}_{12}=\Lambda^\text{u}_{13}=\Lambda^\text{u}_{23}=\Lambda^\text{u}_{21}=\Lambda^\text{u}_{31}=\Lambda^\text{u}_{32}(=\lambda^\text{u}_1).
\end{flalign}
In particular, evaluation of $\lambda^{\text{u}}_{0},\lambda^{\text{u}}_{1}$ using a linearized fermion spectrum gives
\bea
\lambda^{\text{u}}_{0}&=& 0.14,\nn\\
\lambda^{\text{u}}_{1}&=& -0.07.
\label{eq:lambda_u_value}
\eea

RG flow equations for the scaling fields can be written down from Eq. (\ref{eq:diagram_1_expr}). 
Consider bond $\gamma$ in Fig. \ref{fig:bonds} (b).
Denote $m_\gamma,\overline{m}_\gamma$ as the two sites (modulo $3$) connected by bond $\gamma$, i.e., $\gamma=\langle m_\gamma\overline{m}_\gamma \rangle$,
where $\overline{m}_\gamma=m_\gamma\pm 1$.
Let $n_\gamma$ be the third site (modulo $3$) other than $m_\gamma,\overline{m}_\gamma$ in a unit cell.
We note that the values of $m_\gamma,\overline{m}_\gamma,n_\gamma$ are all within the range $\{1,2,3\}$.
Since the interaction on $\gamma$ bond involves sites $m_\gamma$ and $\overline{m}_\gamma$,
it is clear from the contraction structure in Fig. \ref{fig:diagrams} (a) that $h_{\text{u},m_\gamma}^\gamma$ and $h_{\text{u},\overline{m}_\gamma}^\gamma$ are renormalized by $h_{\text{u},l}^\gamma$ ($l=1,2,3$),
whereas  $h_{\text{u},n_\gamma}^\gamma$ is not renormalized by other scaling fields. 
More explicitly, the flow equation for $h_{\text{u},m_\gamma}^\gamma,h_{\text{u},n_\gamma}^\gamma$ is
\bea
\frac{d h_{\text{u},m_\gamma}^\gamma}{d\ln b}&=&(1-2S \delta_\Gamma \frac{1}{t}\Lambda^\text{u}_{\overline{m}_\gamma m_\gamma}) h_{\text{u},m_\gamma}^\gamma\nn\\
&&-2S \delta_\Gamma \frac{1}{t}\Lambda^\text{u}_{\overline{m}_\gamma\overline{m}_\gamma} h_{\text{u},\overline{m}_\gamma}^\gamma\nn\\
&&-2S \delta_\Gamma\frac{1}{t} \Lambda^\text{u}_{\overline{m}_\gamma n_\gamma} h_{\text{u},n_\gamma}^\gamma, \nn\\
\frac{d h_{\text{u},n_\gamma}^\gamma}{d\ln b}&=&h_{\text{u},n_\gamma}^\gamma,
\label{eq:flow_KG_compact}
\eea
and the flow equation for $h_{\text{u},\overline{m}_\gamma}^\gamma$ can be obtained from that of $h_{\text{u},m_\gamma}^\gamma$ by switching $m_\gamma$ and $\overline{m}_\gamma$.
Notice that Eq. (\ref{eq:flow_KG_compact}) covers the RG flow of all spin operators: 
There are three bond types of $\gamma$ in Fig. \ref{fig:bonds} (b),
and for each bond type $\gamma$, there are three flow equations in Eq. (\ref{eq:flow_KG_compact}) 
corresponding to sites $m_\gamma,\overline{m}_\gamma,n_\gamma$,
which matches with the total number (i.e., nine) of scaling fields.
Explicit expressions for the RG flow equations are included in Appendix \ref{app_sub:flow_KG_explicit}.

In the $\Delta_\Gamma\ll1$ limit, it is enough to keep up to $O(\Delta_\Gamma)$ terms in the solutions $h_{\text{u},i}^\alpha(b)$ of the RG flow equations of the scaling fields.   
The results are
\bea
h_{\text{u},m_\gamma}^\gamma(b)&=&b[(1-2S \delta_\Gamma \frac{1}{t}\lambda^\text{u}_1\ln b) h_{m_\gamma}^{(0)\gamma}\nn\\
&&-2S \delta_\Gamma \frac{1}{t}\lambda^\text{u}_0 \ln b \,h_{\overline{m}_\gamma}^{(0)\gamma}\nn\\
&&-2S \delta_\Gamma \frac{1}{t}\lambda^\text{u}_1 \ln b\, h_{n_\gamma}^{(0)\gamma}]\nn\\
h_{n_\gamma}^\gamma(b)&=&bh_{n_\gamma}^{(0)\gamma},
\label{eq:sol_RG}
\eea
in which $h_{m_\gamma}^{(0)\gamma}$, $h_{\overline{m}_\gamma}^{(0)\gamma}$, and $h_{n_\gamma}^{(0)\gamma}$ denote the bare values of  the scaling fields at the scale $\Lambda_0\sim1/a$ where RG begins.

\subsubsection{RG flows for staggered components}

For the staggered components of the scaling fields, Fig. \ref{fig:diagrams} (a) contributes the following term
\bea
\delta_\Gamma\frac{1}{t}\lambda^{(S)}_{\text{s},jl}\delta_{\alpha\gamma} \ln b\cdot \int d\tau \sum_n h_{\text{s},l}^\alpha (\tau,n) S_{\text{s},i}^\gamma(\tau,n),
\label{eq:diagram_1_expr_s}
\eea
in which $\lambda^{(S)}_{\text{s},jl}$ is given by
\bea
\lambda^{(S)}_{\text{s},jl}=n_c \Lambda_{jl}^{\text{s}}.
\label{eq:Lambda_S_s}
\eea
The expression of $\Lambda^\text{s}_{jl}$ in Eq. (\ref{eq:Lambda_S_s}) is given by
\bea
\frac{1}{t}\Lambda^\text{s}_{jl}\ln b&=&-\frac{a}{6} \sum_{m=1}^3 e^{-i\frac{2\pi}{3}(m+\frac{1}{2})(j-l)}\nn\\
&&\times\int_{\Lambda/(b+\Delta b)}^{\Lambda/b} d^2k  \mathcal{G}(k) \mathcal{G}(k+\frac{2\pi }{3a} (m+\frac{1}{2}) \hat{x}).\nn\\
\label{eq:Lam_jl_s}
\eea

As can be seen from Eq. (\ref{eq:Lambda_S_s}), the value of $\Lambda^\text{s}_{jl}$ is the same as $\lambda^{(S)}_{\text{s},jl}$ when $S=1/2$, 
and the result is 
\bea
\frac{1}{t}\Lambda^\text{s}_{jl}=\frac{1}{t}  \sum_{m=0,\pm1} e^{-i\frac{2\pi}{3}(m+\frac{1}{2})(j-l)} a_m,
\label{eq:Lambda_jl_s}
\eea
where $j,l\in\{1,2,3\}$.
The values of $a_m$ can be evaluated numerically as
\bea
a_0&=&-0.069, \nn\\
a_{+1}&=&0.053,\nn\\ 
a_{-1}&=&-0.069.
\eea
Since $a_0=a_{-1}$, there are only two independent  values of  $\Lambda^\text{s}_{jl}$ as a result of Eq. (\ref{eq:Lambda_jl_s}), namely,
\begin{flalign}
&\Lambda^\text{s}_{11}=\Lambda^\text{s}_{22}=\Lambda^\text{s}_{33}(=\lambda^\text{s}_{0})\nn\\
&\Lambda^\text{s}_{12}=\Lambda^\text{s}_{13}=\Lambda^\text{s}_{23}=\Lambda^\text{s}_{21}=\Lambda^\text{s}_{31}=\Lambda^\text{s}_{32}(=\lambda^\text{s}_1).
\end{flalign}
Evaluation of $\lambda^{\text{s}}_{0},\lambda^{\text{s}}_{1}$ using a linearized fermion spectrum gives
\bea
\lambda^{\text{s}}_{0}&=& -0.04,\nn\\
\lambda^{\text{s}}_{1}&=& 0.06.
\label{eq:lambda_s_value}
\eea

The flow equations for $h_{\text{s},m_\gamma}^\gamma,h_{\text{s},n_\gamma}^\gamma$ are
\bea
\frac{d h_{\text{s},m_\gamma}^\gamma}{d\ln b}&=&(1-2S \delta_\Gamma \frac{1}{t}\Lambda^\text{s}_{\overline{m}_\gamma m_\gamma}) h_{\text{s},m_\gamma}^\gamma\nn\\
&&-2S \delta_\Gamma\frac{1}{t} \Lambda^\text{s}_{\overline{m}_\gamma \overline{m}_\gamma} h_{\text{s},\overline{m}_\gamma}^\gamma\nn\\
&&-2S \delta_\Gamma \frac{1}{t}\Lambda^\text{s}_{\overline{m}_\gamma n_\gamma} h_{\text{s},n_\gamma}^\gamma, \nn\\
\frac{d h_{\text{s},n_\gamma}^\gamma}{d\ln b}&=&h_{\text{s},n_\gamma}^\gamma,
\label{eq:flow_KG_compact_s}
\eea
and the flow equation for $h_{\text{s},\overline{m}_\gamma}^\gamma$ can be obtained from that of $h_{\text{s},m_\gamma}^\gamma$ by switching $m_\gamma$ and $\overline{m}_\gamma$.

In the $\Delta_\Gamma\ll1$ limit, it is enough to keep up to the $O(\Delta_\Gamma)$ terms in the solutions $h_{\text{s},i}^\alpha(b)$ of the RG flow equations of the scaling fields.   
The results are
\bea
h_{\text{s},m_\gamma}^\gamma(b)&=&b[(1-2S \delta_\Gamma \frac{1}{t}\lambda^\text{s}_1\ln b) h_{m_\gamma}^{(0)\gamma}\nn\\
&&-2S \delta_\Gamma \frac{1}{t}\lambda^\text{s}_0 \ln b \,h_{\overline{m}_\gamma}^{(0)\gamma}\nn\\
&&-2S \delta_\Gamma\frac{1}{t} \lambda^\text{s}_1 \ln b\, h_{n_\gamma}^{(0)\gamma}]\nn\\
h_{n_\gamma}^\gamma(b)&=&bh_{n_\gamma}^{(0)\gamma}.
\label{eq:sol_RG_s}
\eea

\subsection{Connection to bosonization coefficients}
\label{subsec:connect_boson_coeff_KG}

When the RG flows in Sec. \ref{subsec:RG_KG} terminate at scale $\Lambda_s\sim \Lambda_0/b_s$,  
the system is still at rather high energy scales,
where the discreteness of the lattice remains visible. 
When the cutoff is further lowered to $\Lambda_f$ where a linearization of the dispersion around the Fermi points is legitimate,
the fermion model in Eq. (\ref{eq:fermion}) enters the low energy regime,
separating into left and right movers. 
At energy scale $\Lambda_f$, the already smeared spin operators $S^\alpha(j+3n)$ can be separated to a uniform part and staggered part,
originating from intra-Fermi-point and inter-Fermi-point contributions, respectively. 
Namely, 
\bea
S^\alpha(x)=S_{\text{u}}^\alpha(x)+(-)^x S_{\text{s}}^\alpha(x),
\eea
where $x=i+3n$, and
\bea
S_\text{u}^\alpha(x)&=&\sum_{d=1}^{n_c} \sum_{\sigma\sigma^\prime} [c^\dagger_{L, d\sigma} \frac{1}{2} \sigma^\alpha_{\sigma\sigma^\prime} c_{L, d\sigma}+c^\dagger_{R, d\sigma} \frac{1}{2} \sigma^\alpha_{\sigma\sigma^\prime} c_{R, d\sigma}],\nn\\
S_\text{s}^\alpha(x)&=&\sum_{d=1}^{n_c} \sum_{\sigma\sigma^\prime} [c^\dagger_{L, d\sigma} \frac{1}{2} \sigma^\alpha_{\sigma\sigma^\prime} c_{R, d\sigma}+c^\dagger_{R, d\sigma} \frac{1}{2} \sigma^\alpha_{\sigma\sigma^\prime} c_{L, d\sigma}].\nn\\
\eea
We note that as the energy scale is lowered into the infrared limit where the charge and color sectors are gapped, 
$S_{\text{u}}^\alpha$ and $S_{\text{s}}^\alpha(x)$ become the WZW current operator $J^{\prime\alpha}_L+J^{\prime\alpha}_R$ and primary field operator   $\text{tr}(g\sigma^\alpha)$, respectively. 

At the energy scale $\Lambda_f$, the expression of the coupling of spin operators to scaling fields
can be rewritten as 
\bea
-\sum_{l=1}^{3}\sum_\alpha \sum_n (h_{\text{u}}^\alpha S_u^\alpha+(-)^{l+3n}h_{\text{s}}^\alpha S_\text{s}^\alpha),
\label{eq:coupling_SuSs}
\eea
in which 
\bea
h_\text{u}^\alpha &=& h_{\text{u},1}^\alpha (b_f)+h_{\text{u},2}^\alpha (b_f)+h_{\text{u},3}^\alpha (b_f)\nn\\
h_\text{s}^\alpha &=& h_{\text{s},1}^\alpha (b_f)+h_{\text{s},2}^\alpha (b_f)+h_{\text{s},3}^\alpha (b_f).
\label{eq:expr_h_us_KG}
\eea
Using Eq. (\ref{eq:sol_RG}), at scale $\Lambda_0/b<\Lambda_0/b_f$, $h_\text{u}^\alpha$ and $h_\text{s}^\alpha$ can be expressed in terms of the bare fields $h^{(0)\alpha}_l$ ($l=1,2,3$) as
\bea
h_\text{u}^\gamma&=&b^\prime\big[ (1-\frac{1}{t}A_\text{u}^{(S)} \ln b_s) (h^{(0)\gamma}_{\text{u},M_\gamma}+h^{(0)\gamma}_{\text{u},\overline{M}_\gamma} ) \nn\\
&&+ (1-\frac{1}{t}B_\text{u}^{(S)}\ln b_s) h_{\text{u},N_\gamma}^{(0)\gamma} \big],\nn\\
h_\text{s}^\gamma&=&b^\prime\big[ (1-\frac{1}{t}A_\text{s}^{(S)} \ln b_s) (h^{(0)\gamma}_{\text{s},M_\gamma}+h^{(0)\gamma}_{\text{s},\overline{M}_\gamma} ) \nn\\
&&+ (1-\frac{1}{t}B_\text{s}^{(S)}\ln b_s) h_{\text{s},N_\gamma}^{(0)\gamma} \big],
\label{eq:h_bf}
\eea
in which $b^\prime=b^\prime(b)$ is a function of $b$, $U$ and $\delta_\Gamma$;
 $(M_\gamma,\overline{M}_\gamma,N_\gamma)$ is in an ascending order of sequence before taking the modulo-3 operation; 
 and the coefficients $A_\text{u}^{(S)}$, $A_\text{s}^{(S)}$ and $B_\text{u}^{(S)}$ are given by
\bea
A_\text{u}^{(S)}&=&2S \delta_\Gamma (\lambda^\text{u}_0+\lambda^\text{u}_1)\nn\\
B_\text{u}^{(S)}&=&4S \delta_\Gamma \lambda^\text{u}_1\nn\\
A_\text{s}^{(S)}&=& 2S\delta_\Gamma (\lambda^\text{s}_0+\lambda^\text{s}_1)\nn\\
B_\text{s}^{(S)}&=&4S \delta_\Gamma \lambda^\text{s}_1.
\label{eq:ABus}
\eea

Notice that in Eq. (\ref{eq:h_bf}), it is $\ln b_s$ not $\ln b$ that appear in  the SU(2) breaking coefficients,
since the RG flow in Sec. \ref{subsec:RG_KG} has already terminated at energy scale $b_s$.
In addition, we have denoted the overall factor in Eq. (\ref{eq:h_bf}) as the function $b^\prime=b^\prime(b)$ rather than $b$ itself,
since $b^\prime$ contains SU(2) invariant renormalization contributions arising from the Hund's coupling $U$ as well as the $\Delta_\Gamma$ coupling below $\Lambda_s$.
Since the precise difference between $b^\prime $ and $b$ only amounts to an overall factor in the bosonization formulas, not affecting the ratios $C_1/C_2$ and $D_1/D_2$,
we do not calculate the explicit expressions of the function  $b^\prime$ and neglect the difference between $b^\prime$ and $b$ henceforth. 

Spin correlation functions can be calculated by taking partial derivatives $\partial/\partial h_l^{(0)\alpha}$ of the free energy.
The chain rule is able to convert $\partial/\partial h_l^{(0)\alpha}$ into $\partial/\partial h_\text{u}^{\alpha}$ and $\partial/\partial h_\text{s}^{\alpha}$,
\bea
\frac{\partial}{\partial h_l^{(0)\alpha}}=\frac{\partial h_\text{u}^\alpha}{\partial h_{l}^{(0)\alpha}}\frac{\partial }{\partial h_\text{u}^\alpha}+\frac{\partial h_\text{s}^\alpha}{\partial h_{l}^{(0)\alpha}}\frac{\partial }{\partial h_\text{s}^\alpha}.
\label{eq:partial_deriv_a}
\eea
Notice that $\partial/\partial h_\text{u}^{\alpha}$ and $\partial/\partial h_\text{s}^{\alpha}$ pull down $S_\text{u}^\alpha=J_L^\alpha+J_R^\alpha$ and $S_\text{s}^\alpha=\text{tr}(g\sigma^\alpha)$ in the path integral calculation, respectively.
Therefore, Eq. (\ref{eq:partial_deriv_a}) gives exactly the nonsymmorphic nonabelian bosonization formulas in Eq. (\ref{S_c}). 
In particular, we can identify the bosonization coefficients as ($j=1,2,3$)
\bea
C_j^\alpha&=&\frac{\partial h_\text{s}^\alpha}{\partial h_j^{(0)\alpha}}\nn\\
D_j^\alpha&=&\frac{\partial h_\text{u}^\alpha}{\partial h_j^{(0)\alpha}},
\eea
which yield
\bea
C_1&=&b(1-A_\text{s}^{(S)} \ln b_s)\nn\\
C_2&=&b(1-B_\text{s}^{(S)} \ln b_s).
\label{eq:C1C2_KG_RG}
\eea
and 
\bea
D_1&=&b(1-\frac{1}{t}A_\text{u}^{(S)} \ln b_s)\nn\\
D_2&=&b(1-\frac{1}{t}B_\text{u}^{(S)} \ln b_s).
\eea

In practice, the precise values of $C_1,C_2$ (or $D_1,D_2$) are determined by the normalization conditions of SU(2)$_1$ WZW current operators and primary fields,
which can differ from the RG results by an overall factor.
To remove such overall factors, we calculate the ratios $C_1/C_2$ and $D_1/D_2$.
Plugging Eq. (\ref{eq:ABus}) into the expressions of $C_1,C_2,D_1,D_2$, and keeping to $O(\Delta_\Gamma)$ terms, we obtain
\bea
\frac{C_1}{C_2}&=&1+ \frac{\ln b_s}{S+1}(\lambda^\text{s}_0-\lambda^\text{s}_1)  \frac{\Delta_\Gamma}{t} +O((\Delta_\Gamma)^2)\nn\\
\frac{D_1}{D_2}&=&1+\frac{\ln b_s}{S+1}(\lambda^\text{u}_0-\lambda^\text{u}_1)   \frac{\Delta_\Gamma}{t}+O((\Delta_\Gamma)^2),
\label{eq:C1C2_D1D2_S}
\eea
in which $\Delta_\Gamma=-(K+\Gamma)=|K|-|\Gamma|$ in the $(K<0,\Gamma>0)$ regime. 

In Eq. (\ref{eq:ham_F}), we have used $H_f$ to mimic the Heisenberg term $\Gamma \sum_i\vec{S}\cdot \vec{S}_{i+1}$.
However, because of the $S$-dependence of the coupling constant in Eq. (\ref{eq:Jw_expression}), the actual Heisenberg model mimicked by $H_f$ is 
$\frac{\Gamma}{2S(S+1)} \sum_i\vec{S}\cdot \vec{S}_{i+1}$, where $\Gamma=\pi t$.
Plugging $t=\Gamma/\pi$ into Eq. (\ref{eq:C1C2_D1D2_S}), we obtain
\begin{flalign}
&\frac{C_1}{C_2}=1+ \frac{\pi \ln b_s}{S+1}(\lambda^\text{s}_0-\lambda^\text{s}_1)  \frac{|K|-\Gamma}{\Gamma} +O\left(\frac{|K|-\Gamma}{\Gamma}\right)^2\nn\\
&\frac{D_1}{D_2}=1+ \frac{\pi \ln b_s}{S+1}(\lambda^\text{u}_0-\lambda^\text{u}_1)   \frac{|K|-\Gamma}{\Gamma} +O\left(\frac{|K|-\Gamma}{\Gamma}\right)^2.
\label{eq:C1C2_D1D2_S_b}
\end{flalign}
Using Eq. (\ref{eq:lambda_u_value},\ref{eq:lambda_s_value}) and retaining only the leading order terms, we obtain Eq. (\ref{eq:1overS_CD_predict}).
Notice that  Eq. (\ref{eq:1overS_CD_predict}) is a leading-order expansion, which is valid only
near $\Delta_\Gamma=0$, i.e. for $|\Delta_\Gamma|/\Gamma \ll 1$.

In particular, notice from Eq. (\ref{eq:C1C2_D1D2_S_b}) that the deviations of $C_1/C_2$ and $D_1/D_2$ away from $1$ scale as $1/S$ in the large-$S$ limit. 
Therefore, in the semiclassical limit $S\gg 1$, both $C_1/C_2$ and $D_1/D_2$ approach $1$. 

\begin{figure*}
\includegraphics[width=0.7\textwidth]{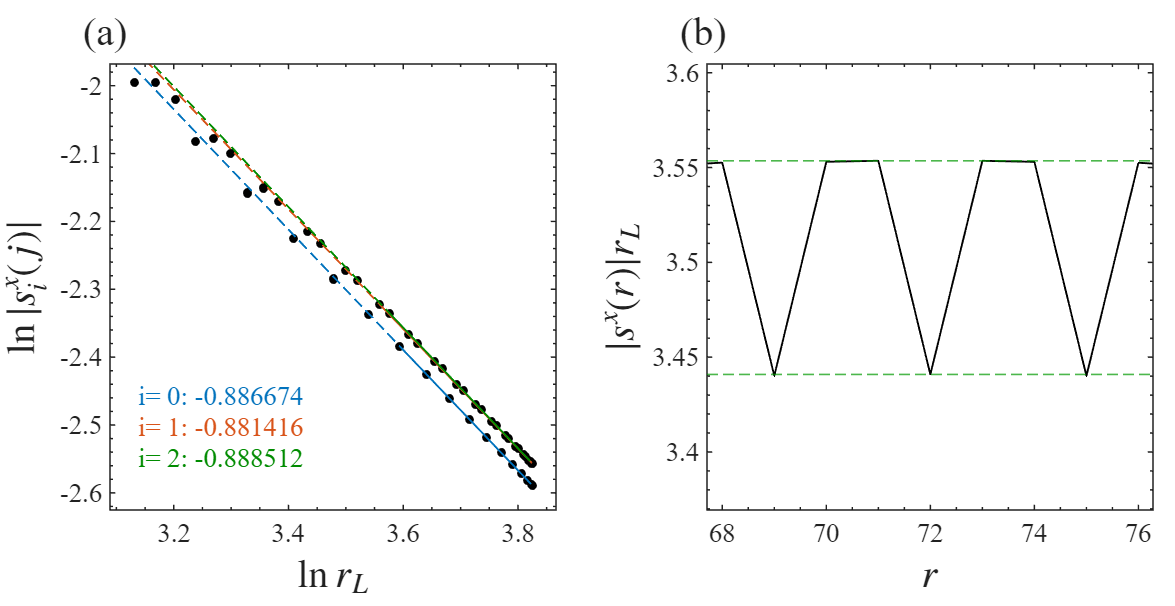}
\caption{(a) $s^x_i(j)$ as a function of $r_L$ on a log-log scale ($i=0,1,2$) and (b)  $s^x(r)r_L$ as a function of $r$ obtained from DMRG numerics at $\phi=0.8\pi$ for $S=3/2$ Kitaev-Gamma chain,
where $K=\cos(\phi)$, $\Gamma=\sin(\phi)$.  
DMRG numerics are performed on systems of $L=144$ sites under periodic boundary conditions.
Bond dimension $m$ and truncation error $\epsilon$ in DMRG numerics are taken as $m=3000$, $\epsilon=10^{-7}$.
}
\label{fig:C12_a}
\end{figure*}

\subsection{DMRG numerics }

Spin correlation functions  in the long distance limit in the spin-$S$ Kitaev-Gamma model can be analytically calculated using the nonsymmorphic nonabelian bosonization formulas in Eq. (\ref{S_c}).
For example, the zero-temperature equal-time correlation function $\langle S^x_{1} S^x_{1+r} \rangle$ ($r\gg 1$) along the $x$-direction is predicted to be 
\bea
\langle S^x_{1} S^x_{1+3j} \rangle &=& \frac{(D_2)^2}{r_L^2} +(-)^j (C_2)^2\frac{\sqrt{\ln r_L} }{r_L} \nn\\
\langle S^x_{1} S^x_{2+3j} \rangle &=&\frac{(D_2)^2}{r_L^2} +(-)^{1+j} (C_2)^2\frac{\sqrt{\ln r_L} }{r_L} \nn\\
\langle S^x_{1} S^x_{3+3j} \rangle &=& \frac{D_1D_2}{r_L^2} +(-)^{j} C_1C_2\frac{\sqrt{\ln r_L} }{r_L},
\label{eq:corr_Sx}
\eea
in which $\langle ... \rangle$ denotes ground state expectation value, and
$r_L$ is given by
\bea
r_L=\frac{L}{\pi} \sin(\frac{\pi r}{L}),
\eea
in accordance with conformal field theory in finite size systems. 
Correlation functions $\langle S^y_{1} S^y_{1+r} \rangle$ and $\langle S^z_{1} S^z_{1+r} \rangle$ along $y$- and $z$-directions can be similarly obtained. 

We have calculated spin correlation functions $f^\alpha(r)=\langle S^\alpha_{1} S^\alpha_{1+r} \rangle$ ($\alpha=x,y,z$) for spin-$3/2$ Kitaev-Gamma chain using DMRG numerics on systems of $L=144$ sites using periodic boundary conditions,
where the bond dimension $m$ and truncation error $\epsilon$ in the DMRG calculations are taken as $m=3000$ and $\epsilon=10^{-7}$, respectively. 
In accordance with Eq. (\ref{eq:corr_Sx}), we separate $f^\alpha(r)$ into three sets $f_i^\alpha (j)$ ($i=0,1,2$) as
\bea
f_i^\alpha (j)=\langle S^\alpha_{1} S^\alpha_{1+i+3j} \rangle. 
\eea
Each $f_i^\alpha (j)$ ($i=0,1,2$)  as a function of $j$ is expected to contain a slowly varying part $u^\alpha_{i} (j)$ and a fast oscillating part $s^\alpha_{i} (j)$ as 
\bea
f_i^\alpha (j)=u^\alpha_{i} (j)+(-)^js^\alpha_{i} (j),
\eea
in which both $u^\alpha_{i} (j)$ and $s^\alpha_{i} (j)$ are smooth functions of $j$,
and can be numerically extracted using a three-point formula as 
\bea
u^\alpha_i(j)&=&\frac{1}{4}f_i^\alpha(j-1)+\frac{1}{2}f_i^\alpha(j)+\frac{1}{4}f_i^\alpha(j+1)\nn\\
s^\alpha_i(j)&=&(-1)^j\left[-\frac{1}{4}f_i^\alpha(j-1)+\frac{1}{2}f_i^\alpha(j)-\frac{1}{4}f_i^\alpha(j+1)\right].\nn\\
\eea

Fig. \ref{fig:C12_a} (a) shows $s_i^x(j)$ ($i=0,1,2$) as functions of $r_L$ ($r=3j$) on a log-log scale,
in which DMRG numerics are performed for $\phi=0.8\pi$ on systems of $L=144$ sites under periodic boundary conditions,
where $K=\cos(\phi)$, $\Gamma=\sin(\phi)$.
In Fig. \ref{fig:C12_a} (a), the $i=1$ and $i=2$ lines of $s_i^x(j)$ nearly coincide,
which is in agreement with Eq. (\ref{eq:corr_Sx}).
The three exponents for $s_i^x(j)$ ($i=0,1,2$) are numerically obtained to be $0.887$, $0.881$, and $0.889$ for $i=0$, $i=1$, and $i=2$, respectively,  
which are all close to $1$ as predicted by SU(2)$_1$ WZW model in Eq. (\ref{eq:corr_Sx}).  
The slight deviation away from $1$ originates from the $\sqrt{\ln r_L}$ factor in Eq. (\ref{eq:corr_Sx}).

The three functions $s^\alpha_i(j)$ ($i=0,1,2$) can be combined into a single function $s^x(r)$ in the following way
\bea
s^\alpha(r)&=& s^\alpha_i(j), \,\, r\equiv i \mod 3,
\eea
where $r=i+3j$. 
Taking $\alpha=x$ as an example, according to Eq. (\ref{eq:corr_Sx}), $s^x(r)$ is predicted to be
\bea
s^x(3j)r_L&=& \sqrt{\ln r_L} (C_2)^2,\,\,r=3j\nn\\
s^x(1+3j)r_L&=& \sqrt{\ln r_L} (C_2)^2,\,\,r=1+3j\nn\\ 
s^x(2+3j)r_L&=& \sqrt{\ln r_L} C_1C_2, \,\,r=2+3j.
\label{eq:s_r}
\eea
Fig. \ref{fig:C12_a} (b) shows $s^x(r)$ as a function of $r$ in the middle region of the spin-$3/2$ Kitaev-Gamma chain for $\phi=0.8\pi$,
which is indeed compatible with the analytical predictions in Eq. (\ref{eq:s_r}). 
The extracted value of $C_1/C_2$ is
$C_1/C_2=0.968$. 

\begin{figure}[!htbp]
\includegraphics[width=0.4\textwidth]{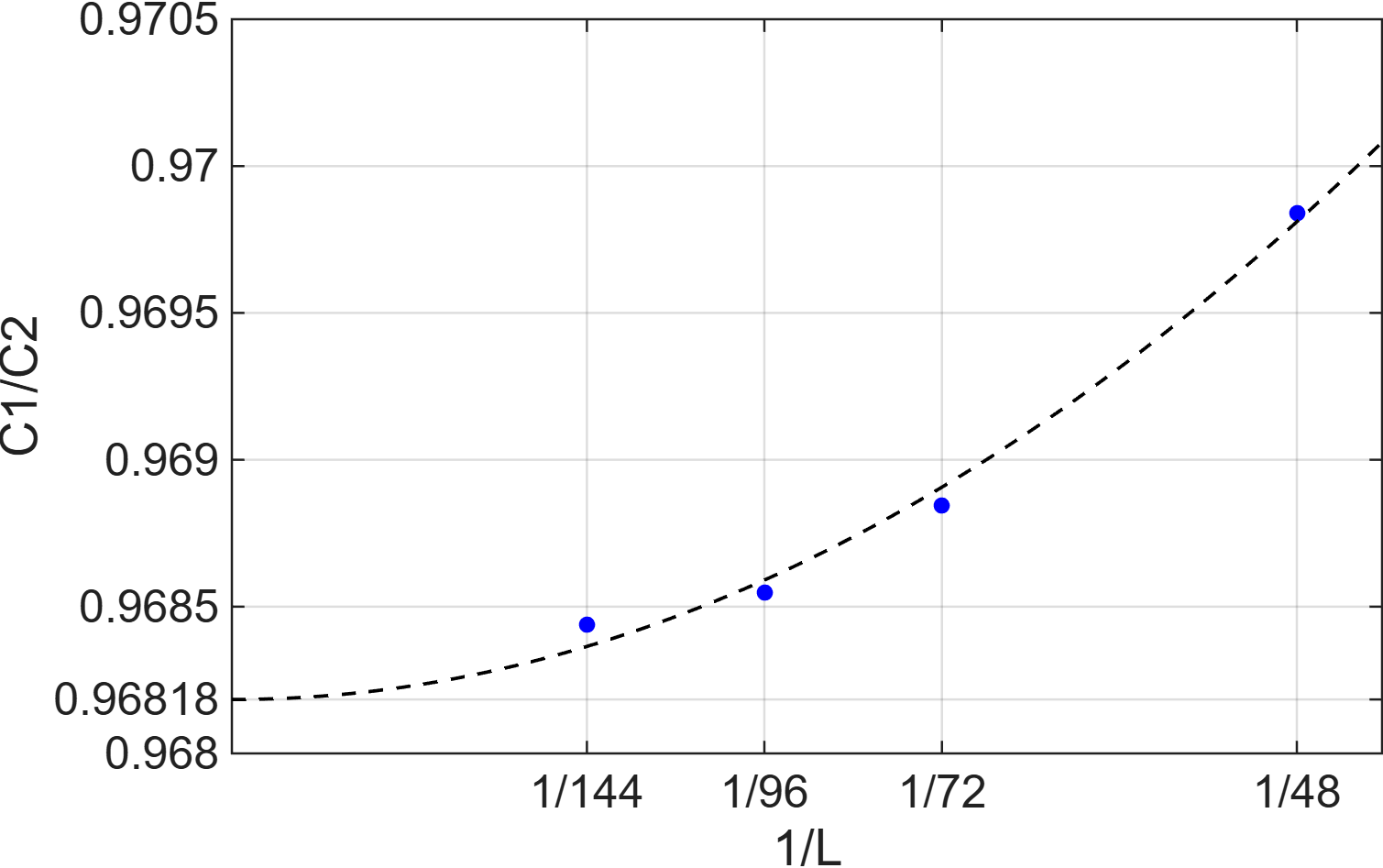}
\caption{System-size dependence of $C_1/C_2$ at $\phi = 0.8\pi$ for different values of $L$.}
\label{fig:L}
\end{figure}

To test the stability of the extracted value of $C_1/C_2$ , we further compare the results for several system sizes $L$ at $\phi=0.8\pi$ shown in Fig. \ref{fig:L}. We find that the variation of $C_1/C_2$ remains small as the system size increases and exhibits a well-defined convergence behavior. In particular, for the larger sizes, the difference between neighboring values becomes very weak. This indicates that the finite-size correction to this ratio is limited, and hence the value of $C_1/C_2$ extracted from the present $L=144$ DMRG data is stable and trustworthy. 

We note that the $\sqrt{\ln r_L}$ factor is slow-varying, which is the reason why the green dashed lines form nearly flat plateaus in Fig. \ref{fig:C12_a} (b).
On the other hand, the uniform components $u_i^x(j)$ ($i=0,1,2$) turn out to be very small compared with the truncation error $\epsilon=10^{-7}$ in DMRG calculations,
making it not possible to extract reliable values of $D_1,D_2$.
Reducing the truncation error can improve the numerical accuracy, and in principle, $u_i^x(j)$ can be extracted by increasing DMRG accuracy.
However, it turns out to be numerically too costly to determine $u_i^x(j)$ for $S=3/2$. 

We have numerically determined the ratio $C_1/C_2$ for a variety of $\phi$'s for spin-$3/2$ Kitaev-Gamma chain as shown by the blue solid circles in Fig. \ref{fig:C12_phi} (c).
The orange solid circles are the $C_1/C_2$ ratios  for spin-$1/2$ Kitaev-Gamma chain obtained in Ref. \onlinecite{Yang2020PRL}.
Notice that according to Fig. \ref{fig:C12_phi}, the deviation away from the $C_1/C_2=1$ flat line in the spin-$3/2$ case is less significant than the spin-$1/2$ case,
which is in qualitative agreement with  Eq. (\ref{eq:1overS_CD_predict}),
where  a $1/S$ suppression of the deviation is predicted in the large-$S$ limit. 
In addition, the slopes in Fig. \ref{fig:C12_phi} (c) are negative for both spin-$1/2$ and spin-$3/2$, consistent with the prediction in Eq. (\ref{eq:1overS_CD_predict}).

\section{Low energy theory of  spin-$S$ Kitaev-Heisenberg-Gamma  chains}
\label{sec:Low_KHG}

In this section, we derive the low energy theory of spin-$S$ Kitaev-Heisenberg-Gamma chain  in the $(K<0,J>0,\Gamma>0)$ region in a perturbative manner,
taking the SU(2)$_1$ low energy theory of the Kitaev-Gamma chain established in Sec. \ref{sec:Low_KG} as the perturbative starting point.

\subsection{Symmetries}

It is useful to further perform the following global spin rotation superimposed on the six-sublattice rotation for the Kitaev-Heisenberg-Gamma chain, 
\begin{equation}
\label{eq:O-def-correct}
\begin{pmatrix}
S_i^{\prime\prime x} & S_i^{\prime\prime y} & S_i^{\prime\prime z}
\end{pmatrix}
=
\begin{pmatrix}
S_i^{\prime x} & S_i^{\prime y} & S_i^{\prime z}
\end{pmatrix}
O,
\end{equation}
where the $3\times 3$ matrix $O$ is defined in Eq. (\ref{eq:def_O_mat}).
We will refer to the $\vec{S}_i$, $\vec{S}^\prime_i$, and $\vec{S}^{\prime\prime}_i$ as the original, $U_6$, and $OU_6$ frames, respectively. 
Notice that $O$ rotates the $(1,1,1)$-direction in the $U_6$ frame to the $z^{\prime\prime}$-axis in the $OU_6$ frame. 

After performing the $O$-transformation, the Hamiltonian becomes
\bea
H^{\prime\prime}_{KJ\Gamma}=\Gamma \sum_i \vec{S}^{\prime\prime}_i\cdot \vec{S}^{\prime\prime}_{i+1}+\Delta H^{\prime\prime},
\label{eq:Hpp_KJG}
\eea
in which $\Delta H^{\prime\prime}$ is 
\bea
\Delta H^{\prime\prime}=\sum_{j=1}^3\sum_n \vec{S}^{\prime\prime T}_{j+3n} A_j \vec{S}^{\prime\prime}_{j+1+3n},
\label{eq:DeltaHpp}
\eea
where $\vec{S}^{\prime\prime}_i = (S_i^{\prime\prime x},S_i^{\prime\prime y},S_i^{\prime\prime z})^T$, 
and the matrices $A_j$ ($j=1,2,3$) are given by
\bea
A_j=M_z A_{j-1}M_Z^T,
\eea
where $M_z$ is given in Eq. (\ref{eq:Mz_mat}), and 
\begin{flalign}
&A_{3}= \begin{pmatrix}
-\frac{2}{3}(K+\Gamma)-J & 0 & -\frac{2}{3\sqrt2}(K+\Gamma)\\
0 & J & 0\\
-\frac{2}{3\sqrt2}(K+\Gamma) & 0 & -\frac{1}{3}(K+\Gamma)-J
\end{pmatrix}.
\end{flalign}

It can be checked that  $H^{\prime\prime}_{KJ\Gamma}$ is invariant under the symmetry transformations
$T$, $R(\hat{y}^{\prime\prime},\pi)I$, $R(\hat{z}^{\prime\prime},-2\pi/3)T_a$,
in which $T$ is time reversal; 
$T_a$ is the spatial translation by one lattice site;
$I$ is the spatial inversion around the site $2$ in Fig. \ref{fig:bonds} (b);
$R(\hat{n},\beta)$ represents a global spin rotation by angle $\beta$ and
\begin{eqnarray}
\hat{z}^{\prime\prime}&=&\frac{1}{\sqrt{3}}(\hat{x}^\prime+\hat{y}^\prime+\hat{z}^\prime) \nn\\
\hat{y}^{\prime\prime}&=&\frac{1}{\sqrt{2}}(-\hat{x}^\prime+\hat{z}^\prime),
\label{eq:yz_prime}
\end{eqnarray}
where $\hat{\alpha}^\prime$ ($\alpha=x,y,z$) are unit vectors in spin space in the $U_6$ frame.
Compared with spin-$S$ Kitaev-Gamma chain, the symmetry group is reduced to the nonsymmorphic $D_3$ group, 
where the normal direction of the planar $D_3$ group is along the $z^{\prime\prime}$-direction. 
More explicitly, the symmetry group $G^{\prime\prime}_{KJ\Gamma}$ of $H^{\prime\prime}_{KJ\Gamma}$ in the $OU_6$ frame satisfies the following short exact sequence
\bea
1\rightarrow \langle T_{3a}\rangle \rightarrow G^{\prime\prime}_{KJ\Gamma} \rightarrow D_3 \rightarrow 1.
\eea
Notice that the special role of the $(1,1,1)$-direction can be understood from the nonsymmorphic symmetry operation $R(\hat{z}^{\prime\prime},2\pi/3)T_a$.

\subsection{Luttinger liquid theory for half-odd-integer-spin Kitaev-Heisenberg-Gamma chains}

The low energy theory of the spin-$1/2$ Kitaev-Heisenberg-Gamma chain has been derived in Ref. \onlinecite{Yang2020PRR}.
The derivation can be directly generalized to arbitrary half-odd-integer spins. 
The strategy is to take the Heisenberg term in Eq. (\ref{eq:6rotated_KJG}) in the $U_6$ frame as a perturbation on the SU(2)$_1$ low energy theory of the Kitaev-Gamma model. 
Because of the nonsymmorphic planar $D_3$ symmetry, the symmetry-allowed relevant and marginal terms are all U(1) invariant, and the U(1) breaking terms only arise at irrelevant levels which can be neglected at low energies. 

Treating Heisenberg term as a perturbation in the $OU_6$ frame, and plugging nonsymmorphic bosonization formulas of the Kitaev-Gamma chain into the Heisenberg term, we obtain the following low energy Hamiltonian,
\begin{eqnarray}
H&=&\frac{2\pi}{3}v_0\int dx
\left(
\vec{J}_L^{\prime\prime}\cdot\vec{J}_L^{\prime\prime}
+
\vec{J}_R^{\prime\prime}\cdot\vec{J}_R^{\prime\prime}
\right)\nn\\
&&+u_1\int dx
\left(
J_L^{\prime\prime z}-J_R^{\prime\prime z}
\right)
\nn\\
&&
-g_c\int dx\,
\vec{J}_L^{\prime\prime}\cdot\vec{J}_R^{\prime\prime}
+u_2\int dx\,
J_L^{\prime\prime z}J_R^{\prime\prime z},
\label{eq:Low_ham_KJG}
\end{eqnarray}
in which 
\begin{flalign}
&u_1=4\pi\sqrt{3}J[(C_2)^2-(C_1)^2]\nn\\
&u_2=\frac{J}{a}\left(\frac{3}{2}[(D_1)^2+(D_2)^2]
+12\pi^2[(C_1)^2+(C_2)^2]\right),
\end{flalign}
where $C_1$ and $C_2$ are the nonsymmorphic bosonization coefficients in the Kitaev-Gamma chain. 

When $J>0$, the system has an easy-plane anisotropy such that the low energy physics of the system can be described by the following Luttinger liquid Hamiltonian with emergent U(1) symmetry (i.e., U(1) rotational symmetry around $z^{\prime\prime}$-axis)
\bea
H_{LL}=\frac{v}{2}\int dx
\left[
\kappa^{-1}(\nabla\tilde{\varphi})^2
+\kappa(\nabla\tilde{\theta})^2
\right]
+\lambda\int dx\,\partial_x\tilde{\theta},
\label{eq:H_LL}
\eea
in which $v$ is the velocity; $\kappa$ is the Luttinger parameter; and the fields $\tilde{\theta},\tilde{\varphi}$ satisfy
\bea
[\tilde{\varphi}(x),\tilde{\theta}(x^\prime)]
=
\frac{i}{2}\text{sgn}(x^\prime-x).
\eea
The parameters in Eq. (\ref{eq:H_LL}) are related to those in Eq. (\ref{eq:Low_ham_KJG}) by
\begin{align}
v &=
\sqrt{
v_0^2-\left(\frac{g_c-u_2}{4\pi}\right)^2
}\\[2mm]
\kappa &=
\sqrt{
\frac{v_0+\dfrac{g_c-u_2}{4\pi}}
     {v_0-\dfrac{g_c-u_2}{4\pi}}
}\\[2mm]
\lambda &= \frac{u_1}{\sqrt{2\pi}}.
\end{align}
In abelian bosonization, the uniform and staggered fluctuations
$J^{\prime\prime\alpha}$, $N^{\prime\prime\alpha}$
($\alpha=x,y,z$) are related to the $\tilde{\theta},\tilde{\varphi}$ fields by
\bea
J^{\prime\prime x}
&=&
\frac{1}{a}
\cos(\sqrt{4\pi}\tilde{\varphi})
\cos(\sqrt{\pi}\tilde{\theta}),
\nn\\
J^{\prime\prime y}
&=&
\frac{1}{a}
\cos(\sqrt{4\pi}\tilde{\varphi})
\sin(\sqrt{\pi}\tilde{\theta}),
\nn\\
J^{\prime\prime z}
&=&
-\frac{1}{\sqrt{\pi}}\nabla\tilde{\varphi},
\label{eq:bosonize_J_b}
\eea
and 
\bea
N^{\prime\prime x}
&=&
\frac{1}{a}\cos(\sqrt{\pi}\tilde{\theta}),
\nn\\
N^{\prime\prime y}
&=&
\frac{1}{a}\sin(\sqrt{\pi}\tilde{\theta}),
\nn\\
N^{\prime\prime z}
&=&
\frac{1}{a}\sin(\sqrt{4\pi}\tilde{\varphi}),
\label{eq:bosonize_N_b}
\eea

The $\partial_x\tilde{\theta}$ term in Eq. (\ref{eq:H_LL}) can be eliminated by the following transformation,
\bea
\varphi(x)&=&\tilde{\varphi}(x),\nn\\
\theta(x)&=&\tilde{\theta}(x)+\frac{\lambda}{v\kappa}x.
\eea
The transformed Hamiltonian becomes
\begin{flalign}
&H_{LL}=
\frac{v}{2}\int dx\,
\left[
\kappa^{-1}(\partial_x\varphi)^2
+
\kappa(\partial_x\theta)^2
\right]
-
\frac{L\lambda^2}{2v\kappa},
\end{flalign}
in which $L$ is the system length, and the constant term
$\frac{L\lambda^2}{2v\kappa}$ is a shift of the energy, therefore inessential to the low energy physics. 
Defining $\mathcal{J}^\alpha$, $\mathcal{N}^\alpha$ ($\alpha=x,y,z$) as
\begin{align}
\mathcal{J}^x
&=
\frac{1}{a}
\cos\!\left(\sqrt{4\pi}\,\varphi\right)
\cos\!\left(\sqrt{\pi}\,\theta\right),
\nn\\
\mathcal{J}^y
&=
\frac{1}{a}
\cos\!\left(\sqrt{4\pi}\,\varphi\right)
\sin\!\left(\sqrt{\pi}\,\theta\right),
\nn\\
\mathcal{J}^z
&=
-\frac{1}{\sqrt{\pi}}\partial_x\varphi,
\end{align}
and
\begin{align}
\mathcal{N}^x
&=
\frac{1}{a}
\cos\!\left(\sqrt{\pi}\,\theta\right),
\nn\\
\mathcal{N}^y
&=
\frac{1}{a}
\sin\!\left(\sqrt{\pi}\,\theta\right),
\nn\\
\mathcal{N}^z
&=
\frac{1}{a}
\sin\!\left(\sqrt{4\pi}\,\varphi\right),
\end{align}
the operators $J^{\prime\prime\alpha}$ and $N^{\prime\prime\alpha}$
($\alpha=x,y,z$) can be expressed through
$\mathcal{J}^\alpha$ and $\mathcal{N}^\alpha$ as
\bea
(J^{\prime\prime x},J^{\prime\prime y},J^{\prime\prime z})
=
(\mathcal{J}^x,\mathcal{J}^y,\mathcal{J}^z)Q,
\eea
and
\bea
(N^{\prime\prime x},N^{\prime\prime y},N^{\prime\prime z})
=
(\mathcal{N}^x,\mathcal{N}^y,\mathcal{N}^z)Q,
\eea
where 
\bea
Q=\begin{pmatrix}
\cos(q_{\text{ic}}x) & -\sin(q_{\text{ic}}x) & 0\\
\sin(q_{\text{ic}}x) & \cos(q_{\text{ic}}x) & 0\\
0 & 0 & 1
\end{pmatrix},
\label{eq:define_Q_mat}
\eea
in which the incommensurate wave vector $q_{\text{ic}}$ is 
\bea
q_{\text{ic}}=\frac{\sqrt{\pi}\lambda}{v\kappa}.
\eea

When $J<0$, the anisotropy is of the easy-axis type, and the system becomes magnetically ordered.
We will be interested in the $J>0$ region in this work.
It is worth noting that the above analysis holds for small values of $J$. 
There may be phase transitions that go beyond the Luttinger liquid phase when $J$ becomes large enough.

\subsection{Nonsymmorphic abelian  bosonization coefficients}

Although the system has an emergent U(1) symmetry when $J>0$,
the bosonization formulas only respect the exact nonsymmorphic $D_3$ group, not the emergent U(1) symmetry.
In this section, we derive the expressions of the nonsymmorphic abelian bosonization formulas in the $OU_6$ frame. 


The most general bosonization formulas are given by
\begin{flalign}
&(S^{\prime\prime x}_{i+3n}~S^{\prime\prime y}_{i+3n}~S^{\prime\prime z}_{i+3n})=\nn\\
&(J^{\prime\prime x}~J^{\prime\prime y}~J^{\prime\prime z})F_i
+(-)^{i+n}(N^{\prime\prime x}~N^{\prime\prime y}~N^{\prime\prime z})E_i.
\label{eq:abelian_LL1_matrix_pp}
\end{flalign}
The formulas in Eq. (\ref{eq:abelian_LL1_matrix}) can be obtained from Eq. (\ref{eq:abelian_LL1_matrix_pp}) by further performing the $Q$ and $O^{-1}$ transformation where $Q$ is given by Eq. (\ref{eq:define_Q_mat}), and $O$ is given by Eq. (\ref{eq:def_O_mat}),
since the former and latter apply to the $U_6$ and $OU_6$ frames, respectively, differing by an $O$ transformation. 

Notice that the low energy fields $J^{\prime\prime\alpha}$, $N^{\prime\prime\alpha}$ ($\alpha=x,y,z$) (defined in Eqs. (\ref{eq:bosonize_J},\ref{eq:bosonize_N})) remain in the low energy sector 
when a symmetry operation is performed.
As a result, the bosonization coefficients are not all independent, 
since the left and right hand sides of Eq. (\ref{eq:abelian_LL1_matrix_pp}) have to be covariant under symmetry transformations of the system. 
First consider the symmetry operation $R(\hat{z}^{\prime\prime},-\frac{2\pi}{3})T_a$.
Using the transformation properties 
\begin{flalign}
&[R(\hat{z}^{\prime\prime},-\frac{2\pi}{3})T_a](S^{\prime\prime x}_{i+3n}~S^{\prime\prime y}_{i+3n}~S^{\prime\prime z}_{i+3n})[R(\hat{z}^{\prime\prime},-\frac{2\pi}{3})T_a]^{-1}\nn\\
&=(S^{\prime\prime x}_{i+1+3n}~S^{\prime\prime y}_{i+1+3n}~S^{\prime\prime z}_{i+1+3n})M_z,
\end{flalign}
and
\begin{flalign}
&[R(\hat{z}^{\prime\prime},-\frac{2\pi}{3})T_a]
(N^{\prime\prime x}~N^{\prime\prime y}~N^{\prime\prime z})
[R(\hat{z}^{\prime\prime},-\frac{2\pi}{3})T_a]^{-1}\nn\\
&=(-)(N^{\prime\prime x}~N^{\prime\prime y}~N^{\prime\prime z})M_z,
\label{eq:abe_Rz}
\end{flalign}
we obtain
\bea
E_1&=&M_z^{-1}E_2M_z\nn\\
E_3&=&M_zE_2M_z^{-1},
\label{eq:Cs_z_2}
\eea
in which $M_z$ is given in Eq. (\ref{eq:Mz_mat}).
We note that there is an overall minus sign in the right hand side of the second equation in Eq. (\ref{eq:abe_Rz}) since $N^{\prime\prime\alpha}$ changes sign under $T_a$.

Then consider the symmetry operation $R(\hat{y}^{\prime\prime},\pi)I$.
Using the transformation properties
\begin{flalign}
&[R(\hat{y}^{\prime\prime},\pi)I](S^{\prime\prime x}_{i+3n}~S^{\prime\prime y}_{i+3n}~S^{\prime\prime z}_{i+3n})[R(\hat{y}^{\prime\prime},\pi)I]^{-1}\nn\\
&=(S^{\prime\prime x}_{10-i-3n}~S^{\prime\prime y}_{10-i-3n}~S^{\prime\prime z}_{10-i-3n})M_y,
\end{flalign}
and
\begin{flalign}
&[R(\hat{y}^{\prime\prime},\pi)I]
(N^{\prime\prime x}(r)~N^{\prime\prime y}(r)~N^{\prime\prime z}(r))
[R(\hat{y}^{\prime\prime},\pi)I]^{-1}\nn\\
&=(N^{\prime\prime x}(-r)~N^{\prime\prime y}(-r)~N^{\prime\prime z}(-r))M_y,
\end{flalign}
we obtain
\bea
E_1&=&M_yE_3M_y^{-1},\nn\\
E_2&=&M_yE_2M_y^{-1},
\label{eq:Cs_y_2}
\eea
in which the coordinate in $N^{\prime\prime\alpha}$ is $r=(i+3n)a$ and the matrix $M_y$ is
\bea
M_y=\left(\begin{array}{ccc}
-1 & 0 & 0\\
0 & 1 & 0\\
0&0&-1
\end{array}
\right).
\eea
We note that because of the relation $(M_yM_z)^2=1$, Eqs. (\ref{eq:Cs_z_2},\ref{eq:Cs_y_2}) leads to a single independent constraint:
\bea
E_2=M_yE_2M_y^{-1},
\eea
which can be easily solved and 
constrains $E_2$ to be of the form in Eq. (\ref{eq:E2_F2}).
Then the matrices $E_1$ and $E_3$ can be obtained from Eq. (\ref{eq:Cs_z_2}).
The discussions on the matrices $F_i$ ($i=1,2,3$) are exactly similar, yielding the expressions in 
Eqs. (\ref{eq:E2_F2},\ref{eq:EF_mat_13}).

There are in total ten free parameters $\lambda_\Lambda$, $\nu_\Lambda$, $\sigma_\Lambda$, $\rho_\Lambda$, $\delta_\Lambda$ ($\Lambda=C,D$) in the nonsymmorphic bosonization formulas,
which can in principle be determined by comparing numerical results on correlation functions with analytical predictions, 
though very difficult in practice.
We expect the values of these bosonization coefficients to depend on both the couplings $K,J,\Gamma$ and the spin value $S$.

\section{RG analysis of spin-$S$ Kitaev-Heisenberg-Gamma chain}
\label{sec:RG_KHG}

In this section, we perform an RG calculation to obtain the ten nonsymmorphic bosonization coefficients $\lambda_\Lambda$, $\nu_\Lambda$, $\sigma_\Lambda$, $\rho_\Lambda$, $\delta_\Lambda$ ($\Lambda=C,D$) in Eq. (\ref{eq:abelian_LL1_matrix_pp}) for spin-$S$ Kitaev-Heisenberg-Gamma model where $S$ is a half-odd integer. 

\subsection{The fermionic model}
\label{subsec:fermion_KJG}

Similar to the RG analysis in Kitaev-Gamma chain,
we will use a fermionic model that mimics the low energy physics of the Kitaev-Heisenberg-Gamma chain.
What is different from the Kitaev-Gamma case is that it is more useful to take the $OU_6$ frame in the Kitaev-Heisenberg-Gamma case,
so that the direction of the axis for emergent U(1) symmetry is along the $z^{\prime\prime}$-direction. 

In the $OU_6$ frame, the spin Hamiltonian for the spin-$S$ Kitaev-Heisenberg-Gamma chain is given by $H^{\prime\prime}_{KJ\Gamma}$ in Eq. (\ref{eq:Hpp_KJG}).
The first term in Eq. (\ref{eq:Hpp_KJG}) is the AFM Heisenberg model which can be mimicked by the fermion model in Eq. (\ref{eq:fermion}),
and the second term $\Delta H^{\prime\prime}$ will be taken as a perturbation. 
Introducing scaling fields for uniform and staggered components of the spin operators as in the Kitaev-Gamma case, the resulting fermionic Hamiltonian is 
\bea
\tilde{H}_F&=&H_f+\Delta H^{\prime\prime}-\sum_{l=1}^3\sum_n \sum_\alpha h_{\text{u},l}^\alpha(\tau,n) S_{\text{u}}^{\prime\prime\alpha}(l+3n)\nn\\
&&-\sum_{l=1}^3\sum_n \sum_\alpha (-)^{l+3n} h_{\text{s},l}^\alpha(\tau,n) S_{\text{s}}^{\prime\prime\alpha}(l+3n).
\label{eq:ham_F_KJG}
\eea
It is helpful to separate $\Delta H^{\prime\prime}$ into two parts, 
one part with coupling $\Delta_\Gamma=-(K+\Gamma)$, the other part with coupling $J$.
Namely,
\bea
\Delta H^{\prime\prime}&=&\Delta H^{\prime\prime}_{K\Gamma}+\Delta H^{\prime\prime}_J,
\eea
in which according to Eq. (\ref{eq:DeltaHpp}), $\Delta H^{\prime\prime}_{K\Gamma}$ and $\Delta H^{\prime\prime}_J$ are given by
\bea
\Delta H^{\prime\prime}_{K\Gamma}&=&\delta_\Gamma\sum_{j=1}^3\sum_n \vec{S}_{j+3n}^{\prime\prime T} A^{(K\Gamma)}_j \vec{S}^{\prime\prime}_{j+1+3n},\nn\\
\Delta H^{\prime\prime}_{J}&=&\delta_J\sum_{j=1}^3\sum_n \vec{S}_{j+3n}^{\prime\prime T} A^{(J)}_j \vec{S}^{\prime\prime}_{j+1+3n},
\label{eq:DeltaH_JG_J}
\eea
where 
\bea
\delta_\Gamma&=&\frac{\Delta_\Gamma}{2S(S+1)},\nn\\
\delta_J&=&\frac{J}{2S(S+1)}.
\eea
In Eq. (\ref{eq:DeltaH_JG_J}), the matrices $A^{(K\Gamma)}_j,A^{(J)}_j$ are given by
\bea
A^{(K\Gamma)}_j&=&M_z A^{(K\Gamma)}_{j-1}M_z^T,\nn\\
A^{(J)}_j&=&M_z A^{(J)}_{j-1}M_z^T,
\label{eq:R_transform_A_KG_J}
\eea
and 
\bea
A^{(K\Gamma)}_3&=&\left(\begin{array}{ccc}
\frac{2}{3}& 0 & \frac{2}{3\sqrt2}\\
0 & 0 & 0\\
\frac{2}{3\sqrt2}& 0 & -\frac{1}{3}
\end{array}\right)\nn\\
A^{(J)}_3&=&\left(\begin{array}{ccc}
-1&  & \\
 & 1 & \\
&  & -1
\end{array}\right).
\eea
Notice that  $\Delta H^{\prime\prime}_{K\Gamma}$ is just the $\delta_\Gamma$ term in Eq. (\ref{eq:ham_F}) in the Kitaev-Gamma case, albeit in the $OU_6$ rather than the $U_6$ frame.

\subsection{RG flow equations for scaling fields}
\label{subsec:RG_KJG}

The RG flow equations for the uniform and staggered scaling fields can be derived as ($\eta=\text{u},\text{s}$)
\bea
\frac{d h_{\eta,l}^\alpha}{d\ln b}&=&h_{\eta,l}^\alpha-n_c\sum_{\beta=x,y,z}\sum_{j}\big[ 
(\tilde{A}_l)_{\alpha\beta}\Lambda^{\eta}_{l+1,j} \nn\\
&&+(\tilde{A}_{l-1})_{\beta\alpha} \Lambda^{\eta}_{l-1,j}
\big]h_{\eta,j}^\beta.
\eea
in which $\tilde{A}_l$ ($l=1,2,3$) is 
\bea
\tilde{A}_l=\delta_\Gamma A^{(K\Gamma)}_l+\delta_J A^{(J)}_l.
\label{eq:tildeA_l}
\eea

For $\Delta_\Gamma,J\ll 1$ and $b<b_s$, it is enough to keep up to $O(\Delta_\Gamma,J)$ terms in the solutions of the RG flow equations. 
Under this approximation, the solutions are ($\eta=\text{u},\text{s}$)
\bea
h_{\eta,l}^\alpha(b)&=&b\big( h_{l}^{(0)\alpha}-2S\ln b_s
\sum_{\beta}\sum_{j}\big[ 
(\tilde{A}_l)_{\alpha\beta}\Lambda^{\eta}_{l+1,j} \nn\\
&&+(\tilde{A}_{l-1})_{\beta\alpha} \Lambda^{\eta}_{l-1,j}
\big]h_{j}^{(0)\beta}\big),
\label{eq:sol_RG_hu}
\eea
in which we have neglected the SU(2) invariant renormalization effects that can change $b$ to $b^\prime=b'(b)$ as discussed in Sec. \ref{subsec:connect_boson_coeff_KG}.

\subsection{Connection to bosonization coefficients}
\label{subsec:RG_coeff_KJG}

Similar to Sec. \ref{subsec:connect_boson_coeff_KG}, at energy scales below $\Lambda_f$, 
the coupling to scaling fields reduces to Eq. (\ref{eq:coupling_SuSs}),
in which $h_{\text{u}}^\alpha,h_{\text{s}}^\alpha$ are given in Eq. (\ref{eq:expr_h_us_KG}).
Plugging Eq. (\ref{eq:sol_RG_hu}) in Eq. (\ref{eq:expr_h_us_KG}),
we obtain 
\bea
h^\alpha_{\eta}&=&\sum_{i=1}^3\sum_{\beta=x,y,z}(M^\eta_i)_{\alpha\beta} h^{(0)\beta}_{i},
\eea
in which the $3\times 3$ matrices $M^\eta_i$ ($i=1,2,3$) are given by
\begin{flalign}
M^\eta_i=I_3-2S\ln b_s \sum_{l=1}^3 (\Lambda_{l+1,i}^\eta \tilde{A}_l + \Lambda_{l-1,i}^\eta \tilde{A}_{l-1}),
\end{flalign}
in which $I_3$ is the $3\times 3$ identity matrix, 
and the matrices $\tilde{A}_l,\tilde{A}_{l-1}$ are given in Eq. (\ref{eq:tildeA_l}).
Notice that because of the relations in Eq. (\ref{eq:R_transform_A_KG_J}), $M^\eta_i$ satisfies
\bea
M^\eta_i=M_zM^\eta_{i-1}M_z^T.
\eea

We can relate $M^\eta_i$ ($i=1,2,3$) to the matrices of bosonization coefficients $E_i,F_i$ in Eq. (\ref{eq:abelian_LL1_matrix_pp}).
Using the chain rule for partial derivatives
\bea
\frac{\partial}{\partial h_l^{(0)\alpha}}=\sum_\beta\frac{\partial h_\text{u}^\beta}{\partial h_{l}^{(0)\alpha}}\frac{\partial }{\partial h_\text{u}^\beta}+\sum_\beta\frac{\partial h_\text{s}^\beta}{\partial h_{l}^{(0)\alpha}}\frac{\partial }{\partial h_\text{s}^\beta},
\label{eq:partial_deriv}
\eea
we obtain
\bea
\frac{\partial}{\partial h_l^{(0)\alpha}}=\sum_\beta(M_i^{\text{u}})_{\beta\alpha}\frac{\partial }{\partial h_\text{u}^\beta}+\sum_\beta(M_i^{\text{s}})_{\beta\alpha}\frac{\partial }{\partial h_\text{s}^\beta}. 
\eea
Comparing with Eq. (\ref{eq:abelian_LL1_matrix_pp}), we obtain
\bea
E_i&=&(M_i^{\text{s}})^T,\nn\\
F_i&=&(M_i^{\text{u}})^T.
\eea

The bosonization coefficients can be determined as 
\bea
\lambda_\eta&=&b\left[1-\frac{\pi \ln b_s}{S+1} \left(\frac{\lambda_0^{\eta}+5\lambda_1^{\eta}}{3}\frac{\Delta_\Gamma}{\Gamma}+(\lambda_0^{\eta}-\lambda_1^{\eta})\frac{J}{\Gamma}\right) \right]\nn\\
\sigma_\eta&=&b\frac{\pi \ln b_s}{S+1}\frac{\sqrt{2}}{3} (\lambda_0^{\eta}-\lambda_1^{\eta}) \frac{\Delta_\Gamma}{\Gamma} \nn\\
\delta_\eta&=&-b\frac{\pi \ln b_s}{S+1} \frac{2}{3}(\lambda_0^{\eta}-\lambda_1^{\eta})\frac{\Delta_\Gamma-3J}{\Gamma} \nn\\
\nu_\eta&=& b\left[1-\frac{\pi \ln b_s}{S+1}(\lambda_0^{\eta}+2\lambda_1^{\eta})\left(\frac{2}{3}  \frac{\Delta_\Gamma}{\Gamma}-2\frac{J}{\Gamma}\right) \right]\nn\\
\rho_\eta&=&\sigma_\eta,
\label{eq:U1_boson_C_expr}
\eea
in which $\eta=D,C$ on the left correspond to $\eta=\text{u},\text{s}$ on the right;
and $O(\frac{\Delta_\Gamma^2}{\Gamma^2})$, $O(\frac{\Delta_\Gamma J}{\Gamma^2})$, $O(\frac{J^2}{\Gamma^2})$ terms are neglected. 
Plugging Eqs. (\ref{eq:lambda_u_value},\ref{eq:lambda_s_value}) into Eq. (\ref{eq:U1_boson_C_expr}), we obtain Eqs. (\ref{eq:U1_boson_C_expr_b},\ref{eq:U1_boson_D_expr_b}).
The above expressions are obtained by retaining terms up to linear order in
$\Delta_\Gamma/\Gamma$ and $J/\Gamma$, with
$O(\Delta_\Gamma^2/\Gamma^2)$, $O(\Delta_\Gamma J/\Gamma^2)$, and
$O(J^2/\Gamma^2)$ terms neglected, and are therefore controlled only near
$(\Delta_\Gamma,J)=(0,0)$, i.e. for $|\Delta_\Gamma|/\Gamma \ll 1$ and
$0<J/\Gamma \ll 1$.
Detailed derivations of Eq. (\ref{eq:U1_boson_C_expr}) are included in Appendix \ref{app:explicit_KHG_RG}. 

It is worth noting that according to RG predictions,  $\sigma_C,\rho_C,\nu_D,\sigma_D,\rho_D$ do not contain linear terms in  $J$,
and their dependences on $J$ start with $J^2$;
in addition, $\nu_D$ does not contain linear term in $\Delta_\Gamma$.
Again, the deviations away from U(1) symmetric bosonization coefficients are suppressed by a factor of $1/S$ in the large-$S$ limit. 

Finally, as a consistency check, we note that using the $O$-transformation between the $U_6$ frame for $C_1,C_2$ and the $OU_6$ frame for $\lambda_C,\sigma_C,\delta_C,\nu_C,\rho_C$, there are the relations
\bea
C_1&=&\nu_C(J=0)+\sqrt{2}\rho_C(J=0)\nn\\
C_2&=&\nu_C(J=0)-\frac{1}{\sqrt{2}}\rho_C(J=0) \nn\\
D_1&=&\nu_D(J=0)+\sqrt{2}\rho_D(J=0)\nn\\
D_2&=&\nu_D(J=0)-\frac{1}{\sqrt{2}}\rho_D(J=0). 
\label{eq:relation_CD_nurho}
\eea
It can be verified that Eq. (\ref{eq:C1C2_KG_RG}) and Eq. (\ref{eq:U1_boson_C_expr}) indeed satisfy the relations in Eq. (\ref{eq:relation_CD_nurho}).

\subsection{DMRG numerics}

It turns out that a numerical determination of the ten bosonization coefficients is very difficult.
We have calculated the spin correlation functions on $L=144$ sites for spin-$1/2$ Kitaev-Heisenberg-Gamma model
and tried to compare with the predictions of the nonsymmorphic abelian bosonization formulas in Eq. (\ref{eq:abelian_LL1_matrix}).
However, unlike the Kitaev-Gamma chain, the numerical results  for Kitaev-Heisenberg-Gamma chain cannot yield reliable values of the bosonization coefficients. 
This is possibly because of huge finite size effects in Kitaev-Heisenberg-Gamma chains. 

\section{Summary}
\label{sec:summary}

In this work, we have investigated the bosonization formulas of half-odd-integer-spin Kitaev chains by means of a renormalization-group analysis. 
Although the low-energy theory exhibits emergent continuous symmetry, the bosonization coefficients retain the imprint of the exact discrete symmetries of the microscopic lattice model. 
The  effects of the breaking of emergent continuous symmetries in the bosonization coefficients are found to scale as \(1/S\).
For the Kitaev-Heisenberg-Gamma chain, we have identified  ten independent bosonization coefficients and showed that five of them are independent of Heisenberg coupling  to linear order. 
These results provide a useful basis for analyzing the magnetic ordering patterns in 2D generalized Kitaev models based on a quasi-one-dimensional approach.

\begin{acknowledgments}

J.L. and W.Y.  are supported by the National Natural Science Foundation of
China (Grants No. 12474476)
and the Fundamental Research Funds for the Central Universities (Grant No. 63263108).
C.X. acknowledges the support from Shuimu Tsinghua Scholar program of Tsinghua University. 
The DMRG calculations in this work were performed using the software package ITensor in  Ref. \onlinecite{ITensor}.
Numerical data are available upon request from the authors. 

\end{acknowledgments}

\appendix
\begin{widetext}




\section{Derivation of Heisenberg model in the large-$U$ limit}
\label{app:2nd_perturbation}

To keep the notation fully consistent with the main text, we use $S$ for the on-site spin quantum number and $c_{ia\sigma}$ for the fermion operator throughout this appendix. For each color $a$, we define
\begin{equation}
\vec{S}_{ia}
=
\frac12
\sum_{\alpha,\beta}
c_{ia\alpha}^\dagger \vec{\sigma}_{\alpha\beta} c_{ia\beta},
\qquad
\vec{S}_i=\sum_{a=1}^{2S}\vec{S}_{ia}.
\label{eq:Sia_def}
\end{equation}
Let $P$ denote the projector onto the low-energy subspace in which each site contains $n_c=2S$ electrons and carries the maximal total spin $S$. In this subspace, each color is singly occupied, so that
\begin{equation}
P n_{ia} P = P,
\qquad
n_{ia}\equiv \sum_{\sigma} c_{ia\sigma}^\dagger c_{ia\sigma}.
\label{eq:nia_proj}
\end{equation}
Within the projected low-energy subspace $P\mathcal H$, we identify any operator of the form $P O P$ with its restriction to $P\mathcal H$ and keep the same notation for simplicity.
Moreover, the projection identities become
\begin{equation}
\label{eq:P1P}
P \vec{S}_{ia} P = \frac{1}{2S}\vec{S}_{i},
\end{equation}
and
\begin{equation}
\label{eq:PPP}
P\bigl(\vec{S}_{ia} \cdot \vec{S}_{ja}\bigr)P
= \frac{1}{4S^2}\vec{S}_{i} \cdot \vec{S}_{j}.
\end{equation}

For a fixed nearest-neighbor bond $\langle ij\rangle$, we write
\begin{equation}
H_{1,ij}
=
-t \sum_{a=1}^{2S}\sum_{\sigma}
\left(
c_{ia\sigma}^\dagger c_{ja\sigma}
+
c_{ja\sigma}^\dagger c_{ia\sigma}
\right),
\label{eq:H1ij}
\end{equation}
while $H_0\equiv H_U$. Since a single hopping process changes the local occupation numbers, one has
\begin{equation}
P H_{1,ij} P = 0.
\end{equation}
Therefore, the leading correction is the second-order term
\begin{equation}
H_{\mathrm{eff},ij}^{(2)}
=
P H_{1,ij}\frac{1}{E_0-H_0}(1-P)H_{1,ij}P.
\label{eq:Heff2_start}
\end{equation}
In the unperturbed low-energy manifold, both sites $i$ and $j$ carry spin $S$, and hence the Hund energy on this bond is
\begin{equation}
E_0^{(ij)}=-2U\,S(S+1).
\end{equation}

After one hopping event, one site contains $2S+1$ electrons and the other contains $2S-1$ electrons. On the site with $2S+1$ electrons, one color becomes doubly occupied and must form a spin singlet on that color; the remaining $2S-1$ singly occupied colors therefore carry total spin $S-\frac12$. The site with $2S-1$ electrons likewise has maximal spin $S-\frac12$. Thus the intermediate-state energy is
\begin{equation}
E_{\mathrm{int}}^{(ij)}
=
-2U\left(S-\frac12\right)\left(S+\frac12\right).
\end{equation}
Therefore,
\begin{equation}
E_0^{(ij)}-E_{\mathrm{int}}^{(ij)}
=
-U\left(2S+\frac12\right),
\end{equation}
and Eq.~(\ref{eq:Heff2_start}) becomes
\begin{equation}
H_{\mathrm{eff},ij}^{(2)}
=
\frac{P H_{1,ij}^2 P}{-U\left(2S+\frac12\right)}.
\label{eq:Heff2_general}
\end{equation}

Because the hopping is diagonal in the color index, only virtual processes with the same color $a$ contribute. The two equivalent processes $i\to j\to i$ and $j\to i\to j$ give the same contribution, and hence
\begin{align}
P H_{1,ij}^2 P
&=
2t^2
\sum_{a=1}^{2S}\sum_{\sigma,\sigma'}
P\,
c_{ia\sigma}^\dagger c_{ja\sigma}
c_{ja\sigma'}^\dagger c_{ia\sigma'}
\,P
\nonumber\\
&=
2t^2
\sum_{a=1}^{2S}
P\left[
n_{ia}
-
\sum_{\sigma,\sigma'}
c_{ia\sigma}^\dagger c_{ia\sigma'}
c_{ja\sigma'}^\dagger c_{ja\sigma}
\right]P .
\label{eq:H1sq_step1}
\end{align}
To simplify the quartic term, we use
\begin{equation}
c_{ia\alpha}^\dagger c_{ia\beta}
=
\frac12 n_{ia}\delta_{\alpha\beta}
+
\bigl(\vec{S}_{ia}\cdot\vec{\sigma}\bigr)_{\alpha\beta},
\end{equation}
which implies 
\begin{equation}
\sum_{\sigma,\sigma'}
c_{ia\sigma}^\dagger c_{ia\sigma'}
c_{ja\sigma'}^\dagger c_{ja\sigma}
=
\frac12 n_{ia}n_{ja}
+
2\,\vec{S}_{ia}\cdot\vec{S}_{ja}.
\label{eq:fierz_identity}
\end{equation}
Substituting Eq.~(\ref{eq:fierz_identity}) into Eq.~(\ref{eq:H1sq_step1}) and using Eq.~(\ref{eq:nia_proj}), we obtain
\begin{align}
P H_{1,ij}^2 P
&=
2t^2
\sum_{a=1}^{2S}
P\left[
1-\left(\frac12+2\,\vec{S}_{ia}\cdot\vec{S}_{ja}\right)
\right]P
\nonumber\\
&=
4t^2
\sum_{a=1}^{2S}
P\left(
\frac14-\vec{S}_{ia}\cdot\vec{S}_{ja}
\right)P .
\label{eq:H1sq_step2}
\end{align}

Finally, applying Eqs.~(\ref{eq:P1P}) and (\ref{eq:PPP}), one finds
\begin{align}
P H_{1,ij}^2 P
&=
4t^2
\sum_{a=1}^{2S}
\left(
\frac14 
-
\frac{1}{4S^2}\vec{S}_i\cdot\vec{S}_j
\right)
\nonumber\\
&=
4t^2
\left(
\frac{S}{2}
-
\frac{1}{2S}\vec{S}_i\cdot\vec{S}_j
\right).
\label{eq:H1sq_step3}
\end{align}
Substituting Eq.~(\ref{eq:H1sq_step3}) into Eq.~(\ref{eq:Heff2_general}) yields
\begin{align}
H_{\mathrm{eff},ij}^{(2)}
&=
-\frac{4t^2}{U\left(2S+\frac12\right)}
\left(
\frac{S}{2}
-
\frac{1}{2S}\vec{S}_i\cdot\vec{S}_j
\right)
\nonumber\\
&=
\frac{2t^2}{US\left(2S+\frac12\right)}
\vec{S}_i\cdot\vec{S}_j
-
\frac{2St^2}{U\left(2S+\frac12\right)}
\nonumber\\
&=
\frac{4t^2}{US(4S+1)}
\vec{S}_i\cdot\vec{S}_j
-
\frac{4St^2}{U(4S+1)} .
\label{eq:Heffij_final}
\end{align}
Summing over all nearest-neighbor bonds, we arrive at
\begin{equation}
H_{\mathrm{eff}}
=
J\sum_{\langle ij\rangle}\vec{S}_i\cdot\vec{S}_j
+\mathrm{const},
\qquad
J=\frac{4t^2}{US(4S+1)},
\end{equation}
which reproduces Eqs.~(\ref{eq:H_eff_Heisenberg}) and (\ref{eq:J_expresssion}).

The $1/S^2$ scaling of $J$ can be easily understood. The number of allowed virtual hopping channels on a given bond grows as $n_c=2S$, giving an overall factor of order $S$ in the second-order matrix element. On the other hand, after one electron hops, one site contains $n_c-1$ electrons, whose maximal spin is $S-1/2$. The corresponding Hund energy denominator therefore scales as $\Delta E \sim U [S(S+1)-(S-1/2)(S+1/2)]\sim US $. Hence the second-order energy shift on a bond is of order $\Delta E^{(2)} \sim t^2S/(US)\sim t^2/U$. Since $\vec{S}_i\cdot\vec{S}_j\sim O(S^2)$ in the spin-$S$ manifold,  consistency  with the effective form $J\,\vec{S}_i\cdot\vec{S}_j$ implies $J \sim t^2/(US^2)$, in agreement with the explicit result in Eq. (\ref{eq:J_expresssion}). 





\section{One-loop flow equations in spin-$S$ Kitaev-Gamma chains}

In this appendix, 
we derive the flow equations of scaling fields in spin-$S$ Kitaev-Gamma chains by calculating the one-loop diagrams in Fig. \ref{fig:diagrams} (a,b).
We will focus on the renormalization effects induced by the SU(2) breaking $\delta_\Gamma$ term. 

\subsection{Diagram in Fig. \ref{fig:diagrams} (a)}
\label{app:Lambda_derive}


In this appendix, we evaluate the diagram in Fig. \ref{fig:diagrams} (a). 
We focus on the uniform component of the scaling field,
and the treatment for the staggered component is exactly similar. 
The calculation for RG flow equations of staggered components is exactly similar.

We first rewrite the lattice interaction and the external-field coupling in the frequency-momentum space, and then evaluate the one-loop correction by integrating out the fast modes. 
Because of the three-site periodicity of the lattice, the resulting renormalization generally mixes different sublattices.
In the following derivation, we use the notations
\[
\text{site in unit cell: } {i}, {j}, {l}
\qquad
\text{color: }a,b,c...
\qquad
\text{spinor: }\alpha,\beta;\ \sigma,\delta;\ \theta,\gamma;...
\qquad
\text{spin direction: } M,N,P,
\]
where $1\leq {i}, {j}, {l}\leq 3$, and instead of Greek letters $\alpha,\mu,\nu$ in Fig. \ref{fig:diagrams} (a), we use $M,N,P$ to denote the spin directions $x,y,z$.

The corresponding four-fermion vertex is shown in Fig \ref{fig:vertex} (a). 
Using
\begin{equation}
\frac{1}{\beta}\int d\tau \,
e^{-i(-\omega_1+\omega_2-\omega_3+\omega_4)\tau}
=
\delta_{\omega_1-\omega_2+\omega_3-\omega_4,\,0}
\end{equation}
\begin{equation}
\frac{1}{L}\sum_n
e^{\,i(k_1-k_2+k_3-k_4)3na}
=
\frac{1}{3}\sum_{m=1}^{3}
\delta_{\vec{k}_1-\vec{k}_2+\vec{k}_3-\vec{k}_4,\,
\frac{2\pi}{3a}m},
\end{equation}
rewriting the spin operators in terms of fermion bilinears, and Fourier transforming to momentum space, the interaction vertex can be expressed as
\bea
&&\delta_\Gamma \int d\tau \sum_{n}
   S^{M}_{ {i}+3n}(\tau)\,S^{M}_{ {j}+3n}(\tau) \nn\\ 
&=& \delta_\Gamma \int d\tau \sum_{n}\sum_{ab}\sum_{\alpha\beta\gamma\delta}
   c^{\dagger}_{i+3n,a\alpha}(\tau)\,
   \frac{(\sigma^{M})_{\alpha\beta}}{2}\,
   c_{i+3n,a\beta}(\tau)\;
   c^{\dagger}_{j+3n,b\gamma}(\tau)\,
   \frac{(\sigma^{M})_{\gamma\delta}}{2}\,
   c_{j+3n,b\delta}(\tau) \nn\\
&=& \frac{\delta_\Gamma}{L^2 \beta^2}
   \sum_{n}\sum_{\vec{k}_1\vec{k}_2\vec{k}_3\vec{k}_4}\sum_{\omega_1\omega_2\omega_3\omega_4}\sum_{ab}\sum_{\alpha\beta\gamma\delta}
   \int d\tau \,
   c^{\dagger}_{a\alpha}(\vec{k}_{1},\omega_1)\,
   \frac{(\sigma^{M})_{\alpha\beta}}{2}\,
   c_{a\beta}(\vec{k}_{2},\omega_2)\;
   c^{\dagger}_{b\gamma}(\vec{k}_{3},\omega_3)\,
   \frac{(\sigma^{M})_{\gamma\delta}}{2}\,
   c_{b\delta}(\vec{k}_{4},\omega_4) \nn\\
&&\quad\times
   e^{-i(\omega_{1}-\omega_{2}+\omega_{3}-\omega_{4})\tau}\,
   e^{i(\vec{k}_{1}-\vec{k}_{2})\cdot( {i}+3n)a\hat{x}}\,
   e^{i(\vec{k}_{3}-\vec{k}_{4})\cdot( {j}+3n)a\hat{x}} \nn\\
&=& \frac{\delta_\Gamma}{3L \beta}
   \sum_{\vec{k}\vec{p}\vec{q}}\sum_{\omega_1\omega_2\omega_3\omega_4}\sum_{ab}\sum_{\alpha\beta\gamma\delta}\sum_{m}
   c^{\dagger}_{a\alpha}(\vec{k}+\vec{q},\omega_1)\,
   \frac{(\sigma^{M})_{\alpha\beta}}{2}\,
   c_{a\beta}(\vec{k},\omega_2)\;
   c^{\dagger}_{b\gamma}(\vec{p}-\vec{q},\omega_3)\,
   \frac{(\sigma^{M})_{\gamma\delta}}{2}\,
   c_{b\delta}(\vec{p}-\frac{2\pi m}{3a}\hat{x},\omega_4) \nn\\
&&\quad\times
   e^{i\vec{q}\cdot( {i}- {j})a\hat{x}}\,
   e^{i\frac{2\pi m  {j}}{3}}\delta_{\omega_1-\omega_2+\omega_3-\omega_4,\,0}.
\label{eq:1}
\eea

To track the renormalization of the spin operators, we next introduce external scaling fields coupled linearly to the local spins. 
The corresponding vertex is shown in Fig \ref{fig:vertex} (b), and its expression in the frequency-momentum space form is given by
\begin{flalign}
&\int d\tau \sum_{n}
   h_{\text{u}, {l}}^N(\tau,n)\,S_{ \text{u},{l}+3n}^N(\tau)  \nn\\
&= \frac{1}{L\beta}
   \int d\tau \sum_{n}
   \sum_{\vec{k}_5,\vec{k}_6,\vec{k}_7}\sum_{\omega_5,\omega_6,\omega_7}\sum_{c}\sum_{\theta\eta} e^{-i(\omega_{6}-\omega_{7})\tau}e^{i(\vec{k}_6-\vec{k}_7)\cdot( {l}+3n)a\hat{x}}
   \nn\\
&\quad\times h_{\text{u}, {l}}^N(\vec{k}_5,\omega_5)\,
   e^{-i\omega_5\tau+i\vec{k}_5\cdot 3na\hat{x}}
   c^{\dagger}_{c\theta}(\vec{k}_{6},\omega_6)\,
   \frac{(\sigma^{N})_{\theta\eta}}{2}\,
   c_{c\eta}(\vec{k}_{7},\omega_7) 
    \nn\\
&= \frac{1}{3}
   \sum_{\vec{k},\vec{q}}\sum_{\omega_5,\omega_6,\omega_7}\sum_{c}\sum_{\theta\eta}\sum_{m'}
   e^{i\vec{q}\cdot{l}a\hat{x}}\,e^{i\frac{2\pi}{3}m' {l}}
   h_{\text{u}, {l}}^N(-\vec{q},\omega_5)
   c^{\dagger}_{c\theta}\left(\vec{k}+\frac{2\pi}{3a}m'\hat{x},\omega_6\right)
   \frac{(\sigma^{N})_{\theta\eta}}{2}\,
   c_{c\eta}(\vec{k}-\vec{q},\omega_7)\,
   \delta_{\omega_5-\omega_6+\omega_7,\,0}.
\label{eq:2}
\end{flalign}
Notice that for uniform component of the scaling field, both $\vec{q}$ and $\omega_5$ are small. 

We consider the one-loop correction to the scaling field generated by the $\delta_\Gamma$ term in Fig. \ref{fig:diagrams} (a). 
Contracting the fast modes connects the interaction vertex to the external-field vertex, and the momentum-conservation conditions fix the internal variables accordingly. Keeping the terms that reproduce the original field insertion, one obtains
\bea
F_{\text{loop,}a}&=&\frac{\delta_\Gamma}{9L\beta}
\sum_{m,m'}\sum_{\vec{k},\vec{p},\vec{q},\vec{k}^\prime,\vec{q}^\prime}\sum_{\omega_1,\omega_2,\omega_3,\omega_4,\omega_5,\omega_6,\omega_7}\sum_{abc}\sum_{\alpha\beta\gamma\delta\theta\eta}
e^{i\frac{2\pi}{3}m {j}}
e^{i\vec{q}\cdot ( {i}- {j})a\hat{x}}\,
e^{i\vec{q}'\cdot {l}a\hat{x}}e^{i\frac{2\pi}{3}m' {l}}\nn\\
&&\times h_{ \text{u},{l}}^N(-\vec{q}',\omega_5)\,
c^{\dagger}_{a\alpha}(\vec{k}+\vec{q},\omega_1)
\frac{(\sigma^{M})_{\alpha\beta}}{2}
c_{a\beta}(\vec{k},\omega_2)
\nn\\
&&\times
\left\langle
c^{\dagger}_{b\gamma}(\vec{p}-\vec{q},\omega_3)
\frac{(\sigma^{M})_{\gamma\delta}}{2}
c_{b\delta}\!\left(\vec{p}-\frac{2\pi m}{3a}\hat{x},\omega_4\right)
c^{\dagger}_{c\theta}\!\left(\vec{k}'+\frac{2\pi m'}{3a}\hat{x},\omega_6\right)
\frac{(\sigma^{N})_{\theta\eta}}{2}
c_{c\eta}(\vec{k}'-\vec{q}',\omega_7)
\right\rangle_f
\nn\\
&&\times
\delta_{\omega_1-\omega_2+\omega_3-\omega_4,\,0}\,
\delta_{\omega_5-\omega_6+\omega_7,\,0}.
\label{eq:Fa_loop}
\eea

Using Wick's theorem, the fast-mode contraction gives
\bea
&&\sum_{\gamma\delta\theta\eta}\left\langle
c^{\dagger}_{b\gamma}(\vec{p}-\vec{q},\omega_3)
\frac{(\sigma^{M})_{\gamma\delta}}{2}
c_{b\delta}\!\left(\vec{p}-\frac{2\pi m}{3a}\hat{x},\omega_4\right)
c^{\dagger}_{c\theta}\!\left(\vec{k}'+\frac{2\pi m'}{3a}\hat{x},\omega_6\right)
\frac{(\sigma^{N})_{\theta\eta}}{2}
c_{c\eta}(\vec{k}'-\vec{q}',\omega_7)
\right\rangle_f
\nn\\[4pt]
&=&
-\frac14 \sum_{\gamma\delta\theta\eta}
(\sigma^{M})_{\gamma\delta}(\sigma^{N})_{\theta\eta}
\left\langle
c_{b\delta}\!\left(\vec{p}-\frac{2\pi m}{3a}\hat{x},\omega_4\right)
c^{\dagger}_{c\theta}\!\left(\vec{k}'+\frac{2\pi m'}{3a}\hat{x},\omega_6\right)
\right\rangle_f
\left\langle
c_{c\eta}(\vec{k}'-\vec{q}',\omega_7)
c^{\dagger}_{b\gamma}(\vec{p}-\vec{q},\omega_3)
\right\rangle_f
\nn\\[4pt]
&=&
-\frac14
\text{tr}(\sigma^M\sigma^N)\,
\delta_{bc}\,
\delta_{\,\vec{p}-\frac{2\pi m}{3a}\hat{x},\,\vec{k}'+\frac{2\pi m'}{3a}\hat{x}}\,
\delta_{\,\vec{p}-\vec{q},\,\vec{k}'-\vec{q}'}\,
\delta_{\omega_4,\omega_6}\,
\delta_{\omega_3,\omega_7}\,
G\!\left(\vec{p}-\frac{2\pi m}{3a}\hat{x},\omega_4\right)
G(\vec{p}-\vec{q},\omega_3)
\nn\\[4pt]
&=&
-\frac12\delta_{MN}\,
\delta_{bc}\,
\delta_{\,\vec{p}-\frac{2\pi m}{3a}\hat{x},\vec{k}'+\frac{2\pi m'}{3a}\hat{x}}\,
\delta_{\,\vec{p}-\vec{q},\,\vec{k}'-\vec{q}'}\,
\delta_{\omega_4,\omega_6}\,
\delta_{\omega_3,\omega_7}\,
G\!\left(\vec{p}-\frac{2\pi m}{3a}\hat{x},\omega_4\right)
G(\vec{p}-\vec{q},\omega_3),
\label{f6}
\eea
with
\begin{equation}
\vec{p}-\vec{q} = \vec{k}'-\vec{q}',\qquad
\vec{p}-\frac{2\pi m}{3a}\hat{x}=\vec{k}'+\frac{2\pi m'}{3a}\hat{x},\qquad
\omega_4=\omega_6,\qquad
\omega_3=\omega_7.
\end{equation}
Therefore,
\begin{equation}
\vec{q}=\vec{q}'+\frac{2\pi}{3a}(m+m')\hat{x}.
\end{equation}

Plugging Eq.~\eqref{f6} into Eq. (\ref{eq:Fa_loop}), we obtain
\begin{flalign}
&F_{\text{loop,}a}=\frac{\delta_\Gamma}{18L\beta}\delta_{MN}n_c\sum_{m,m'}\sum_{\vec{k},\vec{p},\vec{q}^\prime}\sum_{\omega_1,\omega_2,\omega_3,\omega_4}\sum_{a}\sum_{\alpha\beta}
e^{i\frac{2\pi}{3}mj } e^{i\vec{q}^\prime \cdot (i-j)a\hat{x} } e^{i \frac{2\pi}{3} (m+m^\prime) (i-j) } e^{i\vec{q}^\prime \cdot la\hat{x}} e^{i\frac{2\pi}{3}m^\prime l}\nn\\
&\qquad\times h_{\text{u},l}^N(-\vec{q}^\prime,\omega_1-\omega_2) c^\dagger_{a\alpha} (\vec{k}+\vec{q}'+\frac{2\pi}{3}(m+m')\hat{x},\omega_1)  \frac{(\sigma^M)_{\alpha\beta}}{2} c_{a\beta} (\vec{k},\omega_2)\nn\\
&\qquad\times G(\vec{p}-\frac{2\pi}{3a}m\hat{x},\omega_3+\omega_1-\omega_2) G(\vec{p}-\vec{q}'-\frac{2\pi}{3a}(m+m')\hat{x},\omega_3),
\label{eq:F_loop_a_2}
\end{flalign}
in which the factor of $n_c$ comes from $\sum_{bc}\delta_{bc}$.
Since $h_l^N$ is slowly varying, both $\vec{q}^\prime$ and $\omega_1-\omega_2$ are small. 
Therefore, we can neglect $\vec{q}^\prime$ and $\omega_1-\omega_2$ in fermion propagator $G$. 
Rearranging and recombining the terms, Eq. (\ref{eq:F_loop_a_2}) becomes
\begin{flalign}
&F_{\text{loop,}a}=\frac{\delta_\Gamma}{18L\beta}\delta_{MN}n_c\nn\\
&\qquad \times \sum_{mm'}\sum_{\omega_1\omega_2\omega_3}\sum_{\vec{k}\vec{p}\vec{q}^\prime}\sum_a\sum_{\alpha\beta}
   e^{i\vec{q}'\cdot{i}a\hat{x}}\,e^{i\frac{2\pi}{3}(m+m') {i}} h_{\text{u},l}^N(-\vec{q}^\prime,\omega_1-\omega_2) c^\dagger_{a\alpha} (\vec{k}+\vec{q}'+\frac{2\pi}{3}(m+m')\hat{x},\omega_1)  \frac{(\sigma^M)_{\alpha\beta}}{2} c_{a\beta} (\vec{k},\omega_2)\nn\\
&\qquad\times e^{-i\vec{q}' (j-l)a\hat{x} }  e^{-i\frac{2\pi}{3}m^\prime (j-l)  } G(\vec{p}-\frac{2\pi}{3a}m\hat{x},\omega_3) G(\vec{p}-\frac{2\pi}{3a}(m+m')\hat{x},\omega_3).
\end{flalign}
Approximating $e^{-i\vec{q}' (j-l)a\hat{x} }$ as $1$ (since $|\vec{q}'|\ll 1$), redefining $\vec{p}-\frac{2\pi}{3a}m\hat{x}$ as $\vec{p}'$, and denoting $m+m^\prime$ as $\bar{m}$, we obtain 
\begin{flalign}
&F_{\text{loop,}a}=
\frac{1}{3}\sum_{\bar{m}}\sum_{\omega_1\omega_2}\sum_{\vec{k}\vec{q}'}\sum_a\sum_{\alpha\beta}
e^{i\vec{q}'\cdot{i}a\hat{x}}\,e^{i\frac{2\pi}{3}\bar{m} {i}} h_{\text{u},l}^N(-\vec{q}^\prime,\omega_1-\omega_2) c^\dagger_{a\alpha} (\vec{k}+\vec{q}'+\frac{2\pi}{3}\bar{m}\hat{x},\omega_1)  \frac{(\sigma^M)_{\alpha\beta}}{2} c_{a\beta} (\vec{k},\omega_2)\nn\\
&\qquad \times\frac{\delta_\Gamma}{6L\beta}\delta_{MN}n_c\sum_{m'}\sum_{\omega_3}\sum_{\vec{q}'} e^{-i\frac{2\pi}{3}m^\prime (j-l)  } G(\vec{p}',\omega_3) G(\vec{p}'-\frac{2\pi}{3a}m'\hat{x},\omega_3).
\label{eq:F_loop_a_rearrange}
\end{flalign}
Using Eq. (\ref{eq:2}), Eq. (\ref{eq:F_loop_a_rearrange}) can be written as 
\begin{flalign}
&F_{\text{loop,}a}=\delta_{MN}\delta_\Gamma\lambda_{\text{u}, {j} {l}}^{(S)}\ln b\int d\tau \sum_{n}
   h_{ \text{u},{l}}^N(\tau,n)\,S_{ {i}+3n}^N(\tau), 
\end{flalign}
in which $\lambda_{\text{u}, {j} {l}}^{(S)}$ is given by
\bea
\lambda_{\text{u}, {j} {l}}^{(S)}\ln b
=
-n_c\frac{a}{6}\sum_m e^{-i\frac{2\pi}{3}m( {j}- {l})}
\int_{\Lambda/b}^{\Lambda} d\omega d\vec{k}\,G(\vec{k},\omega)\,G\!\left({\vec{k}+\frac{2\pi}{3a}m\hat{x}},\omega\right),
\eea
where $k=(i\omega,\vec{k})$ in accordance with the convention in the main text. 


\subsection{Diagram in Fig. \ref{fig:diagrams} (b)}
\label{app:proof_vanish}

Next we consider the one-loop correction to the scaling field generated by the $\delta_\Gamma$ term in Fig. \ref{fig:diagrams} (b). 
We again focus on the uniform component of the scaling fields.
The analysis for the staggered component is exactly similar.

The contribution $F_{\text{loop},b}$ is given by 
\bea
F_{\text{loop},b}&=&\frac{\delta_\Gamma}{9L\beta}
\sum_{m,m'}\sum_{\vec{k},\vec{p},\vec{q},\vec{k}^\prime,\vec{q}^\prime}\sum_{\omega_1,\omega_2,\omega_3,\omega_4,\omega_5,\omega_6,\omega_7}\sum_{abc}\sum_{\alpha\beta\gamma\delta\theta\eta}
e^{i\frac{2\pi}{3}m {j}}
e^{i\vec{q}\cdot ( {i}- {j})a\hat{x}}\,
e^{i\vec{q}'\cdot {l}a\hat{x}}e^{i\frac{2\pi}{3}m' {l}}   h_{ \text{u},{l}}^P(-\vec{q}',\omega_5) \nn\\
&&\times c^{\dagger}_{a\alpha}(\vec{k}+\vec{q},\omega_1)
\frac{(\sigma^{M})_{\alpha\beta}}{2}
\left\langle c_{a\beta}(\vec{k},\omega_2)c^{\dagger}_{c\theta}\!\left(\vec{k}'+\frac{2\pi m'}{3a}\hat{x},\omega_6\right)
\right\rangle_f
\frac{(\sigma^{P})_{\gamma\delta}}{2}
\left\langle c_{c\eta}(\vec{k}'-\vec{q}',\omega_7) c^{\dagger}_{b\gamma}(\vec{p}-\vec{q},\omega_3)
\right\rangle_f\nn\\
&&\times \frac{(\sigma^{N})_{\theta\eta}}{2} c_{b\delta}\!\left(\vec{p}-\frac{2\pi m}{3a}\hat{x},\omega_4\right)
\delta_{\omega_1-\omega_2+\omega_3-\omega_4,\,0}\,
\delta_{\omega_5-\omega_6+\omega_7,\,0}.
\label{eq:F_loop_b_1}
\eea
The contractions impose the constraints
\bea
\vec{k}=\vec{k}'+\frac{2\pi}{3a} m'\hat{x},\qquad \vec{q}=\vec{p}-\vec{k}'+\vec{q}',\qquad  \omega_6=\omega_2, \qquad \omega_7=\omega_3.
\label{eq:constraint_loop_b}
\eea
Plugging Eq. (\ref{eq:constraint_loop_b}) into Eq. (\ref{eq:F_loop_b_1}), we obtain 
\bea
F_{\text{loop},b}&=&\frac{\delta_\Gamma}{9L\beta}\sum_{m,m'}\sum_{\vec{k}'\vec{q}'\vec{p}}
e^{i(\vec{p}-\vec{k}'+\vec{q}')\cdot(i-j)a\hat{x}} e^{i\frac{2\pi}{3}mj} e^{i\vec{q}'\cdot la\hat{x} } e^{i\frac{2\pi}{3}m'l}\nn\\
&&\times h_{\text{u},l}^P (-\vec{q}',\omega_5)
c^\dagger_{a\alpha}(\vec{p}+\vec{q}'+\frac{2\pi}{3a}m'\hat{x},\omega_1 ) \left(\frac{\sigma^M\sigma^P\sigma^N}{8}\right)_{\alpha\delta} c_{a\delta} (\vec{p}-\frac{2\pi}{3a}m\hat{x},\omega_1-\omega_5 )\nn\\
&& \times G(\vec{k}'+\frac{2\pi}{3a}m'\hat{x},\omega_2) G(\vec{k}'-\vec{q}',\omega_2-\omega_5). 
\eea
Notice that there is no overall factor of $n_c$ in this case. 

Since both $\vec{q}'$ and $\omega_5$ are small, we neglect their dependences in fermion propagator.
Then $F_{\text{loop},b}$ can be simplified to
\bea
F_{\text{loop},b}&=&\frac{\delta_\Gamma}{9L\beta}\sum_{\bar{m}} \sum_{\vec{p}\vec{q}^\prime} \sum_{\omega_1\omega_5}e^{i\vec{q}'\cdot (l-j)a\hat{x}} e^{i(\vec{p}+\vec{q}') \cdot ia\hat{x}} e^{-i\vec{p} \cdot ja\hat{x}} e^{i\frac{2\pi}{3} \bar{m} i} \nn\\
&&\times h_{\text{u},l}^P (-\vec{q}',\omega_5)
c^\dagger_{a\alpha}(\vec{p}+\vec{q}',\omega_1 ) \left(\frac{\sigma^M\sigma^P\sigma^N}{8}\right)_{\alpha\delta} c_{a\delta} (\vec{p}-\frac{2\pi}{3a}\bar{m}\hat{x},\omega_1-\omega_5 )\nn\\
&&\times \sum_{m'} \sum_{\vec{k}'}\sum_{\omega_2} e^{-i \frac{2\pi}{3} m'(i-l)} e^{-i\vec{k}'\cdot (i-j)a\hat{x} } G(\vec{k}'+\frac{2\pi}{3a}m'\hat{x},\omega_2) G(\vec{k}',\omega_2).
\eea
Since $\vec{q}'$ is small, we further approximate the factor $e^{i\vec{q}'\cdot (l-j)a\hat{x}}$ as $1$.
Using the following identity for Fourier transformation
\begin{flalign}
&\int d\tau  \sum_n\sum_a\sum_{\alpha\delta}  h_{\text{u},l}^P(n,\tau)c^\dagger_{i+3n,a\alpha}(\tau) \left(\frac{\sigma^M\sigma^P\sigma^N}{8}\right)_{\alpha\delta} c_{j+3n,a\delta} (\tau)=\nn\\
&\frac{1}{3}\sum_{\bar{m}} \sum_{\vec{p}\vec{q}^\prime} \sum_{\omega_1\omega_5}e^{i\vec{q}'\cdot (l-j)a\hat{x}} e^{i(\vec{p}+\vec{q}') \cdot ia\hat{x}} e^{-i\vec{p} \cdot ja\hat{x}} e^{i\frac{2\pi}{3} \bar{m} i} h_{\text{u},l}^P (-\vec{q}',\omega_5)
c^\dagger_{a\alpha}(\vec{p}+\vec{q}',\omega_1 ) \left(\frac{\sigma^M\sigma^P\sigma^N}{8}\right)_{\alpha\delta} c_{a\delta} (\vec{p}-\frac{2\pi}{3a}\bar{m}\hat{x},\omega_1-\omega_5 ),
\end{flalign}
we arrive at 
\bea
F_{\text{loop},b}&=&\delta_\Gamma \lambda^{(S)}_{\text{u},ijl}  \int d\tau  \sum_n\sum_a\sum_{\alpha\delta}  h_{\text{u},l}^P(n,\tau)c^\dagger_{i+3n,a\alpha}(\tau) \left(\frac{\sigma^M\sigma^P\sigma^N}{2}\right)_{\alpha\delta} c_{j+3n,a\delta} (\tau)
\eea
in which $\lambda^{(S)}_{\text{u},ijl}$ is given by
\bea
\lambda^{(S)}_{\text{u},ijl}\ln b=\frac{a}{12}\sum_m
e^{i\frac{2\pi}{3}m( l- i)}
\int d\vec{k}'d\omega\,
e^{-i\vec{k'}\cdot( i- j)a x}\,
G(\vec{k}'+\frac{2\pi}{3a}m'\hat{x},\omega) G(\vec{k}',\omega).
\label{f0}
\eea

Now we show that Eq.~\eqref{f0} vanishes for the only nontrivial case
$| i-j|=1$.
In the present three-periodic setting, $| i-j|$ can only take the values
$0$ or $1$, so there is no $| i-j|=2$ contribution to consider.
For $| i-j|=1$, under the change of variables
\begin{equation}
\vec{k}' \rightarrow \vec{k}' + \frac{\pi}{a} \hat{x},
\qquad
\omega \rightarrow -\omega,
\end{equation}
the phase factor changes sign,
\begin{equation}
e^{-i(\vec{k}'+\frac{\pi}{a}\hat{x})\cdot( i-j)a \hat{x}}
=
-\,e^{-i\vec{k}'\cdot( i-j)a \hat{x}},
\end{equation}
while, using
\begin{equation}
\epsilon\!\left(\vec{k}+\frac{\pi}{a}\hat{x}\right)=-\epsilon(\vec{k}),
\end{equation}
the product of propagators transforms as
\begin{equation}
G\!\left(\vec{k}'+\frac{2\pi}{3a}m'\hat{x}+\frac{\pi}{a}\hat{x},\omega\right)
G\!\left(\vec{k}'+\frac{\pi}{a}\hat{x},\omega\right)
=
G\!\left(\vec{k}'+\frac{2\pi}{3a}m'\hat{x},-\omega\right)
G(\vec{k}',-\omega).
\end{equation}
Since Eq.~\eqref{f0} contains an integration (or Matsubara sum) over $\omega$,
which is invariant under $\omega\rightarrow -\omega$, the product of propagators
is unchanged inside the loop integral.
Therefore,
\begin{equation}
\lambda_{ \text{u},i l j}^{(S)}
=
-\lambda_{ \text{u},i l j}^{(S)}
=
0,
\qquad | i-j|=1,
\end{equation}
and hence Fig.~\ref{fig:diagrams}(b) does not contribute.

\subsection{Explicit RG flow equations}   
\label{app_sub:flow_KG_explicit}

The RG flow equations of the scaling fields for spin-$S$ Kitaev-Gamma chain are given by ($\eta=$u,s)
\bea
\frac{d h_{\eta,1}^x}{d\ln b} &=& (1-2S \delta_\Gamma \lambda^\eta_1)h_{\eta,1}^x - 2S \delta_\Gamma\lambda^\eta_0 h_{\eta,2}^x- 2S \delta_\Gamma\lambda^\eta_1 h_{\eta,3}^x,\nn\\
\frac{d h_{\eta,2}^x}{d\ln b} &=& (1-2S \delta_\Gamma\lambda^\eta_1)h_{\eta,2}^x - 2S \delta_\Gamma\lambda^\eta_0 h_{\eta,1}^x- 2S \delta_\Gamma\lambda^\eta_1h_{\eta,3}^x,\nn\\
\frac{d h_{\eta,3}^x}{d\ln b} &=& h_{\eta,3}^x,
\label{eq:flow_x}
\eea
\bea
\frac{d h_{\eta,1}^z}{d\ln b} &=&h_{\eta,1}^z ,\nn\\
\frac{d h_{\eta,2}^z}{d\ln b} &=&(1-2S \delta_\Gamma \lambda^\eta_1)h_{\eta,2}^z- 2S \delta_\Gamma\lambda^\eta_0 h_{\eta,3}^z- 2S \delta_\Gamma\lambda^\eta_1 h_{\eta,1}^z,\nn\\
\frac{d h_{\eta,3}^z}{d\ln b} &=&(1-2S \delta_\Gamma \lambda^\eta_1)h_{\eta,3}^z- 2S \delta_\Gamma\lambda^\eta_0 h_{\eta,2}^z- 2S \delta_\Gamma\lambda^\eta_1 h_{\eta,1}^z,
\label{eq:flow_z}
\eea
\bea
\frac{d h_{\eta,1}^y}{d\ln b} &=&(1-2S \delta_\Gamma \lambda^\eta_1)h_{\eta,1}^y- 2S \delta_\Gamma\lambda^\eta_0 h_{\eta,3}^y- 2S \delta_\Gamma\lambda^\eta_1 h_{\eta,2}^y,\nn\\
\frac{d h_{\eta,2}^y}{d\ln b} &=&h_{\eta,2}^y,\nn\\
\frac{d h_{\eta,3}^y}{d\ln b} &=&(1-2S \delta_\Gamma \lambda^\eta_1)h_{\eta,3}^y- 2S \delta_\Gamma\lambda^\eta_0 h_{\eta,1}^y- 2S \delta_\Gamma\lambda^\eta_1 h_{\eta,2}^y.
\label{eq:flow_y}
\eea


\section{Explicit calculations of RG flow equations in Kitaev-Heisenberg-Gamma chain}
\label{app:explicit_KHG_RG}

In the $U_6$ and $OU_6$ frames, we define column vectors $\vec{S}_i^\prime$ and $\vec{S}_i^{\prime\prime}$ as
$\vec{S}_i^\prime=
(S_i^{\prime x},
S_i^{\prime y},
S_i^{\prime z})^T$
and $\vec{S}_i^\prime=O\,\vec{S}_i^{\prime\prime}$,
where $O$ is defined in Eq. (\ref{eq:def_O_mat}). 

In the $U_6$ frame, the Hamiltonians in a unit cell  are given by
\begin{eqnarray}
H'_{12}=\vec{S}_1^{\prime T} M_{12}\vec{S}_2^{\prime},\qquad
H'_{23}=\vec{S}_2^{\prime T} M_{23}\vec{S}_3^{\prime},\qquad
H'_{2=34}=\vec{S}_3^{\prime T} M_{34}\vec{S}_4^{\prime},
\end{eqnarray}
where
\begin{eqnarray}
M_{12}=
\begin{pmatrix}
-K-J&0&0\\
0&\Gamma&-J\\
0&-J&\Gamma
\end{pmatrix},
\qquad
M_{23}=
\begin{pmatrix}
\Gamma&-J&0\\
-J&\Gamma&0\\
0&0&-K-J
\end{pmatrix},
\qquad
M_{34}=
\begin{pmatrix}
\Gamma&0&-J\\
0&-K-J&0\\
-J&0&\Gamma
\end{pmatrix}.
\end{eqnarray}

Define matrices $A_1^{(K\Gamma)},A_2^{(K\Gamma)},A_3^{(K\Gamma)}$ as
\begin{eqnarray}
A_1^{(K\Gamma)}=O^T\operatorname{diag}(1,0,0)O,\qquad
A_2^{(K\Gamma)}=O^T\operatorname{diag}(0,0,1)O,\qquad
A_3^{(K\Gamma)}=O^T\operatorname{diag}(0,1,0)O,
\end{eqnarray}
namely
\begin{eqnarray}
A_1^{(K\Gamma)}=
\begin{pmatrix}
\frac16 & \frac1{2\sqrt3} & -\frac1{3\sqrt2}\\[6pt]
\frac1{2\sqrt3} & \frac12 & -\frac1{\sqrt6}\\[6pt]
-\frac1{3\sqrt2} & -\frac1{\sqrt6} & \frac13
\end{pmatrix},
\quad
A_2^{(K\Gamma)}=
\begin{pmatrix}
\frac16 & -\frac1{2\sqrt3} & -\frac1{3\sqrt2}\\[6pt]
-\frac1{2\sqrt3} & \frac12 & \frac1{\sqrt6}\\[6pt]
-\frac1{3\sqrt2} & \frac1{\sqrt6} & \frac13
\end{pmatrix},
\quad
A_3^{(K\Gamma)}=
\begin{pmatrix}
\frac23 & 0 & \frac{\sqrt2}{3}\\[6pt]
0 & 0 & 0\\[6pt]
\frac{\sqrt2}{3} & 0 & \frac13
\end{pmatrix}.
\end{eqnarray}
Next define
\begin{eqnarray}
B_x=
\begin{pmatrix}
1&0&0\\
0&0&1\\
0&1&0
\end{pmatrix},
\qquad
B_z=
\begin{pmatrix}
0&1&0\\
1&0&0\\
0&0&1
\end{pmatrix},
\qquad
B_y=
\begin{pmatrix}
0&0&1\\
0&1&0\\
1&0&0
\end{pmatrix},
\end{eqnarray}
and
\begin{eqnarray}
A_1^{(J)}=-O^TB_xO,\qquad
A_2^{(J)}=-O^TB_zO,\qquad
A_3^{(J)}=-O^TB_yO,
\end{eqnarray}
so that
\begin{eqnarray}
A_1^{(J)}=
\begin{pmatrix}
\frac12 & -\frac{\sqrt3}{2} & 0\\[6pt]
-\frac{\sqrt3}{2} & -\frac12 & 0\\[6pt]
0&0&-1
\end{pmatrix},
\qquad
A_2^{(J)}=
\begin{pmatrix}
\frac12 & \frac{\sqrt3}{2} & 0\\[6pt]
\frac{\sqrt3}{2} & -\frac12 & 0\\[6pt]
0&0&-1
\end{pmatrix},
\quad
A_3^{(J)}=
\begin{pmatrix}
-1&0&0\\
0&1&0\\
0&0&-1
\end{pmatrix}.
\end{eqnarray}
 
As a result, we have
\begin{eqnarray}
M_{12}&=&\Gamma I+\Delta_\Gamma\operatorname{diag}(1,0,0)-JB_x\nn\\
M_{23}&=&\Gamma I+\Delta_\Gamma\operatorname{diag}(0,0,1)-JB_z\nn\\
M_{34}&=&\Gamma I+\Delta_\Gamma\operatorname{diag}(0,1,0)-JB_y.
\end{eqnarray}
Performing  the $O$ rotation, one obtains
\begin{eqnarray}
O^TM_{12}O&=&\Gamma I+\Delta_\Gamma A_1^{(K\Gamma)}+J A_1^{(J)}\nn\\
O^TM_{23}O&=&\Gamma I+\Delta_\Gamma A_2^{(K\Gamma)}+J A_2^{(J)}\nn\\
O^TM_{34}O&=&\Gamma I+\Delta_\Gamma A_3^{(K\Gamma)}+J A_3^{(J)}.
\end{eqnarray}
Explicitly,
\begin{eqnarray}
O^TM_{12}O&=&
\begin{pmatrix}
\Gamma+\frac{\Delta_\Gamma}{6}+\frac J2 &
\frac{\Delta_\Gamma-3J}{2\sqrt3} &
-\frac{\Delta_\Gamma}{3\sqrt2}
\\[8pt]
\frac{\Delta_\Gamma-3J}{2\sqrt3} &
\Gamma+\frac{\Delta_\Gamma}{2}-\frac J2 &
-\frac{\Delta_\Gamma}{\sqrt6}
\\[8pt]
-\frac{\Delta_\Gamma}{3\sqrt2} &
-\frac{\Delta_\Gamma}{\sqrt6} &
\Gamma+\frac{\Delta_\Gamma}{3}-J
\end{pmatrix},\nn\\
O^TM_{23}O&=&
\begin{pmatrix}
\Gamma+\frac{\Delta_\Gamma}{6}+\frac J2 &
-\frac{\Delta_\Gamma-3J}{2\sqrt3} &
-\frac{\Delta_\Gamma}{3\sqrt2}
\\[8pt]
-\frac{\Delta_\Gamma-3J}{2\sqrt3} &
\Gamma+\frac{\Delta_\Gamma}{2}-\frac J2 &
\frac{\Delta_\Gamma}{\sqrt6}
\\[8pt]
-\frac{\Delta_\Gamma}{3\sqrt2} &
\frac{\Delta_\Gamma}{\sqrt6} &
\Gamma+\frac{\Delta_\Gamma}{3}-J
\end{pmatrix},\nn\\
O^TM_{34}O&=&
\begin{pmatrix}
\Gamma+\frac{2\Delta_\Gamma}{3}-J &
0 &
\frac{\sqrt2}{3}\Delta_\Gamma
\\[8pt]
0 &
\Gamma+J &
0
\\[8pt]
\frac{\sqrt2}{3}\Delta_\Gamma &
0 &
\Gamma+\frac{\Delta_\Gamma}{3}-J
\end{pmatrix}.
\end{eqnarray}

The Hamiltonian in the $OU_6$ frame  can be written as
\begin{eqnarray}
H''=\Gamma\sum_i \vec{S}_i''\cdot \vec{S}_{i+1}''+\Delta H'',
\end{eqnarray}
where
\begin{eqnarray}
\Delta H''=\sum_n\Big[
\vec{S}_{1+3n}^{\prime\prime T}\widehat A_1\vec{S}_{2+3n}''
+
\vec{S}_{2+3n}^{\prime\prime T}\widehat A_2\vec{S}_{3+3n}''
+
\vec{S}_{3+3n}^{\prime\prime T}\widehat A_3\vec{S}_{4+3n}''
\Big],
\end{eqnarray}
and
\begin{eqnarray}
\widehat A_l=\Delta_\Gamma A_l^{(K\Gamma)}+J A_l^{(J)},\qquad l=1,2,3.
\end{eqnarray}
When entering the weak-$U$ fermion RG, the matrices actually used are the rescaled ones
\begin{eqnarray}
\widetilde A_l=\delta_\Gamma A_l^{(K\Gamma)}+\delta_J A_l^{(J)},
\end{eqnarray}
where
\bea
\delta_\Gamma=\frac{\Delta_\Gamma}{2S(S+1)},
\qquad
\delta_J=\frac{J}{2S(S+1)}.
\eea
Here $\widehat A_l$ are the physical coupling matrices in the spin Hamiltonian, whereas $\widetilde A_l$ are the perturbation matrices in the fermion RG.

Denote the uniform and staggered channels by $\eta=\text{u},\text{s}$. 
The source term and the interaction term are taken to be
\begin{eqnarray}
S_h=
-\int d\tau \sum_{l=1}^3\sum_n
\Big[
\vec{h}_{u,l}(\tau,n)\cdot \vec{S}_{u,l+3n}''(\tau)
+
(-)^{l+n}\vec{h}_{s,l}(\tau,n)\cdot \vec{S}_{s,l+3n}''(\tau)
\Big],
\end{eqnarray}
and 
\begin{eqnarray}
S_{\rm int}
=
\int d\tau \sum_{l=1}^3\sum_n
\vec{S}_{l+3n}^{\prime\prime T}\widetilde A_l\,\vec{S}_{l+1+3n}''.
\end{eqnarray}
Here
$\vec{h}_{\eta,l}=
(h_{\eta,l}^x,
h_{\eta,l}^y,
h_{\eta,l}^z)^T$, 
and $l+1$ is understood modulo $3$. 

At one loop, only the cumulant containing one source and one interaction is retained:
\begin{eqnarray}
\delta S_h^{(\eta)}=-\langle S_h^{(\eta)}S_{\rm int}\rangle_>^c.
\end{eqnarray}
After contracting the source with the right leg and the left leg, the correction collected at site $l$ is
\begin{eqnarray}
\delta \vec{h}_{\eta,l}
=
-\frac{2S\ln b}{t}\sum_j
\Big(
\lambda^\eta_{l+1,j}\widetilde A_l
+
\lambda^\eta_{l-1,j}\widetilde A_{l-1}^T
\Big)\vec{h}_{\eta,j}.
\end{eqnarray}
Since both $A_l^{(K\Gamma)}$ and $A_l^{(J)}$ are symmetric matrices, one has $\widetilde A_l^T=\widetilde A_l$, and therefore the RG equation is
\begin{eqnarray}
\frac{d\vec{h}_{\eta,l}}{d\ln b}
=
\vec{h}_{\eta,l}
-
\frac{2S}{t}\sum_j
\Big(
\lambda^\eta_{l+1,j}\widetilde A_l
+
\lambda^\eta_{l-1,j}\widetilde A_{l-1}
\Big)\vec{h}_{\eta,j}.
\end{eqnarray}
To linear order in $O(\delta_\Gamma,\delta_J)$, we have
\begin{eqnarray}
\vec{h}_{\eta,l}(b)
=
b\left[
\vec{h}_{\eta,l}^{(0)}
-
\frac{2S\ln b_s}{t}
\sum_j
\Big(
\lambda^\eta_{l+1,j}\widetilde A_l
+
\lambda^\eta_{l-1,j}\widetilde A_{l-1}
\Big)\vec{h}_{\eta,j}^{(0)}
\right].
\end{eqnarray}

After neglecting the common symmetric overall renormalization, the overall factor is still denoted by $b$. When the RG is integrated to $b_s\sim 3$, the three sublattices are coarse-grained, and we define
\begin{eqnarray}
\vec{h}_\eta=\sum_{l=1}^3\vec{h}_{\eta,l}(b_f)
=\sum_{i=1}^3 M_i^\eta \vec{h}_{\eta,i}^{(0)},
\end{eqnarray}
where
\begin{eqnarray}
M_i^\eta
=
b\left[
I_3
-
\frac{2S\ln b_s}{t}
\sum_{l=1}^3
\Big(
\lambda^\eta_{l+1,i}\widetilde A_l
+
\lambda^\eta_{l-1,i}\widetilde A_{l-1}
\Big)
\right].
\end{eqnarray}

For $i=2$, one has
\begin{eqnarray}
M_2^\eta
=
b\left[
I_3
-
\frac{2S\ln b_s}{t}
\Big(
(\lambda_0^\eta+\lambda_1^\eta)(\widetilde A_1+\widetilde A_2)
+
2\lambda_1^\eta \widetilde A_3
\Big)
\right].
\end{eqnarray}
First compute the matrix combinations:
\begin{eqnarray}
A_1^{(K\Gamma)}+A_2^{(K\Gamma)}
=
\begin{pmatrix}
\frac13 & 0 & -\frac{2}{3\sqrt2}\\[6pt]
0 & 1 & 0\\[6pt]
-\frac{2}{3\sqrt2} & 0 & \frac23
\end{pmatrix},
\end{eqnarray}
thus
\begin{eqnarray}
(\lambda_0^\eta+\lambda_1^\eta)(A_1^{(K\Gamma)}+A_2^{(K\Gamma)})
+2\lambda_1^\eta A_3^{(K\Gamma)}
=
\begin{pmatrix}
\frac{\lambda_0^\eta+5\lambda_1^\eta}{3} & 0 &
-\frac{\sqrt2}{3}(\lambda_0^\eta-\lambda_1^\eta)
\\[8pt]
0 & \lambda_0^\eta+\lambda_1^\eta & 0
\\[8pt]
-\frac{\sqrt2}{3}(\lambda_0^\eta-\lambda_1^\eta) & 0 &
\frac23(\lambda_0^\eta+2\lambda_1^\eta)
\end{pmatrix}.
\end{eqnarray}
One also has
\begin{eqnarray}
A_1^{(J)}+A_2^{(J)}
=
\begin{pmatrix}
1&0&0\\
0&-1&0\\
0&0&-2
\end{pmatrix},
\end{eqnarray}
hence
\begin{eqnarray}
(\lambda_0^\eta+\lambda_1^\eta)(A_1^{(J)}+A_2^{(J)})
+2\lambda_1^\eta A_3^{(J)}
=
\begin{pmatrix}
\lambda_0^\eta-\lambda_1^\eta & 0 & 0
\\[8pt]
0 & -(\lambda_0^\eta-\lambda_1^\eta) & 0
\\[8pt]
0 & 0 & -2(\lambda_0^\eta+2\lambda_1^\eta)
\end{pmatrix}.
\end{eqnarray}
Therefore
\begin{flalign}
&M_2^\eta = b\times \nonumber\\
&
\begin{pmatrix}
1-\frac{2S\ln b_s}{t}
\left[
\frac{\lambda_0^\eta+5\lambda_1^\eta}{3}\,\delta_\Gamma
+
(\lambda_0^\eta-\lambda_1^\eta)\,\delta_J
\right]
&
0
&
\frac{2S\ln b_s}{t}\,
\frac{\sqrt2}{3}(\lambda_0^\eta-\lambda_1^\eta)\,\delta_\Gamma
\\[10pt]
0
&
1-\frac{2S\ln b_s}{t}
\left[
(\lambda_0^\eta+\lambda_1^\eta)\,\delta_\Gamma
-
(\lambda_0^\eta-\lambda_1^\eta)\,\delta_J
\right]
&
0
\\[10pt]
\frac{2S\ln b_s}{t}\,
\frac{\sqrt2}{3}(\lambda_0^\eta-\lambda_1^\eta)\,\delta_\Gamma
&
0
&
1-\frac{2S\ln b_s}{t}
(\lambda_0^\eta+2\lambda_1^\eta)
\left(
\frac23 \delta_\Gamma-2 \delta_J
\right)
\end{pmatrix}.
\end{flalign}

The bosonization coefficients are givenin Eq. (\ref{eq:E2_F2}).
Since $M_2^\eta$ is symmetric,
\begin{eqnarray}
E_2=(M_2^s)^T=M_2^s,\qquad
F_2=(M_2^u)^T=M_2^u.
\end{eqnarray}
Thus 
\begin{eqnarray}
\lambda_\eta
&=&
b\left[
1-
\frac{2S\ln b_s}{t}
\left(
\frac{\lambda_0^\eta+5\lambda_1^\eta}{3}\,\delta_\Gamma
+
(\lambda_0^\eta-\lambda_1^\eta)\,\delta_J
\right)
\right],\nn\\
\delta_\eta
&=&
-\frac{2Sb\ln b_s}{t}\,
\frac23(\lambda_0^\eta-\lambda_1^\eta)(\delta_\Gamma-3\delta_J),\nn\\
\nu_\eta
&=&
b\left[
1-
\frac{2S\ln b_s}{t}
(\lambda_0^\eta+2\lambda_1^\eta)
\left(
\frac23 \delta_\Gamma-2\delta_J
\right)
\right],\nn\\
\sigma_\eta&=&\rho_\eta
=
\frac{2Sb\ln b_s}{t}\,
\frac{\sqrt2}{3}(\lambda_0^\eta-\lambda_1^\eta)\,\delta_\Gamma,
\end{eqnarray}
where ``$C$" for $\eta=\text{s}$ and ``$D$" for $\eta=\text{u}$.

Plugging in the values of $\lambda_0^\eta,\lambda_1^\eta$ ($\eta=\text{u}, \text{s}$), we obtain
\begin{eqnarray}
\lambda_C
&=&
b\left[
1-
\frac{2S\ln b_s}{t}
\left(
\frac{0.26}{3}\,\delta_\Gamma
-0.10\,\delta_J
\right)
\right],\nn\\
\delta_C
&=&
\frac{2Sb\ln b_s}{t}
\left(\frac{2}{3}\times 0.10\right)
(\delta_\Gamma-3\delta_J),\nn\\
\nu_C
&=&
b\left[
1-
\frac{2S\ln b_s}{t}\,
0.08
\left(
\frac23 \delta_\Gamma-2\delta_J
\right)
\right],\nn\\
\sigma_C&=&\rho_C
=
-\frac{2Sb\ln b_s}{t}\,
\frac{0.10\sqrt2}{3}\,\delta_\Gamma,
\end{eqnarray}
and
\begin{eqnarray}
\lambda_D
&=&
b\left[
1+
\frac{2S\ln b_s}{t}
\left(
0.07\,\delta_\Gamma
-0.21\,\delta_J
\right)
\right],\nn\\
\delta_D
&=&
-\frac{2Sb\ln b_s}{t}\,
0.14\,(\delta_\Gamma-3\delta_J),\nn\\
\nu_D&=&b,\nn\\
\sigma_D&=&\rho_D
=
\frac{2Sb\ln b_s}{t}\,
\frac{0.21\sqrt2}{3}\,\delta_\Gamma.
\end{eqnarray}
Further using
$\delta_\Gamma=\frac{\Delta_\Gamma}{2S(S+1)}$,
$\delta_J=\frac{J}{2S(S+1)}$,
$t=\frac{\Gamma}{\pi}$,
we arrive at Eq. (\ref{eq:U1_boson_C_expr_b}) and  Eq. (\ref{eq:U1_boson_D_expr_b}).

\section{Bosonization formulas for Kitaev-Heisenberg-Gamma model}

\subsection{Explicit forms of bosonization formulas}
The nonsymmorphic bosonization formulas in the $OU_6$ frame are given by
\bea
S_{1+3n}^{\prime\prime x}&=&
\left[
\left(\lambda_D+\frac{3}{4}\delta_D\right)\cos(q_{\rm ic}(1+3n))
-\frac{\sqrt{3}}{4}\delta_D\sin(q_{\rm ic}(1+3n))
\right]\mathcal{J}^x
\nn\\
&&+
\left[
\left(\lambda_D+\frac{3}{4}\delta_D\right)\sin(q_{\rm ic}(1+3n))
+\frac{\sqrt{3}}{4}\delta_D\cos(q_{\rm ic}(1+3n))
\right]\mathcal{J}^y
-\frac{1}{2}\rho_D\mathcal{J}^z
\nn\\
&&+(-)^{1+n}\bigg\{
\left[
\left(\lambda_C+\frac{3}{4}\delta_C\right)\cos(q_{\rm ic}(1+3n))
-\frac{\sqrt{3}}{4}\delta_C\sin(q_{\rm ic}(1+3n))
\right]\mathcal{N}^x
\nn\\
&&\hspace{1.7cm}+
\left[
\left(\lambda_C+\frac{3}{4}\delta_C\right)\sin(q_{\rm ic}(1+3n))
+\frac{\sqrt{3}}{4}\delta_C\cos(q_{\rm ic}(1+3n))
\right]\mathcal{N}^y
-\frac{1}{2}\rho_C\mathcal{N}^z
\bigg\},
\nn\\
S_{1+3n}^{\prime\prime y}&=&
\left[
\frac{\sqrt{3}}{4}\delta_D\cos(q_{\rm ic}(1+3n))
-\left(\lambda_D+\frac{1}{4}\delta_D\right)\sin(q_{\rm ic}(1+3n))
\right]\mathcal{J}^x
\nn\\
&&+
\left[
\frac{\sqrt{3}}{4}\delta_D\sin(q_{\rm ic}(1+3n))
+\left(\lambda_D+\frac{1}{4}\delta_D\right)\cos(q_{\rm ic}(1+3n))
\right]\mathcal{J}^y
+\frac{\sqrt{3}}{2}\rho_D\mathcal{J}^z
\nn\\
&&+(-)^{1+n}\bigg\{
\left[
\frac{\sqrt{3}}{4}\delta_C\cos(q_{\rm ic}(1+3n))
-\left(\lambda_C+\frac{1}{4}\delta_C\right)\sin(q_{\rm ic}(1+3n))
\right]\mathcal{N}^x
\nn\\
&&\hspace{1.7cm}+
\left[
\frac{\sqrt{3}}{4}\delta_C\sin(q_{\rm ic}(1+3n))
+\left(\lambda_C+\frac{1}{4}\delta_C\right)\cos(q_{\rm ic}(1+3n))
\right]\mathcal{N}^y
+\frac{\sqrt{3}}{2}\rho_C\mathcal{N}^z
\bigg\},
\nn\\
S_{1+3n}^{\prime\prime z}&=&
\left[
-\frac{1}{2}\sigma_D\cos(q_{\rm ic}(1+3n))
-\frac{\sqrt{3}}{2}\sigma_D\sin(q_{\rm ic}(1+3n))
\right]\mathcal{J}^x
\nn\\
&&+
\left[
-\frac{1}{2}\sigma_D\sin(q_{\rm ic}(1+3n))
+\frac{\sqrt{3}}{2}\sigma_D\cos(q_{\rm ic}(1+3n))
\right]\mathcal{J}^y
+\nu_D\mathcal{J}^z
\nn\\
&&+(-)^{1+n}\bigg\{
\left[
-\frac{1}{2}\sigma_C\cos(q_{\rm ic}(1+3n))
-\frac{\sqrt{3}}{2}\sigma_C\sin(q_{\rm ic}(1+3n))
\right]\mathcal{N}^x
\nn\\
&&\hspace{1.7cm}+
\left[
-\frac{1}{2}\sigma_C\sin(q_{\rm ic}(1+3n))
+\frac{\sqrt{3}}{2}\sigma_C\cos(q_{\rm ic}(1+3n))
\right]\mathcal{N}^y
+\nu_C\mathcal{N}^z
\bigg\}.
\label{eq:bosonize_LL1_S1_a}
\eea

\bea
S_{2+3n}^{\prime\prime x}&=&
\lambda_D\cos(q_{\rm ic}(2+3n))\mathcal{J}^x
+\lambda_D\sin(q_{\rm ic}(2+3n))\mathcal{J}^y
+\rho_D\mathcal{J}^z
\nn\\
&&+(-)^n\big[
\lambda_C\cos(q_{\rm ic}(2+3n))\mathcal{N}^x
+\lambda_C\sin(q_{\rm ic}(2+3n))\mathcal{N}^y
+\rho_C\mathcal{N}^z
\big],
\nn\\
S_{2+3n}^{\prime\prime y}&=&
-(\lambda_D+\delta_D)\sin(q_{\rm ic}(2+3n))\mathcal{J}^x
+(\lambda_D+\delta_D)\cos(q_{\rm ic}(2+3n))\mathcal{J}^y
\nn\\
&&+(-)^n\big[
-(\lambda_C+\delta_C)\sin(q_{\rm ic}(2+3n))\mathcal{N}^x
+(\lambda_C+\delta_C)\cos(q_{\rm ic}(2+3n))\mathcal{N}^y
\big],
\nn\\
S_{2+3n}^{\prime\prime z}&=&
\sigma_D\cos(q_{\rm ic}(2+3n))\mathcal{J}^x
+\sigma_D\sin(q_{\rm ic}(2+3n))\mathcal{J}^y
+\nu_D\mathcal{J}^z
\nn\\
&&+(-)^n\big[
\sigma_C\cos(q_{\rm ic}(2+3n))\mathcal{N}^x
+\sigma_C\sin(q_{\rm ic}(2+3n))\mathcal{N}^y
+\nu_C\mathcal{N}^z
\big].
\label{eq:bosonize_LL1_S2_a}
\eea

\bea
S_{3+3n}^{\prime\prime x}&=&
\left[
\left(\lambda_D+\frac{3}{4}\delta_D\right)\cos(q_{\rm ic}(3+3n))
+\frac{\sqrt{3}}{4}\delta_D\sin(q_{\rm ic}(3+3n))
\right]\mathcal{J}^x
\nn\\
&&+
\left[
\left(\lambda_D+\frac{3}{4}\delta_D\right)\sin(q_{\rm ic}(3+3n))
-\frac{\sqrt{3}}{4}\delta_D\cos(q_{\rm ic}(3+3n))
\right]\mathcal{J}^y
-\frac{1}{2}\rho_D\mathcal{J}^z
\nn\\
&&+(-)^{1+n}\bigg\{
\left[
\left(\lambda_C+\frac{3}{4}\delta_C\right)\cos(q_{\rm ic}(3+3n))
+\frac{\sqrt{3}}{4}\delta_C\sin(q_{\rm ic}(3+3n))
\right]\mathcal{N}^x
\nn\\
&&\hspace{1.7cm}+
\left[
\left(\lambda_C+\frac{3}{4}\delta_C\right)\sin(q_{\rm ic}(3+3n))
-\frac{\sqrt{3}}{4}\delta_C\cos(q_{\rm ic}(3+3n))
\right]\mathcal{N}^y
-\frac{1}{2}\rho_C\mathcal{N}^z
\bigg\},
\nn\\
S_{3+3n}^{\prime\prime y}&=&
\left[
-\frac{\sqrt{3}}{4}\delta_D\cos(q_{\rm ic}(3+3n))
-\left(\lambda_D+\frac{1}{4}\delta_D\right)\sin(q_{\rm ic}(3+3n))
\right]\mathcal{J}^x
\nn\\
&&+
\left[
-\frac{\sqrt{3}}{4}\delta_D\sin(q_{\rm ic}(3+3n))
+\left(\lambda_D+\frac{1}{4}\delta_D\right)\cos(q_{\rm ic}(3+3n))
\right]\mathcal{J}^y
-\frac{\sqrt{3}}{2}\rho_D\mathcal{J}^z
\nn\\
&&+(-)^{1+n}\bigg\{
\left[
-\frac{\sqrt{3}}{4}\delta_C\cos(q_{\rm ic}(3+3n))
-\left(\lambda_C+\frac{1}{4}\delta_C\right)\sin(q_{\rm ic}(3+3n))
\right]\mathcal{N}^x
\nn\\
&&\hspace{1.7cm}+
\left[
-\frac{\sqrt{3}}{4}\delta_C\sin(q_{\rm ic}(3+3n))
+\left(\lambda_C+\frac{1}{4}\delta_C\right)\cos(q_{\rm ic}(3+3n))
\right]\mathcal{N}^y
-\frac{\sqrt{3}}{2}\rho_C\mathcal{N}^z
\bigg\},
\nn\\
S_{3+3n}^{\prime\prime z}&=&
\left[
-\frac{1}{2}\sigma_D\cos(q_{\rm ic}(3+3n))
+\frac{\sqrt{3}}{2}\sigma_D\sin(q_{\rm ic}(3+3n))
\right]\mathcal{J}^x
\nn\\
&&+
\left[
-\frac{1}{2}\sigma_D\sin(q_{\rm ic}(3+3n))
-\frac{\sqrt{3}}{2}\sigma_D\cos(q_{\rm ic}(3+3n))
\right]\mathcal{J}^y
+\nu_D\mathcal{J}^z
\nn\\
&&+(-)^{1+n}\bigg\{
\left[
-\frac{1}{2}\sigma_C\cos(q_{\rm ic}(3+3n))
+\frac{\sqrt{3}}{2}\sigma_C\sin(q_{\rm ic}(3+3n))
\right]\mathcal{N}^x
\nn\\
&&\hspace{1.7cm}+
\left[
-\frac{1}{2}\sigma_C\sin(q_{\rm ic}(3+3n))
-\frac{\sqrt{3}}{2}\sigma_C\cos(q_{\rm ic}(3+3n))
\right]\mathcal{N}^y
+\nu_C\mathcal{N}^z
\bigg\}.
\label{eq:bosonize_LL1_S3_a}
\eea

\subsection{Reducing to the $J=0$ case}
\label{app:reducing_Oh}

Notice that the bosonization fields $\mathcal{J}^\alpha,\mathcal{N}^\alpha$ ($\alpha=x,y,z$) are defined in the $OU_6$ frame.
Let's rotate $\mathcal{J}^\alpha,\mathcal{N}^\alpha$ to the $U_6$ frame,
by defining $J^{\prime\alpha},N^{\prime\alpha}$ as
\bea
(\mathcal{J}^{x}\,\mathcal{J}^{y}\,\mathcal{J}^{z})
&=&
(J^{\prime x}\,J^{\prime y}\,J^{\prime z})O,\nn\\
(\mathcal{N}^{x}\,\mathcal{N}^{y}\,\mathcal{N}^{z})
&=&
(N^{\prime x}\,N^{\prime y}\,N^{\prime z})O,
\eea
where the matrix $O$ is defined in Eq. (\ref{eq:def_O_mat}).
Then we have
\bea
(S_{1+3n}^{\prime x}\,S_{1+3n}^{\prime y}\,S_{1+3n}^{\prime z})
&=&
(J^{\prime x}\,J^{\prime y}\,J^{\prime z})O\mathcal{D}_1O^{-1}
-(-)^{n}(N^{\prime x}\,N^{\prime y}\,N^{\prime z})O\mathcal{C}_1O^{-1},\nn\\
(S_{2+3n}^{\prime x}\,S_{2+3n}^{\prime y}\,S_{2+3n}^{\prime z})
&=&
(J^{\prime x}\,J^{\prime y}\,J^{\prime z})O\mathcal{D}_2O^{-1}
+(-)^{n}(N^{\prime x}\,N^{\prime y}\,N^{\prime z})O\mathcal{C}_2O^{-1},\nn\\
(S_{3+3n}^{\prime x}\,S_{3+3n}^{\prime y}\,S_{3+3n}^{\prime z})
&=&
(J^{\prime x}\,J^{\prime y}\,J^{\prime z})O\mathcal{D}_3O^{-1}
-(-)^{n}(N^{\prime x}\,N^{\prime y}\,N^{\prime z})O\mathcal{C}_3O^{-1}.
\eea
The structures of the $O\Lambda_iO^{-1}$ ($\Lambda=C,D$, $i=1,2,3$) matrices are
\begin{flalign}
O\Lambda_1O^{-1}=\left(\begin{array}{ccc}
\frac{1}{6}[\lambda+3\mu+2\nu-\sqrt{2}(\rho+\sigma)]
&
\frac{1}{6}[\lambda-3\mu+2\nu-\sqrt{2}(\rho+\sigma)]
&
\frac{1}{6}[-2\lambda+2\nu+\sqrt{2}(2\rho-\sigma)]
\\
\frac{1}{6}[\lambda-3\mu+2\nu-\sqrt{2}(\rho+\sigma)]
&
\frac{1}{6}[\lambda+3\mu+2\nu-\sqrt{2}(\rho+\sigma)]
&
\frac{1}{6}[-2\lambda+2\nu+\sqrt{2}(2\rho-\sigma)]
\\
\frac{1}{6}[-2\lambda+2\nu-\sqrt{2}(\rho-2\sigma)]
&
\frac{1}{6}[-2\lambda+2\nu-\sqrt{2}(\rho-2\sigma)]
&
\frac{1}{6}[2\lambda+\nu+\sqrt{2}(\rho+\sigma)]
\end{array}\right),\nn\\
\end{flalign}
\begin{flalign}
O\Lambda_2O^{-1}=\left(\begin{array}{ccc}
\frac{1}{6}[\lambda+3\mu+2\nu-\sqrt{2}(\rho+\sigma)]
&
\frac{1}{6}[-2\lambda+2\nu+\sqrt{2}(2\rho-\sigma)]
&
\frac{1}{6}[\lambda-3\mu+2\nu-\sqrt{2}(\rho+\sigma)]
\\
\frac{1}{6}[-2\lambda+2\nu-\sqrt{2}(\rho-2\sigma)]
&
\frac{1}{6}[2\lambda+\nu+\sqrt{2}(\rho+\sigma)]
&
\frac{1}{6}[-2\lambda+2\nu-\sqrt{2}(\rho-2\sigma)]
\\
\frac{1}{6}[\lambda-3\mu+2\nu-\sqrt{2}(\rho+\sigma)]
&
\frac{1}{6}[-2\lambda+2\nu+\sqrt{2}(2\rho-\sigma)]
&
\frac{1}{6}[\lambda+3\mu+2\nu-\sqrt{2}(\rho+\sigma)]
\end{array}\right),\nn\\
\end{flalign}
\begin{flalign}
O\Lambda_3O^{-1}=\left(\begin{array}{ccc}
\frac{1}{6}[2\lambda+\nu+\sqrt{2}(\rho+\sigma)]
&
\frac{1}{6}[-2\lambda+2\nu-\sqrt{2}(\rho-2\sigma)]
&
\frac{1}{6}[-2\lambda+2\nu-\sqrt{2}(\rho-2\sigma)]
\\
\frac{1}{6}[-2\lambda+2\nu+\sqrt{2}(2\rho-\sigma)]
&
\frac{1}{6}[\lambda+3\mu+2\nu-\sqrt{2}(\rho+\sigma)]
&
\frac{1}{6}[\lambda-3\mu+2\nu-\sqrt{2}(\rho+\sigma)]
\\
\frac{1}{6}[-2\lambda+2\nu+\sqrt{2}(2\rho-\sigma)]
&
\frac{1}{6}[\lambda-3\mu+2\nu-\sqrt{2}(\rho+\sigma)]
&
\frac{1}{6}[\lambda+3\mu+2\nu-\sqrt{2}(\rho+\sigma)]
\end{array}\right),
\end{flalign}
in which $\mu_\Lambda$ is defined as
\bea
\mu_\Lambda=\lambda_\Lambda+\delta_\Lambda,
\eea
and the subscripts $\Lambda=C,D$ in the coefficients $\lambda_\Lambda$, $\nu_\Lambda$, $\rho_\Lambda$, $\delta_\Lambda$, and $\sigma_\Lambda$ are dropped for simplifications of notations.

When $J=0$, by requiring the $R(\hat{\alpha}^\prime,\pi)$ ($\alpha=x,y,z$) symmetries, we obtain ($\Lambda=C,D$)
\bea
\lambda_\Lambda&=&\nu_\Lambda+\frac{1}{\sqrt{2}}\rho_\Lambda,\nn\\
\delta_\Lambda&=&-\sqrt{2}\rho_\Lambda,\nn\\
\sigma_\Lambda&=&\rho_\Lambda.
\eea
Then we get
\bea
O\Lambda_1O^{-1}\rightarrow\left(\begin{array}{ccc}
\nu_\Lambda-\frac{1}{\sqrt{2}}\rho_\Lambda&0&0\\
0&\nu_\Lambda-\frac{1}{\sqrt{2}}\rho_\Lambda&0\\
0&0&\nu_\Lambda+\sqrt{2}\rho_\Lambda
\end{array}\right),
\eea
\bea
O\Lambda_2O^{-1}\rightarrow\left(\begin{array}{ccc}
\nu_\Lambda-\frac{1}{\sqrt{2}}\rho_\Lambda&0&0\\
0&\nu_\Lambda+\sqrt{2}\rho_\Lambda&0\\
0&0&\nu_\Lambda-\frac{1}{\sqrt{2}}\rho_\Lambda
\end{array}\right),
\eea
\bea
O\Lambda_3O^{-1}\rightarrow\left(\begin{array}{ccc}
\nu_\Lambda+\sqrt{2}\rho_\Lambda&0&0\\
0&\nu_\Lambda-\frac{1}{\sqrt{2}}\rho_\Lambda&0\\
0&0&\nu_\Lambda-\frac{1}{\sqrt{2}}\rho_\Lambda
\end{array}\right).
\eea
Comparing the above expressions with Eq. (\ref{S_c}), it can be verified that the parameters $C_j$, $D_j$ ($j=1,2$) are related to $\nu_\Lambda$, $\rho_\Lambda$ ($\Lambda=C,D$) by Eq. (\ref{eq:relation_CD_nurho}).

\end{widetext}



\end{document}